\documentclass[a4paper,fleqn]{cas-dc}

\usepackage[authoryear]{natbib}

\graphicspath{{Figures/}}

\usepackage{setspace}
\usepackage[OT1]{fontenc}
\usepackage{hyperref}

\def\tsc#1{\csdef{#1}{\textsc{\lowercase{#1}}\xspace}}
\tsc{WGM}
\tsc{QE}

\begin{document}
\let\WriteBookmarks\relax
\def\floatpagepagefraction{1}
\def\textpagefraction{.001}

\shorttitle{Thermal and magnetic evolution of an Earth-like planet with a basal magma ocean}

\shortauthors{V. Lherm, M. Nakajima, E.G. Blackman}

\title [mode = title]{Thermal and magnetic evolution of an Earth-like planet with a basal magma ocean}



%

\author[1,2]{Victor Lherm}[orcid=0000-0001-5814-0637]

\cormark[1]


\ead{vlherm@ur.rochester.edu}


\credit{Conceptualization, Methodology, Software, Formal analysis, Writing - Original Draft}

\affiliation[1]{organization={Department of Earth and Environmental Sciences, University of Rochester},
    addressline={227 Hutchison Hall},
    city={Rochester},
    citysep={}, 
    postcode={14627},
    state={NY},
    country={USA}}

\affiliation[2]{organization={Department of Physics and Astronomy, University of Rochester},
    addressline={206 Bausch \& Lomb Hall},
    city={Rochester},
    citysep={}, 
    postcode={14627},
    state={NY},
    country={USA}}

\author[1,2]{Miki Nakajima}[orcid=0000-0001-5014-0448]


\ead{mnakajima@rochester.edu}


\credit{Conceptualization, Methodology, Formal analysis, Writing - Original Draft}

\author[2,3]{Eric G. Blackman}[orcid=0000-0002-9405-8435]


\ead{blackman@pas.rochester.edu}


\credit{Conceptualization, Methodology, Formal analysis, Writing - Original Draft}

\affiliation[3]{organization={Laboratory for Laser Energetics, University of Rochester},
    addressline={250 E River Rd},
    city={Rochester},
    citysep={}, 
    postcode={14623},
    state={NY},
    country={USA}}

\cortext[1]{Corresponding author}



\begin{abstract}
    Earth's geodynamo has operated for over 3.5 billion years. The magnetic field is currently powered by thermocompositional convection in the outer core, which involves the release of light elements and latent heat as the inner core solidifies. However, since the inner core nucleated no more than 1.5 billion years ago, the early dynamo could not rely on these buoyancy sources. Given recent estimates of the thermal conductivity of the outer core, an alternative mechanism may be required to sustain the geodynamo prior to nucleation of the inner core. One possibility is a silicate dynamo operating in a long-lived basal magma ocean.
    Here, we investigate the structural, thermal, buoyancy, and magnetic evolution of an Earth-like terrestrial planet. Using modern equations of state and melting curves, we include a time-dependent parameterization of the compositional evolution of an iron-rich basal magma ocean. We combine an internal structure integration of the planet with energy budgets in a coupled core, basal magma ocean, and mantle system. We determine the thermocompositional convective stability of the core and the basal magma ocean, and assess their respective dynamo activity using entropy budgets and magnetic Reynolds numbers.
    Our conservative nominal model predicts a transient basal magma ocean dynamo followed by a core dynamo after 1 billion years. The model is sensitive to several parameters, including the initial temperature of the core-mantle boundary, the parameterization of mantle convection, the composition of the basal magma ocean, the radiogenic content of the planet, as well as convective velocity and magnetic scaling laws. We use the nominal model to constrain the range of basal magma ocean electrical conductivity and core thermal conductivity that sustain a dynamo.
    This highlights the importance of constraining the parameters and transport properties that influence planetary evolution using experiments and simulations conducted at pressure, temperature, and composition conditions found in planetary interior, in order to reduce model degeneracies.
\end{abstract}


\maketitle

\section{Introduction}
\label{sec:introduction}

How planetary magnetic fields may protect planetary atmospheres from erosion by stellar winds is an active area of investigation \citep{lundin_2007,tarduno_2010,blackman_2018,gunell_2018,egan_2019}. In turn, magnetic fields may also influence the emergence of life and habitability \citep{tarduno_2014,camprubi_2019}.
Paleomagnetic evidence indicates the presence of a geomagnetic field for over 3.5 billion years (Gyr), potentially extending to 4.2 Gyr \citep{tarduno_2010,biggin_2011,tarduno_2015,borlina_2020,tarduno_2020,bono_2022,taylor_2023}.

The geomagnetic field is currently powered by thermal and compositional convection in the outer core, including the release of light elements at the inner core boundary (ICB) upon crystallization of the inner core \citep{jones_2015}. Since inner core nucleation depends on many factors, including the thermal evolution model of the planet, the current heat flow at the core-mantle boundary (CMB), and the radiogenic content of the core, the age of inner core formation is debated, but is typically less than 1.5 billion years ago (Ga) \citep{biggin_2015,labrosse_2015,nimmo_2015a}. Therefore, an early geodynamo could not rely on buoyancy sources produced by the release of latent heat and light elements at the ICB.

Another key parameter that controls whether the outer core can sustain an early dynamo is the thermal conductivity of the core \citep{williams_2018}. If the thermal conductivity is excessively large, the heat of the core would be transferred by conduction rather than convection, leaving insufficient energy to power a dynamo.
The thermal conductivity of the core was initially estimated in the range $20-50~\mathrm{W.m^{-1}.K^{-1}}$ \citep{stacey_2001,stacey_2007}, with updated investigations confirming the low values \citep{basu_2020,hsieh_2020}. Recent experiments and simulations indicate values between $70-250~\mathrm{W.m^{-1}.K^{-1}}$ \citep{dekoker_2012,ohta_2016,pozzo_2012,pozzo_2022}. The conflict between the upward revision of the thermal conductivity and the existence of a geomagnetic field throughout most of Earth's history is called the "new core paradox" \citep{olson_2013}.

A dynamo mechanism independent of core crystallization may be necessary to solve the "new core paradox" and sustain an early geomagnetic field \citep{landeau_2022}. The exsolution of light elements in the core, \textit{e.g.} the precipitation of $\mathrm{MgO}$ or $\mathrm{SiO_2}$ due to a decrease in solubility, could produce compositional convection and facilitate an early dynamo \citep{orourke_2016,hirose_2017,badro_2018}. However, the core must cool relatively quickly for this buoyancy source to overcome any early thermal stratification \citep{du_2017,du_2019}.
Mechanical forcing involving precession \citep{malkus_1968,stacey_1973,loper_1975,rochester_1975,tilgner_2005} and tides \citep{lebars_2015,grannan_2017,reddy_2018} can also produce turbulent flows that could help generate an early magnetic field. Although tidal flows could have contributed to an early dynamo before 1.5 Ga, when the Earth's rotation rate and tidal deformation were higher, flows driven by precession seem too weak to drive the geodynamo \citep{landeau_2022}.
Radioactive heating, rapid cooling, a colder solidus, a shallower core adiabat \citep{nimmo_2004} and a high initial core temperature  \citep{driscoll_2023} may also have powered an early dynamo.

Alternatively, the early geomagnetic field may have been powered by a dynamo in a layer of molten silicates above the CMB \citep{labrosse_2007,ziegler_2013,stixrude_2020}, known as a basal magma ocean (BMO).
Several seismic and geochemical observations may be interpreted as remnants of an early BMO on Earth \citep{labrosse_2015a}. This includes seismic heterogeneities above the CMB \citep{garnero_2008,lay_2015}, including Large Low Shear Velocity Provinces (LLSVP) \citep{dziewonski_1977,lekic_2012} and Ultra-Low Velocity Zones (ULVZ) \citep{garnero_1993,thorne_2004,ferrick_2023}, as well as isotopic signatures showing distinct geochemical reservoirs formed early in Earth's history \citep{boyet_2005,mukhopadhyay_2012,mundl_2017}.

A primitive BMO could have been produced through several scenarios \citep{labrosse_2015a}.
At the end of accretion, a giant impact may form a global magma ocean, which could crystallize downwards and form a BMO if the intersection between the adiabatic temperature profile and the mantle liquidus occurs at mid-depth \citep{labrosse_2007}. Even if the intersection between the adiabat and the liquidus occurs at the CMB, crystals formed out of the magma ocean are expected to be less dense than the melt. They should migrate upwards, leaving behind a denser iron-rich melt above the CMB \citep{boukare_2015,nabiei_2021}.
Alternatively, if the impact only melts a fraction of the mantle \citep{tonks_1993,nakajima_2021}, a dense metal silicate emulsion could accumulate at the bottom of a local magma ocean \citep{rubie_2015}. The destabilization of this dense pond by diapirism \citep{karato_1997,samuel_2010} or diking \citep{rubin_1995,stevenson_2003} could then entrain molten silicates toward the core \citep{olson_2008}, possibly forming a BMO.
Finally, depending on the mechanisms of metal-silicate equilibration during \citep{kendall_2016,landeau_2021,lherm_2022,lherm_2023} and after \citep{rubie_2003,deguen_2014,lherm_2018} a planetary impact, the heat from core formation could be preferentially partitioned into the core and super-heat it. This could induce mantle melting above the CMB, resulting in a BMO.

After the formation of the BMO, iron, which is incompatible, is preferentially partitioned in the liquid as crystallization proceeds \citep{nomura_2011,andrault_2012,nabiei_2021}. Fractional crystallization is therefore responsible for the progressive enrichment of the BMO in iron, independently of the formation scenario. The density of the melt necessarily increases over time, leading to a gravitationally stable, long-lived BMO \citep{labrosse_2007}.

In order to understand planetary evolution, which includes processes operating at multiple scales, various models are required. For example, 3D models can capture turbulence, rotation, and dynamo effects in convective layers \citep{christensen_2015}, 2D models can assess the role of solid-liquid multi-phase flow on the crystallisation of the BMO \citep{boukare_2017}, and fully 1D models can determine the thermal evolution of the planet using self-consistent radial distributions of thermodynamic and transport properties \citep{zhang_2022}. Although these models accurately capture a variety of processes, they are computationally expensive. Alternatively, planetary evolution can be assessed using semi-analytical "box models" involving energy and entropy budgets  \citep{nimmo_2015} averaged over each "box" (\textit{e.g.} core, BMO, and mantle).

In an adiabatic (isentropic) and well-mixed layer such as a liquid core or a BMO, that involves a rapidly rotating, vigorously convective and electrically conductive fluid, the production of a self-sustained magnetic field requires that (i) thermocompositional convection supplies sufficient power to balance ohmic dissipation \citep{buffett_1996,lister_2003} and that (ii) magnetic induction is significantly larger than magnetic diffusion \citep{christensen_2006}. The power available to sustain a dynamo can be obtained by calculating energy and entropy budgets, while the induction criterion is assessed with a magnetic Reynolds number.

The theoretical framework governing energy and entropy budgets was originally developed for the Earth's core \citep{verhoogen_1961,backus_1975,hewitt_1975,loper_1978,braginsky_1995,lister_1995,buffett_1996,gubbins_2003,gubbins_2004,nimmo_2015} to assess its thermal and magnetic evolution. In particular, previous work investigated the age of the Earth's inner core \citep{labrosse_1997,labrosse_2001}, the influence of radiogenic heating \citep{labrosse_2003,nimmo_2004}, the effect of light elements exsolution \citep{orourke_2016,orourke_2017,mittal_2020}, the influence of a high thermal conductivities \citep{pozzo_2012,gomi_2013,davies_2015,labrosse_2015,driscoll_2023}, the role of stably-stratified layers in the core \citep{gomi_2013,laneuville_2018,greenwood_2021} and the magnetic evolution of super-Earth exoplanets \citep{blaske_2021,bonati_2021}.
In addition to planetary cores, energy and entropy budgets have also been used to determine the thermal and magnetic evolution of a BMO on Earth \citep{labrosse_2007,ziegler_2013,laneuville_2018,blanc_2020,stixrude_2020,ferrick_2023}, Venus \citep{orourke_2020} and the Moon \citep{scheinberg_2018,hamid_2023}.

The magnetic Reynolds number of a convective and electrically conductive layer depends on the system length scale $L$, the typical convective velocity $U$, the magnetic constant $\mu_0$ and the electrical conductivity $\sigma$
\begin{equation}
    R_m=LU\mu_0\sigma.
    \label{eq:magnetic_reynolds_number_intro}
\end{equation}
When magnetic induction is significantly larger than magnetic diffusion, \textit{i.e.} $R_m \gg 1$, a dynamo can operate and sustain a magnetic field. In practice, planetary dynamo simulations suggest $R_m = 40$ as a critical value \citep{christensen_2006} for large scale fields, consistent with kinematic helical convection dynamo calculations in a sphere \citep{chen_2018}.

In contrast to the core, where the dynamo is limited by the thermal conductivity, dynamo operation in a BMO is critically limited by the low electrical conductivity of the silicates. Assuming $L=300~\mathrm{km}$ and $U=1~\mathrm{cm.s^{-1}}$ \citep{ziegler_2013,stixrude_2020}, the electrical conductivity required to reach the critical magnetic Reynolds number is typically $\sigma=10^4~\mathrm{S.m^{-1}}$. This exceeds the electrical conductivity of silicate melts at low pressure and temperature by at least two orders of magnitude \citep{ni_2015}. However, high pressure and temperature conditions \citep{scipioni_2017}, as well as the addition of metal, such as $\mathrm{Fe}$ and $\mathrm{Ni}$ \citep{holmstrom_2018,soubiran_2018,stixrude_2020}, increase significantly the electrical conductivity of the silicates, potentially allowing a BMO to host a dynamo.

In this paper, we investigate the structural, thermal, buoyancy and magnetic evolution of a planet including a coupled solid mantle, a long-lived BMO and a solid and/or liquid core.
Our goal is to obtain an evolution scenario for an Earth-like planet with a BMO and to assess the sensitivity of the model to key parameters.
First, we establish an Earth-like nominal model using conservative parameters. We focus on the time evolution of the size of the BMO and the liquid core, as well as temperatures, compositions, heat flows, thermocompositional convective stability, rates of entropy production, magnetic Reynolds numbers and magnetic field intensities involved in these layers.
We then analyse the sensitivity of this nominal model to specific parameters, including the initial temperature at the CMB, mantle convection parameterization, the composition of the BMO and the radiogenic content of the planet. We also investigate the influence of the thermal and electrical conductivities on the evolution of the planet, by constraining the existence and efficiency of dynamos hosted in the BMO and the core.

\section{Model}
\label{sec:model}

We model the structural, thermal, buoyancy and magnetic evolution of an Earth-like planet with a mass $M_P=5.972 \times 10^{24}~\mathrm{kg}$ and a core mass fraction $\mathrm{CMF}=0.32$. The variables of the model are summarized in Table \ref{tab:variables} and the parameters are defined in Tables S1-S8.

\begin{figure}[!ht]
    \centering
    \includegraphics[width=\linewidth]{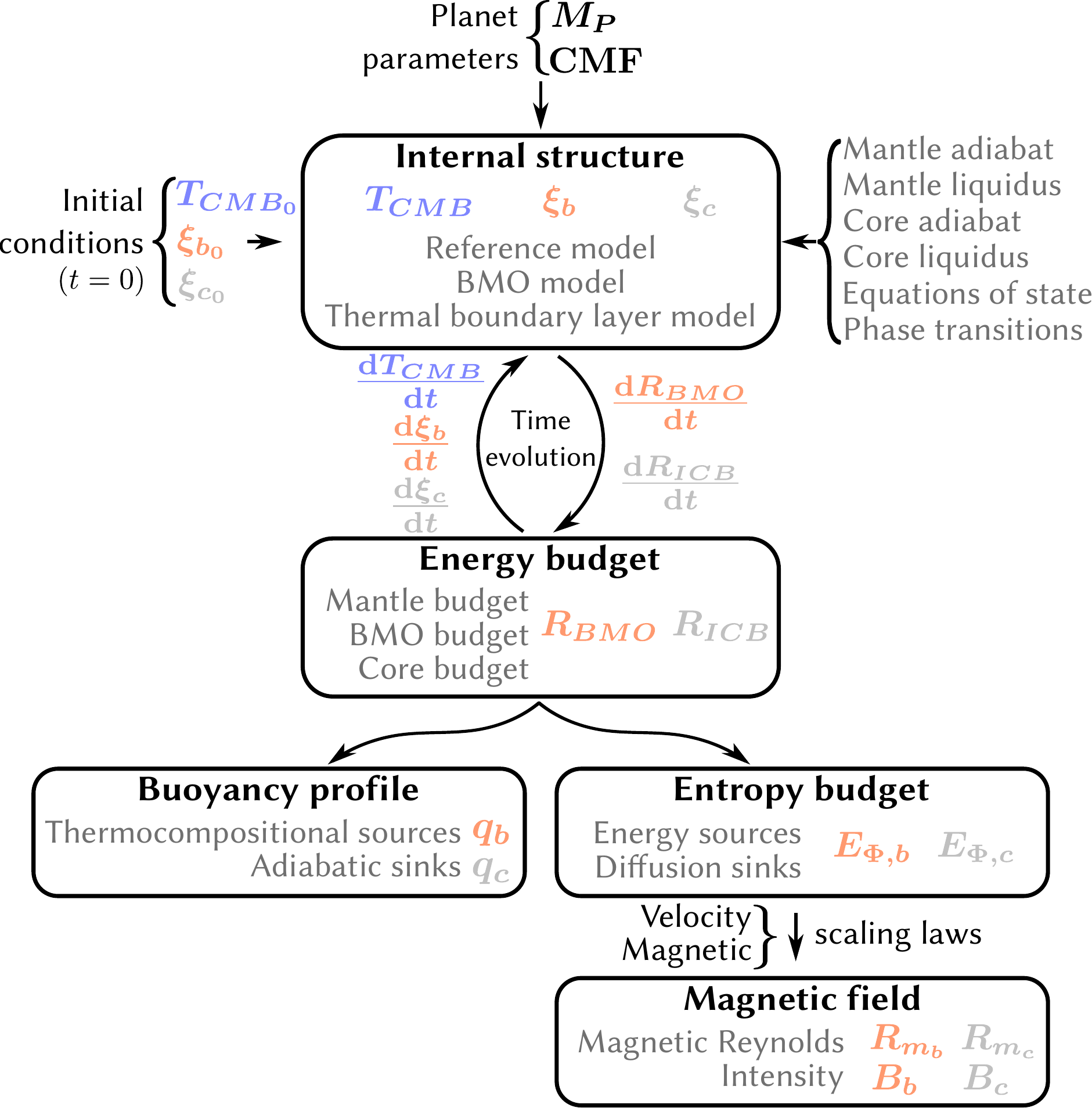}
    \caption{Description of the model workflow. The internal structure (§\ref{sec:model_structure}) and energy budgets  (§\ref{sec:model_thermal}) determine concomitantly the structural and thermal evolution of the planet over time. The convective stability of the BMO and the core is obtained at each time step by calculating buoyancy profiles (§\ref{sec:model_buoyancy}). Similarly, the magnetic evolution is obtained by calculating entropy budgets (§\ref{sec:model_magnetic_entropy}), magnetic Reynolds numbers (§\ref{sec:model_magnetic_reynolds_number}) and magnetic field intensities (§\ref{sec:model_magnetic_field_intensity}) at each time step. The physical quantities are defined in the text and Table \ref{tab:variables}.}
    \label{fig:workflow_sketch}
\end{figure}

We adopt the following workflow (Fig. \ref{fig:workflow_sketch}). First, we pre-compute the internal structure over a wide range of imposed CMB temperatures (§\ref{sec:model_structure}). This pre-computation is motivated because computation efficiency is predominantly limited by the integration of the internal structure. We then determine the structural and thermal evolution of the planet by computing the evolution of the temperature at the CMB with an energy budget (§\ref{sec:model_thermal}). Next, we determine the buoyancy profile evolution of the planet by computing buoyancy sources and sinks in the BMO and the core, \textit{i.e.} the thermocompositional convective stability of both layers, including potential stably-stratified layers (§\ref{sec:model_buoyancy}). Finally, we obtain the magnetic evolution of the planet by computing entropy budgets, magnetic Reynolds numbers, and field intensities in the potential dynamo layers, \textit{i.e.} the BMO and the outer core (§\ref{sec:model_magnetic}).

The model is parameterized to generate an Earth-like evolution converging toward a structure similar to the Earth at present time, including properties such as the inner core radius, temperatures at the CMB and at the ICB, heat flows at the CMB and at the surface, and the surface magnetic field.

\begin{table}[!ht]
    \caption{Summary of model variables. In the text, the physical quantities at the surface, at the top of the BMO, at the CMB, at the ICB, and at the center are respectively identified with the subscripts $P$, $BMO$, $CMB$, $ICB$ and $C$. The subscripts $m$, $b$ and $c$, $oc$ and $ic$ refer to the mantle, the BMO, the core, the outer core and the inner core. The subscripts $R$, $S$, $G$, $L$ and $A$ refer to radiogenic heating, secular cooling, gravitational energy, latent heat and adiabatic terms.}
    \begin{scriptsize}
        \begin{tabular*}{\tblwidth}{@{}LLL@{}}
            \toprule
            Symbol & Units & Definition \\
            \midrule
            $B$ & $\mathrm{T}$ & Magnetic field intensity \\
            $b_{dip}$ & $\mathrm{-}$ & Internal to outer-boundary dipole ratio \\
            $c$ & $\mathrm{ppm}$ & Radiogenic element concentration \\
            $C$ & $\mathrm{kg.m^{-3}}$ & Codensity \\
            $c_P$ & $\mathrm{J.kg^{-1}.K^{-1}}$ & Specific heat capacity \\
            $D$ & $\mathrm{-}$ & Partition coefficient \\
            $E$ & $\mathrm{W.K^{-1}}$ & Rate of entropy production \\
            $\hat{E}$ & $\mathrm{J.mol^{-1}}$ & Molar internal energy \\
            $f$ & $\mathrm{-}$ & Inner-boundary mass anomaly fraction \\
            $F$ & $\mathrm{kg.s^{-1}}$ & Mass anomaly flow rate \\
            $\hat{F}$ & $\mathrm{J.mol^{-1}}$ & Molar Helmholtz free energy \\
            $g$ & $\mathrm{m.s^{-2}}$ & Gravity \\
            $h$ & $\mathrm{W.kg^{-1}}$ & Specific heat production rate \\
            $k$ & $\mathrm{W.m^{-1}.K^{-1}}$ & Thermal conductivity \\
            $K_S$ & $\mathrm{Pa}$ & Isentropic bulk modulus \\
            $K_T$ & $\mathrm{Pa}$ & Isothermal bulk modulus \\
            $L$ & $\mathrm{m}$ & Convective system length scale \\
            $m$ & $\mathrm{kg}$ & Mass distribution \\
            $M$ & $\mathrm{kg}$ & Mass \\
            $\mathcal{M}$ & $\mathrm{A.m^2}$ & True dipole moment \\
            $P$ & $\mathrm{Pa}$ & Pressure \\
            $q$ & $\mathrm{kg.m^{-2}.s^{-1}}$ & Mass anomaly flux \\
            $Q$ & $\mathrm{W}$ & Heat flow \\
            $r$ & $\mathrm{m}$ & Radius \\
            $R_m$ & $\mathrm{-}$ & Magnetic Reynolds number \\
            $Ro$ & $\mathrm{-}$ & Rossby number \\
            $\hat{S}$ & $\mathrm{J.K^{-1}.mol^{-1}}$ & Molar entropy \\
            $t$ & $\mathrm{s}$ & Time \\
            $T$ & $\mathrm{K}$ & Temperature \\
            $T_l$ & $\mathrm{K}$ & Liquidus temperature \\
            $u$ & $\mathrm{m.s^{-1}}$ & Radial contraction velocity \\
            $U$ & $\mathrm{m.s^{-1}}$ & Convective velocity \\
            $\mathbf{v}$ & $\mathrm{m.s^{-1}}$ & Flow velocity \\
            $V$ & $\mathrm{m^3}$ & Volume \\
            $\hat{V}$ & $\mathrm{m^3.mol^{-1}}$ & Molar volume \\
            $\alpha$ & $\mathrm{K^{-1}}$ & Thermal expansion coefficient \\
            $\alpha_c$ & $\mathrm{-}$ & Compositional expansion coefficient \\
            $\chi$ & $\mathrm{-}$ & Mole fraction \\
            $\epsilon$ & $\mathrm{kg.m^{-3}.s^{-1}}$ & Volumetric buoyancy source \\
            $\gamma$ & $\mathrm{-}$ & Grüneisen parameter \\
            $\kappa$ & $\mathrm{m^2.s^{-1}}$ & Thermal or mass diffusivity \\
            $\mu$ & $\mathrm{Pa.s}$ & Dynamic viscosity \\
            $\Omega$ & $\mathrm{s^{-1}}$ & Rotation rate \\
            $\Phi$ & $\mathrm{W}$ & Convective power \\
            $\psi$ & $\mathrm{m^2.s^{-2}}$ & Gravitational potential \\
            $\rho$ & $\mathrm{kg.m^{-3}}$ & Density \\
            $\sigma$ & $\mathrm{S.m^{-1}}$ & Electrical conductivity \\
            $\Theta$ & $\mathrm{K}$ & Einstein temperature \\
            $\xi$ & $\mathrm{-}$ & Mass fraction \\
            $\zeta$ & $\mathrm{J.mol^{-1}.K^{-2}}$ & Thermo-electronic heat capacity \\
            \bottomrule
        \end{tabular*}
    \end{scriptsize}
    \label{tab:variables}
\end{table}

\subsection{Structural evolution}
\label{sec:model_structure}

\begin{figure}[!ht]
    \centering
    \includegraphics[width=\linewidth]{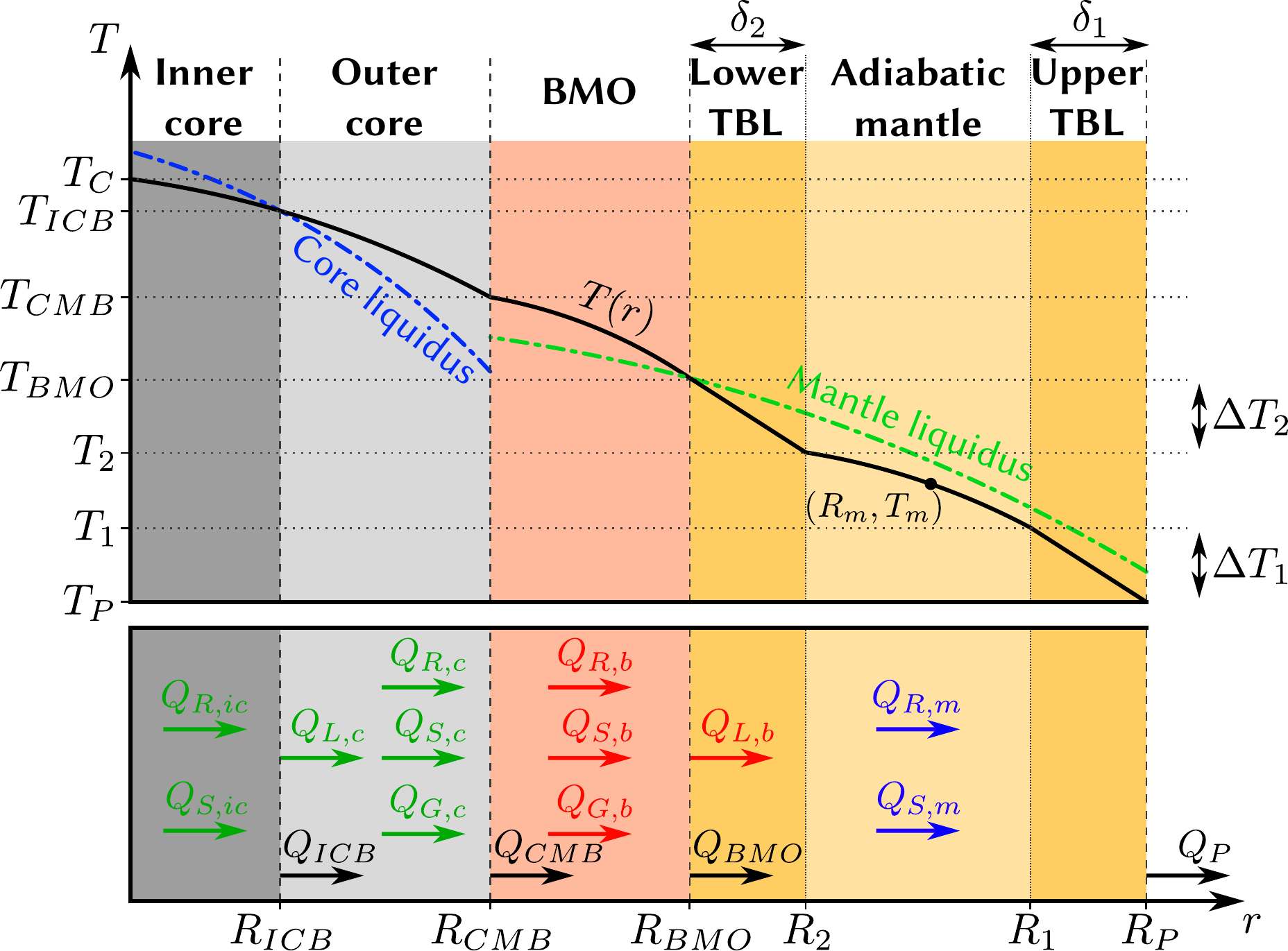}
    \caption{Definition of the layers and interfaces that can constitute the internal structure of the planet, \textit{i.e.} a solid mantle, a liquid BMO, a liquid outer core and a solid inner core. The mantle is divided into an adiabatic section located between two thermal boundary layers (TBL). The extents of the BMO and the outer core are determined by the intersection of the adiabatic temperature profile anchored at the CMB and the appropriate liquidus. The arrows depict the heat flows $Q$ involved in the energy budget. The radiogenic heating, secular cooling, gravitational energy and latent heat terms are denoted with the subscripts $R$, $S$, $G$ and $L$.}
    \label{fig:structure_sketch}
\end{figure}

We model a differentiated spherically symmetric Earth-like planet with a silicate mantle overlying a metal core (Fig. \ref{fig:structure_sketch}). We first define a reference structure with a solid $\mathrm{MgSiO_3}$ mantle and a $\mathrm{Fe}$ core (§\ref{sec:model_structure_reference}). The mantle includes several solid phase transitions and the core may be liquid or solid. The reference structure is calculated for imposed temperatures at the CMB, with discrete integer degrees for $3000 \le T_{CMB} \le 8000~\mathrm{K}$.
The internal structure is then modified by redefining the temperature profile of the mantle. First, we evaluate the possible formation of a liquid BMO, depending on the intersection between the temperature profile anchored at the imposed temperature of the CMB and the mantle liquidus (§\ref{sec:model_structure_BMO}). Then, we redefine a consistent temperature profile in the solid upper mantle using a thermal boundary layer model (§\ref{sec:model_structure_TBL}).

The following equations distinguish the BMO as a separate layer, but when the BMO is fully crystallized, the radius of the bottom of the mantle is the same as the CMB and $BMO$ subscripts must be interpreted as $CMB$ subscripts.
Similarly, the notation distinguishes an inner core. However, if the core is fully molten, the radius at the bottom of the core is zero and $ICB$ subscripts must be interpreted as $C$ subscripts.

\subsubsection{Reference model}
\label{sec:model_structure_reference}

Our reference structural model uses \citet{boujibar_2020}, which is based on \citep{valencia_2006,seager_2007,driscoll_2011}. We obtain the mass distribution $m$, gravity $g$, pressure $P$ and density $\rho$ as a function of radius $r$ by integrating a continuity equation, a Poisson's equation, a hydrostatic equilibrium equation, and an Adams-Williamson equation
\begin{eqnarray}
    \label{eq:reference_mass}
    \frac{\mathrm{d} m}{\mathrm{d} r}&=&4\pi r^2\rho,\\
    \label{eq:reference_gravity}
    \frac{\mathrm{d} g}{\mathrm{d} r}&=&4\pi G\rho-2G\frac{m}{r^3},\\
    \label{eq:reference_pressure}
    \frac{\mathrm{d} P}{\mathrm{d} r}&=&-\rho g,\\
    \label{eq:reference_density}
    \frac{\mathrm{d} \rho}{\mathrm{d} r}&=&-\frac{\rho^2g}{K_S},
\end{eqnarray}
where $G$ is the gravitational constant.
The isentropic bulk modulus $K_S$ depends on the temperature $T$, the thermal expansivity $\alpha$ and the Grüneisen parameter $\gamma$
\begin{equation}
    K_S=K_T\left(1 + \alpha \gamma T \right).
    \label{eq:reference_isentropic_bulk_modulus}
\end{equation}
The isothermal bulk modulus $K_T$ is defined using a third order Vinet-Rydberg \citep{vinet_1989} equation of state (EOS), defined in Appendix \ref{app:EOS_mantle} (Eq. \ref{eq:reference_isothermal_bulk_modulus}-\ref{eq:reference_gruneisen}).
We also use an adiabatic temperature profile in both the mantle and the core, with a corresponding temperature gradient defined as
\begin{equation}
    \frac{\mathrm{d}T}{\mathrm{d}r}=-\frac{\rho g \gamma}{K_s}T.
    \label{eq:reference_adiabat}
\end{equation}
Although this is justifiable for the outer core and mantle, we expect a conductive inner core with a different temperature profile.

We consider up to three phases in the solid mantle with peridotite (Pd), perovskite (Pv) and post-perovskite (pPv) \citep{boujibar_2020}. The phase boundaries are parameterized with
\begin{equation}
    P=P_c+\gamma_{c}(T-T_c)
    \label{eq:reference_transition}
\end{equation}
where $P_c$, $T_c$ and $\gamma_c$ are a reference pressure, temperature and a Clapeyron slope, defined in Table S3.
We also define the core liquidus $T_{l}$ \citep{stixrude_2014,boujibar_2020} using a Simon-Glatzel equation \citep{simon_1929}
\begin{equation}
    T_{l}=T_{l_0}\left(\frac{P-P_{l_0}}{a_l}+1\right)^{b_l},
    \label{eq:liquidus}
\end{equation}
where $T_{l_0}$, $P_{l_0}$, $a_l$ and $b_l$ are a reference temperature, pressure and two composition-related parameters, defined in Table S4. To account for the depression of the liquidus from light elements, we multiply Eq. \ref{eq:liquidus} by $[1-\ln(1-\chi_{c_l})]^{-1}$, where $\chi_{c_l}=0.13$ is the fixed mole fraction of light elements \citep{boujibar_2020}. We do not consider a direct influence of the changing light element content on the core liquidus, although this could significantly modify the liquidus, potentially resulting in a temperature change at the ICB exceeding 1000 K \citep{hirose_2021}.

We determine the reference profile by integrating Eq. \ref{eq:reference_mass}-\ref{eq:reference_adiabat}, together with Eq. \ref{eq:reference_isothermal_bulk_modulus}-\ref{eq:reference_gruneisen}, from the surface, at $r=R_P$, toward the center of the planet \citep{boujibar_2020}. The surface boundary conditions are
\begin{equation}
    \begin{array}{c}
        m_{P}=M_P, ~ g_{P}=\frac{G M_P}{R_P^2}, ~ P_{P}=1.013 \times 10^5~\mathrm{Pa}, \\
        \rho_{P}=3226~\mathrm{kg.m^{-3}}, ~ T_{P}=300~\mathrm{K}.
    \end{array}
    \label{eq:reference_BC}
\end{equation}
The density jumps that occur in the mantle upon phase transitions, at the CMB, and at the ICB upon crystallization of the inner core are calculated assuming continuity of the isothermal bulk modulus (Eq. \ref{eq:reference_isothermal_bulk_modulus}).

The mantle temperature profile of the reference model is adiabatic and anchored at the surface temperature, \textit{i.e.} using $T(R_P)=T_P$ as a boundary condition. This approximation allows us to precompute the internal structure as it does not require modeling the temperature profile of the mantle. In contrast, using a thermal boundary layer model in the integration of the internal structure would require to calculate concomitantly the thermal evolution of the planet to determine the depth of the mantle, the thickness of the boundary layers and the temperature drop in these layers. This approximation is justified because temperature is mostly involved in the integration of the
internal structure through Eq. \ref{eq:reference_isentropic_bulk_modulus}, where $\alpha\gamma T \simeq 3 \times 10^{-2} \ll 1$. Furthermore, temperature has only a weak influence on solid phase transitions in the mantle because of the small Clapeyron slopes \citep{boujibar_2020}. As an example, increasing $T_P$ by $2000~\mathrm{K}$ would change the density profile of the mantle by 0.6\% on average.

The core temperature profile is also adiabatic but is anchored at the imposed CMB temperature $T_{CMB}$, \textit{i.e.} the boundary condition is $T(R_{CMB})=T_{CMB}$ \citep{boujibar_2020}. This is inaccurate for the inner core which is conductive. However, the influence on the position of the ICB is relatively weak since the adiabatic profile of the outer core is anchored at the CMB.

Since we impose the planet mass, the reference model is sensitive to the exact radius of the planet. We search for the value of $R_P$ that gives a solution where mass and gravity go to zero at the center and where the pressure, density and temperature remain smooth and finite \citep{driscoll_2011}. This optimization is parameterized to give a typical error of $1~\mathrm{km}$ on $R_P$. Variations of $R_P$ too far from the optimal value may cause the solution to diverge close to the center (\textit{e.g.} Eq. \ref{eq:reference_gravity}).

\subsubsection{Basal magma ocean model}
\label{sec:model_structure_BMO}

In our model, we follow \citet{blanc_2020}, assuming that a BMO may exist (i) if the slope of the BMO adiabatic temperature profile is steeper than that of the mantle liquidus and (ii) if the temperature is larger than the liquidus at the CMB. The BMO then extends from the CMB to the intersection between the BMO adiabat and the mantle liquidus (Fig. \ref{fig:structure_sketch}). This neglects the role of density contrasts between melt and crystals on the evolution of the BMO \citep{boukare_2015,caracas_2019,nabiei_2021}.

The BMO adiabatic temperature profile, anchored at the CMB, is defined by
\begin{equation}
    T(r)=T_{CMB}\exp\left(\int_{\rho_{CMB}^+}^{\rho(r)} \frac{\gamma_{b}}{\rho'}\mathrm{d}\rho'\right),
    \label{eq:adiabat_BMO}
\end{equation}
where $\gamma_{b}$ is an effective thermal Grüneisen parameter and $\rho_{CMB}^+$ is the mantle density at the CMB. The density profile used in Eq. \ref{eq:adiabat_BMO} is a linear fit of the reference density profile obtained in the solid mantle. This facilitates eliminating density jumps that might occur from solid phase transitions in the mantle from the BMO adiabat.
As a result, the BMO density profile is based on that of the solid mantle in the reference model at the same depth. It follows that the mass distribution, gravity, and pressure of the BMO are also set to the value of the reference profile. As such, we neglect the influence of changes in density caused by melting and variations in composition on the adiabatic temperature profile of the BMO and on the mass distribution, gravity, and pressure profiles of the planet. We therefore do not self-consistently integrate the planet structure with the BMO. Determining such a self-consistent profile would require integrating Eq. \ref{eq:reference_mass}-\ref{eq:reference_adiabat} from the surface, using the appropriate EOS for the BMO, and concomitantly calculating the thermal evolution of the planet to determine the BMO thickness.

We determine the thermodynamics quantities of the BMO from the molar Helmholtz free energy $\hat{F}_{b}$ of an effective EOS (Appendix \ref{app:EOS_BMO}). Hereafter, the thermodynamic quantities commonly known as extensive, including free energy and volume, are defined for a mole of matter and are thus actually intensive molar quantities, denoted with a hat. The free energy $\hat{F}_{b}$ corresponds to an ideal mixing model between $\mathrm{MgSiO_3}$ and $\mathrm{FeO}$
\begin{equation}
    \hat{F}_{b}=\hat{F}_\mathrm{MgSiO_3}+(\hat{F}_\mathrm{FeO}-\hat{F}_\mathrm{MgSiO_3})\chi_{b},
    \label{eq:free_energy_BMO}
\end{equation}
where $\chi_{b}$ is the mole fraction of $\mathrm{FeO}$ in the BMO. $\hat{F}_\mathrm{MgSiO_3}$ uses a generalized Rosenfeld-Tarazona EOS \citep{wolf_2018} and $\hat{F}_\mathrm{FeO}$ uses a Vinet-Rydberg EOS with a thermal and an electronic contribution \citep{morard_2022}. This allows us to consider the progressive enrichment of the BMO in iron-rich oxides.

\sloppy

We derive the thermodynamic quantities of the BMO from $\hat{F}_{b}$. This includes the isothermal bulk modulus $K_T=\hat{V}(\partial^2 \hat{F}/\partial \hat{V}^2)_T$, the isentropic bulk modulus $K_S=K_T(1+\alpha \gamma T)$, the heat capacity at constant volume $\hat{C}_V=-T(\partial^2 \hat{F}/\partial T^2)_{\hat{V}}$, the heat capacity at constant pressure $\hat{C}_P=\hat{C}_V(1+\alpha \gamma T)$, and the thermal expansivity $\alpha=-1/K_T(\partial^2 \hat{F}/\partial \hat{V} \partial T)$. Since the internal structure is not integrated self-consistently with a BMO EOS, the density at which the thermodynamics quantities are computed is underestimated due to a lack of iron in the reference profile.

\fussy

We also derive the effective thermal Grüneisen parameter used in the adiabatic temperature profile (Eq. \ref{eq:adiabat_BMO}) from $\hat{F}_{b}$, using $\gamma_b=\alpha \hat{V} K_S/\hat{C}_P$. However, unlike other thermodynamic quantities used in the energy, buoyancy and entropy terms, $\gamma_b$ is defined using fixed temperature and composition, corresponding respectively to the initial temperature at the CMB and the initial composition of the BMO. As a result, the radial profile of the BMO adiabatic temperature gradient is constant over time. This simplification is required to determine the thermal evolution of the BMO (see Eq. \ref{eq:adiabatic_simplification} in §\ref{sec:model_thermal}) assuming that a time-varying Grüneisen parameter has a negligible effect on the adiabatic gradient \citep{gubbins_2003}. In our nominal model, the maximum value of the radially-averaged Grüneisen parameter computed \textit{a posteriori} using time-dependent temperature and composition is only 16.8\% larger than the fixed value based on the initial conditions.

The mantle liquidus also depends on the BMO composition and is expected to shift toward lower temperatures as the BMO is enriched in $\mathrm{FeO}$ \citep{boukare_2015,ballmer_2017,blanc_2020}. We use a linear phase diagram \citep[\textit{e.g.}][]{labrosse_2007} between $\mathrm{MgSiO_3}$ \citep{fei_2021} and $\mathrm{FeO}$ \citep{morard_2022}. The BMO liquidus is therefore defined as
\begin{equation}
    T_{l_{b}}=T_{l_\mathrm{MgSiO_3}}+(T_{l_\mathrm{FeO}}-T_{l_\mathrm{MgSiO_3}})\xi_{b},
    \label{eq:liquidus_BMO}
\end{equation}
where $\xi_b$ is the mass fraction of $\mathrm{FeO}$ in the BMO. The two end-members are defined using Simon-Glatzel equations (Eq. \ref{eq:liquidus}). The liquidus parameters are defined in Table S4.

As with the Grüneisen parameter, the mantle liquidus is computed using the pressure profile of the reference model at initial time, neglecting the effects of temperature variation and inner core crystallization on reference model. This simplification is required to include the correction related to the compositional depression of the liquidus, computed at constant pressure, in the crystallization rate of the BMO (see Appendix \ref{app:crystallization_rate}). In our nominal model, the maximum difference between the radially-averaged liquidus computed \textit{a posteriori} using the time-dependent pressure profile of the reference model and the fixed pressure profile at initial time is only 0.02\%.

\subsubsection{Thermal boundary layer model}
\label{sec:model_structure_TBL}

In the solid part of the mantle, we redefine the temperature profile using a thermal boundary layer model similar to \citet{driscoll_2014}, that includes two boundary layers (Fig. \ref{fig:structure_sketch}). Temperature differences across the upper and lower boundary layers are denoted by $\Delta T_1$ and $\Delta T_2$, with $T_1$ and $T_2$ the temperatures at the bottom of the upper boundary layer and at the top of the lower boundary layer. The
thicknesses of the boundary layers are denoted by $\delta_1$ and $\delta_2$, with $R_1$ and $R_2$ the lower radius of the upper boundary layer and the upper radius of the lower boundary layer.

We assume an adiabatic temperature profile outside of the boundary layers, so that $T_1$ can be expressed as a function of an average mantle temperature $T_m$, giving
\begin{equation}
    \Delta T_1=T_1-T_P=T_m\eta_1-T_P,
    \label{eq:DT1}
\end{equation}
where $\eta_1$ refers to the adiabatic temperature decrease from the average mantle temperature to the bottom of the upper boundary layer. Eq. \ref{eq:DT1} gives
\begin{equation}
    \eta_1=\exp\left(-\int_{\bar{R}_m}^{R_1} \frac{\rho g \gamma}{K_S}\mathrm{d}r\right),
    \label{eq:eta_1}
\end{equation}
where $\bar{R}_m$ is an average radius corresponding to the average mantle temperature.
We assume that $\delta_1 \ll L_m$, where $L_m=R_P-R_{BMO}$ is the thickness of the mantle, which gives $R_1 \sim R_P$. Furthermore, we assume that $\bar{R}_m$ and $T_m$ depend respectively on the outer radius of the BMO $R_{BMO}$ and on the temperature at the top of the BMO $T_{BMO}$
\begin{equation}
    \left\{
    \begin{array}{rcl}
        \bar{R}_m & = & \frac{R_{BMO}+R_P}{2} \\
        T_m       & = & \frac{T_{BMO}+T_P}{2}
    \end{array}
    \right.,
    \label{eq:T_m_R_m}
\end{equation}
which allows us to write $\Delta T_1$ as a function of quantities independent of the boundary layer model, \textit{i.e.} $R_{BMO}$, $T_{BMO}$, $R_{P}$ and $T_{P}$.

We determine the upper boundary layer thickness by assuming that the local Rayleigh number $Ra_1$ equals the free-slip critical value $Ra_{cr}=660$ \citep{howard_1966,solomatov_1995,driscoll_2014}. This gives
\begin{equation}
    Ra_1=\frac{\rho_P\alpha_P g_P \Delta T_1 \delta_1^3}{\kappa_P\mu_P}=Ra_{cr},
    \label{eq:Rayleigh_1}
\end{equation}
where $\rho_P$, $\alpha_P$, $g_P$, $\mu_P$, $\kappa_P=k_P/(\rho_P c_{P_P})$, $k_P$ and $c_{P_P}$ are density, thermal expansivity, gravity, dynamic viscosity, thermal diffusivity, thermal conductivity and specific heat capacity at constant pressure at the surface of the planet. We assume that these quantities vary negligibly within the thermal boundary layer in calculating $\delta_1$.
We assume that the viscosity at the surface of the upper boundary layer is $\mu_P=\mu_m/f_\mu$, where $\mu_m$ is the viscosity of the mantle and $f_\mu=10$ is an arbitrary factor \citep{driscoll_2014}. Following \citet{driscoll_2014}, the mantle viscosity is
\begin{equation}
    \mu_m=\mu_{m_0}\exp\left(\frac{A_\mu}{R\bar{T}_m}\right).
    \label{eq:viscosity_mantle}
\end{equation}
where $\mu_{m_0}=1.1 \times 10^{15}~\mathrm{Pa.s}$ is a reference viscosity, $A_\mu=3 \times 10^5~\mathrm{J.mol^{-1}}$ is an activation energy, and $\bar{T}_m$ is the average temperature of the mantle assuming an adiabatic profile and $T(R_P)=T_P+\Delta T_1$. The parameters are defined in Table S6. In our nominal model, $\mu_m$ and $\mu_P$ are typically in the range $6 \times 10^{20} - 5 \times 10^{21}~\mathrm{Pa.s}$ and $6 \times 10^{19} - 5 \times 10^{20}~\mathrm{Pa.s}$, respectively (Fig. S1e).
The thermal conductivity at the surface $k_P$ is calculated in Appendix \ref{app:k_mantle} \citep{vandenberg_2010} using the pressure at the surface, a temperature equal to $T_P+\Delta T_1/2$ and thermodynamic quantities computed at the surface. In our nominal model, the average value of $k_P$ is $3.0~\mathrm{W.m^{-1}.K^{-1}}$ (Fig S1f).

Using Eq. \ref{eq:Rayleigh_1}, we write a generalized expression of the upper boundary layer thickness by replacing the $1/3$ exponent by a parameter $\beta$
\begin{equation}
    \delta_1=L_m\left(Ra_{cr}\frac{\kappa_P\mu_P}{\rho_P\alpha_P g_P\Delta T_1 L_m^3}\right)^\beta.
    \label{eq:delta_1_beta}
\end{equation}
In this context, we can interpret the surface heat flow in term of a Nusselt number, \textit{i.e.} the ratio between convective and conductive heat transfer \citep{driscoll_2014}.
$Ra_{cr}$ and $\beta$ can therefore be interpreted as the prefactor and the exponent of the $Nu \propto Ra^\beta$ relationships \citep{solomatov_1995,schubert_2001}.

In the upper boundary layer ($R_1 \leq r \leq R_P$), we use a linear temperature profile with a constant temperature gradient
\begin{equation}
    \frac{\mathrm{d}T}{\mathrm{d}r}=-\frac{\Delta T_1}{\delta_1} \quad \mathrm{with} \quad T(R_P)=T_P,
    \label{eq:dTdr_UBL}
\end{equation}
where $\Delta T_1$ and $\delta_1$ are respectively obtained from Eq. \ref{eq:DT1} and \ref{eq:delta_1_beta}.
In the convecting region of the mantle ($R_2 \leq r < R_1$), we use an adiabatic temperature profile (Eq. \ref{eq:reference_adiabat}) with $T(R_1)=T_1$, where $T_1$ is obtained from Eq. \ref{eq:dTdr_UBL}.
In the lower boundary layer ($R_{BMO} \leq r < R_2$), we use a linear temperature profile with a constant temperature gradient
\begin{equation}
    \frac{\mathrm{d}T}{\mathrm{d}r}=-\frac{1}{4\pi R_{BMO}^2}\frac{1}{k_2}Q_{BMO} \quad \mathrm{with} \quad T(R_{BMO})=T_{BMO},
    \label{eq:dTdr_LBL}
\end{equation}
where the thermal conductivity in the lower boundary layer $k_2$ is estimated in Appendix \ref{app:k_mantle} \citep{vandenberg_2010}, using the pressure at the interface with the BMO, a temperature that extends the adiabatic temperature profile defined in the convecting region of the mantle (Eq. \ref{eq:reference_adiabat}) to $R_{BMO}$, and thermodynamic quantities computed at the bottom of the mantle. In our nominal model, the average value of $k_2$ is $4.9 ~\mathrm{W.m^{-1}.K^{-1}}$ (Fig S1f). Since the heat flow $Q_{BMO}$ at the top of the BMO is obtained from the energy budget (§\ref{sec:model_thermal}), we can therefore obtain concomitantly $\Delta T_2$ and $\delta_2$ by computing the intersection between the adiabatic temperature profile in the convecting mantle (Eq. \ref{eq:reference_adiabat}) and the linear temperature profile in the lower boundary layer (Eq. \ref{eq:dTdr_LBL}).

\subsection{Thermal evolution}
\label{sec:model_thermal}

The thermal evolution of the planet is obtained by computing the evolution of the internal structures to different thermal states (§\ref{sec:model_structure}). We follow a method initially developed for the Earth's core \citep{gubbins_2003,gubbins_2004}, and later adapted for the BMO \citep{labrosse_2007,blanc_2020,orourke_2020}, to obtain the time evolution of the temperature at the CMB with coupled energy budgets of the solid mantle (§\ref{sec:model_thermal_mantle}), BMO (§\ref{sec:model_thermal_BMO}) and core (§\ref{sec:model_thermal_core}).

In the energy budgets, we assume that the BMO and the core are adiabatic and well-mixed by convection \citep{gubbins_2003,gubbins_2004}.
In the BMO, this assumption becomes invalid when the melt fraction decreases below 40\% \citep{solomatov_1993,lejeune_1995}, which means that the dynamics of multi-phase flows beyond the rheological transition should be considered \citep{boukare_2017}.
If the thermodynamic quantities of the BMO and the core vary slowly over time, in particular the Grüneisen parameter, the adiabatic temperature profiles allows the following simplification \citep{gubbins_2003},
\begin{equation}
    \frac{\mathrm{D}T}{\mathrm{D}t}=\frac{T}{T_{CMB}}\frac{\mathrm{d}T_{CMB}}{\mathrm{d}t},
    \label{eq:adiabatic_simplification}
\end{equation}
where
\begin{equation}
    \frac{\mathrm{D}\bullet}{\mathrm{D}t}=\frac{\partial\bullet}{\partial t}+\mathbf{u}\cdot\nabla\bullet
    \label{eq:lagrangian_derivative}
\end{equation}
is a Lagrangian derivative accounting for the effect of the thermal contraction. $\mathbf{u}$ is the slow radial velocity of contraction, in the BMO or the core.

\subsubsection{Energy budget of the mantle}
\label{sec:model_thermal_mantle}

For the mantle, we write the energy budget as
\begin{equation}
    Q_P=Q_{R,m}+Q_{S,m}+Q_{BMO},
    \label{eq:energy_budget_mantle}
\end{equation}
where $Q_P$ is the heat flow at the surface of the planet (expressed in $\mathrm{W}$), $Q_{R,m}$ is the radiogenic heating in the mantle, $Q_{S,m}$ is the secular cooling of the mantle and $Q_{BMO}$ is the heat flow at the top of the BMO (Appendix \ref{app:energy_budget}).

The surface heat flow is expressed as the heat flow through the upper boundary layer of the solid mantle \citep{driscoll_2014}
\begin{equation}
    Q_P=4\pi R_P^2k_1\frac{\Delta T_1}{\delta_1},
    \label{eq:heat_flow_surface}
\end{equation}
where the thermal conductivity in the upper boundary layer $k_1$ is estimated in Appendix \ref{app:k_mantle} \citep{vandenberg_2010}, using pressure, temperature and thermodynamic quantities averaged over the upper boundary layer. In our nominal model, the average value of $k_1$ is $3.0~\mathrm{W.m^{-1}.K^{-1}}$ (Fig S1f). In addition, $\Delta T_1$ and $\delta_1$ are respectively obtained from Eq. \ref{eq:DT1} and \ref{eq:delta_1_beta}.
The radiogenic heating $Q_{R,m}$ (Eq. \ref{eq:radio_mantle}) is integrated over the volume of the mantle. We assume an Earth-like radiogenic content where $\mathrm{^{238}U}$, $\mathrm{^{235}U}$, $\mathrm{^{232}Th}$ and $\mathrm{^{40}K}$ are the only isotopes contributing \citep{jaupart_2015}. We simplify the separate contribution of each radioisotope into a unique parameter, equivalent to an effective concentration, defined as the mantle fraction $f_h=0.58$ of bulk silicate Earth (BSE) radiogenic production. This radiogenic term is independent of the CMB cooling rate.
In contrast, the secular cooling $Q_{S,m}$ (Eq. \ref{eq:secular_mantle}) can be expressed as a function of the CMB cooling rate using the adiabatic profile simplification (Eq. \ref{eq:adiabatic_simplification}) in the BMO.

\subsubsection{Energy budget of the basal magma ocean}
\label{sec:model_thermal_BMO}

In the BMO, we write the energy budget as
\begin{equation}
    Q_{BMO}=Q_{R,b}+Q_{S,b}+Q_{G,b}+Q_{L,b}+Q_{CMB}
    \label{eq:energy_budget_BMO}
\end{equation}
where $Q_{R,b}$ is the heat flow corresponding to radiogenic heating in the BMO, $Q_{S,b}$ is the BMO secular cooling, $Q_{G,b}$ is a gravitational energy due to the release of $\mathrm{FeO}$ upon crystallization, $Q_{L,b}$ is the corresponding latent heat and $Q_{CMB}$ is the heat flow at the CMB (Appendix \ref{app:energy_budget}).
We neglect the heat flows associated with the heat of reaction, pressure heating and pressure effects on freezing \citep{blanc_2020,orourke_2020}.

We integrate the radiogenic heating $Q_{R,b}$ (Eq. \ref{eq:radio_BMO}) over the BMO volume using the same specific heat production rate as in the mantle, thereby assuming homogeneous radiogenic heating in the BMO and the mantle.
As with the mantle, this term is independent of the CMB cooling rate.
In contrast, the secular cooling term $Q_{S,b}$ (Eq. \ref{eq:secular_BMO}) can be directly expressed as a function of the CMB cooling rate using the adiabatic profile (Eq. \ref{eq:adiabatic_simplification}) in the BMO.
In addition, the gravitational energy term $Q_{G,b}$ (Eq. \ref{eq:gravitational_BMO}), which is actually a compositional energy due to the release of $\mathrm{FeO}$ \citep{labrosse_2003} upon crystallization, depends on the rate of release of $\mathrm{FeO}$, which can be expressed as a function of the BMO crystallization rate.
Finally, the latent heat term $Q_{L,b}$ (Eq. \ref{eq:latent_BMO}) is proportional to the BMO crystallization rate.

The secular cooling $Q_{S,b}$, the gravitational energy $Q_{G,b}$ and the latent heat $Q_{L,b}$ terms can therefore be expressed as a function of the cooling rate at the CMB. If the BMO is well-mixed, the rate of release of $\mathrm{FeO}$ into the BMO is proportional to the crystallization rate of the BMO \citep{gubbins_2004,blanc_2020}, with
\begin{equation}
    \frac{\mathrm{D}\xi_{b}}{\mathrm{D}t}
    =-\frac{4\pi R_{BMO}^2\rho_{BMO}^-\Delta\xi_{BMO}}{M_{b}}\frac{\mathrm{d}R_{BMO}}{\mathrm{d}t}
    =C_{\xi_{b}}\frac{\mathrm{d}R_{BMO}}{\mathrm{d}t},
    \label{eq:composition_rate_BMO}
\end{equation}
where $\rho_{BMO}^-$ is the density of the BMO at the solid mantle interface, $M_{b}$ is the mass of the BMO, and $C_{\xi_{b}}$ is the proportionality factor. The composition change across the interface is
\begin{equation}
    \Delta\xi_{BMO}=\xi_{b}(1-D_{BMO}),
    \label{eq:composition_change_BMO}
\end{equation}
where $D_{BMO}$ is the iron partition coefficient between liquid and solid.

In order to obtain an Earth-like nominal model with no BMO at present time, we set $D_{BMO}=0.85$. This value is larger than experimental measurements of the iron partition coefficient, which gives values between $D_{BMO}=0.1-0.2$ \citep{nomura_2011,tateno_2014} and $D_{BMO}=0.5-0.6$ \citep{andrault_2012}. However, this discrepancy might be explained by the linear phase diagram assumption (Eq. \ref{eq:liquidus_BMO}). Using a realistic crystallization sequence in the $\mathrm{MgO-FeO-SiO_2}$ system at the pressure and temperature conditions of the BMO \citep{boukare_2015,nabiei_2021} might reconcile an Earth-like evolution with lower values of the partition coefficient.

Adapting the results of \citet{gubbins_2003} to the BMO, in conjunction with the findings of \citet{blanc_2020}, we show that the BMO crystallization rate is proportional to the CMB cooling rate, with a correction corresponding to the depression of the liquidus as the BMO is enriched in iron
\begin{equation}
    \frac{\mathrm{d}R_{BMO}}{\mathrm{d}t}=C_{R_{b}}\frac{\mathrm{d}T_{CMB}}{\mathrm{d}t},
    \label{eq:crystallization_rate_BMO}
\end{equation}
where $C_{R_{b}}$ is the proportionality factor (see Appendix \ref{app:crystallization_rate}).

\subsubsection{Energy budget of the core}
\label{sec:model_thermal_core}

In the core, we write the energy budget as
\begin{equation}
    Q_{CMB}=Q_{R,c}+Q_{S,c}+Q_{G,c}+Q_{L,c}+Q_{ICB},
    \label{eq:energy_budget_core}
\end{equation}
where $Q_{R,c}$ is the heat flow corresponding to radiogenic heating in the outer core, $Q_{S,c}$ is the outer core secular cooling, $Q_{G,c}$ is the gravitational energy due to the release of light elements upon crystallization, $Q_{L,c}$ is the corresponding latent heat and $Q_{ICB}$ is the heat flow at the ICB. The heat flow at the ICB is
\begin{equation}
    Q_{ICB}=Q_{R,ic}+Q_{S,ic},
    \label{eq:energy_budget_inner_core}
\end{equation}
where $Q_{R,ic}$ and $Q_{S,ic}$ are the radiogenic heating and the secular cooling of the inner core, respectively (see Appendix \ref{app:energy_budget}).
As with the BMO, we neglect the heat flows associated with the heat of reaction, pressure heating and pressure effects on freezing \citep{gubbins_2003,gubbins_2004}.

The radiogenic heating terms $Q_{R,c}$ and $Q_{R,ic}$ (Eq. \ref{eq:radio_core} and \ref{eq:radio_inner_core}) are respectively integrated over the volume of the outer and inner core. We assume that $\mathrm{^{40}K}$ is the only radiogenic element contributing to the radiogenic heating and we choose a moderate potassium concentration in the core, \textit{i.e.} $c_c=20~\mathrm{ppm}$ \citep{hirose_2013,labrosse_2015}. We therefore assume that radiogenic production is homogeneous in the inner and outer core.
As with the mantle and the BMO, these terms are independent of the cooling rate at the CMB.
In contrast, the secular cooling terms $Q_{S,c}$ and $Q_{S,ic}$ (Eq. \ref{eq:secular_core} and \ref{eq:secular_inner_core}) can be expressed as a function of the CMB cooling rate using Eq. \ref{eq:adiabatic_simplification} in the outer core, and an infinite thermal conductivity, \textit{i.e.} an isothermal profile where the temperature is equal to the liquidus \citep{orourke_2018}, in the inner core.
The gravitational energy term $Q_{G,c}$ (Eq. \ref{eq:gravitational_core}) depends on the rate of release of light elements \citep{labrosse_2003}, which can be expressed as a function of the crystallization rate of the inner core.
Finally, the latent heat term $Q_{L,c}$ (Eq. \ref{eq:latent_core}) is proportional to the crystallization rate of the inner core.

As with the BMO, the secular cooling $Q_{S,c}$, the gravitational energy $Q_{G,c}$ and the latent heat $Q_{L,c}$ terms can therefore be expressed as a function of the cooling rate at the CMB. If the outer core is well-mixed, the rate of release of the light elements into the outer core is proportional to the crystallization rate of the core \citep{gubbins_2004}, with
\begin{equation}
    \frac{\mathrm{D}\xi_{c}}{\mathrm{D}t}
    =\frac{4\pi R_{ICB}^2\rho_{ICB}^+\xi_{c}}{M_{oc}}\frac{\mathrm{d}R_{ICB}}{\mathrm{d}t}
    =C_{\xi_{c}}\frac{\mathrm{d}R_{ICB}}{\mathrm{d}t},
    \label{eq:composition_rate_core}
\end{equation}
where $\xi_c$ is the mass fraction of light elements in the outer core, $\rho_{ICB}^+$ is the density of the outer core at the ICB, $M_{oc}$ is the mass of the outer core, and $C_{\xi_{c}}$ is the proportionality factor. Here we assume that there are no light elements in the inner core, \textit{i.e.} $\Delta\xi_{ICB}=\xi_c$ \citep{gubbins_2004}.

Following \citet{gubbins_2003}, the crystallization rate of the core is proportional to the cooling rate at the CMB
\begin{equation}
    \frac{\mathrm{d}R_{ICB}}{\mathrm{d}t}=C_{R_{c}}\frac{\mathrm{d}T_{CMB}}{\mathrm{d}t},
    \label{eq:crystallization_rate_core}
\end{equation}
where $C_{R_{c}}$ is the proportionality factor (see Appendix \ref{app:crystallization_rate}).

\subsubsection{Integration of the initial value problem}
\label{sec:model_thermal_solution}

Using the relationships between the release rates, the crystallization rates and the cooling rates in the BMO (Eq. \ref{eq:composition_rate_BMO} and \ref{eq:crystallization_rate_BMO}) and the core (Eq. \ref{eq:composition_rate_core} and \ref{eq:crystallization_rate_core}), we can write the secular cooling terms ($Q_{S,m}$, $Q_{S,b}$, $Q_{S,c}$, $Q_{S,ic}$), the gravitational energy terms ($Q_{G,b}$, $Q_{G,c}$) and the latent heat terms ($Q_{L,b}$, $Q_{L,b}$) as a function of the cooling rate at the CMB. In addition, since we can write explicitly the heat flow at the surface ($Q_P$) and the radiogenic heating terms ($Q_{R,m}$, $Q_{R,b}$, $Q_{R,c}$, $Q_{R,ic}$) as a function of time, we obtain
\begin{equation}
    \begin{array}{rcl}
         &   & \hspace*{-1.5cm} Q_P-Q_{R,m}-Q_{R,b}-Q_{R,c}-Q_{R,ic}                         \\
         & = & \left[\tilde{Q}_{S,m}+\tilde{Q}_{S,b}+\tilde{Q}_{S,c}+\tilde{Q}_{S,ic}\right. \\
         & + & \tilde{Q}_{G,b}+\tilde{Q}_{G,c}
            +\left.\tilde{Q}_{L,b}+\tilde{Q}_{L,c}\right]\frac{\mathrm{d}T_{CMB}}{\mathrm{d}t},
    \end{array}
    \label{eq:energy_budget}
\end{equation}
where the functions $\tilde{Q}_{X}(r)=Q_{X}(r,t)(\mathrm{d}T_{CMB}(t)/\mathrm{d}t)^{-1}$ only depend on $r$.

We obtain the thermal evolution of the planet by solving numerically Eq. \ref{eq:energy_budget} as an initial value problem with an explicit Runge-Kutta method of order 5. Since the internal structure is pre-computed at discrete integer degrees, each integration of Eq. \ref{eq:energy_budget} uses the structure calculated at the closest available temperature to calculate the terms involved in the energy budget.

\sloppy

We determine the initial CMB temperature $T_{CMB_0}=5710~\mathrm{K}$, and therefore the initial BMO thickness $L_{b_0}=952~\mathrm{km}$, by assessing the temperature at which the BMO adiabat (Eq. \ref{eq:adiabat_BMO}) is tangent to the mantle liquidus (Eq. \ref{eq:liquidus_BMO}), both computed at the initial composition.
The initial conditions for the composition of the BMO and the core are $\xi_{b_0}=16~\mathrm{wt\%}$ \citep{caracas_2019,blanc_2020} and $\xi_{c_0}=5.6~\mathrm{wt\%}$ \citep{masters_2003,gubbins_2013,labrosse_2014,labrosse_2015}.

\fussy

In our model, both the thickness and crystallization rate of the BMO tend toward zero (\textit{e.g.} Fig. \ref{fig:evolution_nominal}). As a result, the BMO becomes very thin and never completely solidifies. This asymptotic behavior, similar to the exponential decay of the BMO thickness obtained by \citet{labrosse_2007}, arises from several assumptions used in our model. In particular, we assume that the BMO is adiabatic and well-mixed, and that crystallization follows a linear phase diagram. However, in reality, the BMO is expected to follow a more complex crystallization sequence, leading to a rheological transition and eventual solidification.
Since these features are not explicitly considered in our model, we instead employ a heuristic criterion to determine the crystallization time of the BMO. We define the BMO as fully solidified when the temperature difference across it reaches 1 K, which corresponds to a 1.4 km BMO. This choice allows to obtain an Earth-like nominal model, where the BMO is either fully crystallized at present time, or where the BMO is sufficiently small for solid mantle convection to concentrate the remaining liquid into melt pockets \citep{labrosse_2007,orourke_2020}. This means that the solidification of the BMO is not self-consistently determined by the model, since the choice of this criterion directly influences the crystallization timescale. However, the energy budget is largely unaffected by this criterion, as it becomes involved only when the BMO becomes very thin.

\subsection{Buoyancy evolution}
\label{sec:model_buoyancy}

The competition between thermocompositional sources (\ref{sec:model_buoyancy_sources}) and adiabatic sinks (\ref{sec:model_buoyancy_sinks}) of buoyancy determines the convective stability (\ref{sec:model_buoyancy_stability}) of the BMO and the core.
Within the Boussinesq approximation, thermal and compositional buoyancy sources can be combined into a codensity (density anomaly) field $C$, defined by \citet{braginsky_1995} as
\begin{equation}
    C=\bar{\alpha}\bar{\rho}T'+\alpha_c\bar{\rho}\xi',
    \label{eq:codensity_definition}
\end{equation}
where the temperature $T'$ and the mass fraction $\xi'$ are deviations from an adiabatic well-mixed reference state, $\bar{\rho}$ is the average density of the layer (BMO or core), $\bar{\alpha}$ is the average thermal expansion coefficient and $\alpha_c$ is a compositional expansion coefficient (Eq. \ref{eq:compositional_expansivity_BMO} and \ref{eq:compositional_expansivity_core}). As a result, we assume that the temperature and composition fields have the same effective diffusivity $\kappa$, neglecting double-diffusivity effects.
We can then write a single codensity transport equation
\begin{equation}
    \frac{\partial C}{\partial t}+\mathbf{v}\cdot\nabla C=\kappa \nabla^2 C + \epsilon,
    \label{eq:codensity_equation}
\end{equation}
where $\mathbf{v}$ is the local flow velocity and $\epsilon$ corresponds to the volumetric sources of buoyancy.
The total codensity can be separated into a steady background profile $C_0(r)$, with superimposed fluctuations $C'(r,\theta,\varphi,t)$.
Eq. \ref{eq:codensity_equation} can then be decomposed into a conduction equation
\begin{equation}
    0=\kappa \nabla^2 C_0 + \epsilon,
    \label{eq:codensity_conduction}
\end{equation}
and a perturbation equation
\begin{equation}
    \frac{\partial C'}{\partial t}+\mathbf{v}\cdot\nabla (C_0+C')=\kappa \nabla^2 C'.
    \label{eq:codensity_perturbation}
\end{equation}

\subsubsection{Thermal and compositional sources}
\label{sec:model_buoyancy_sources}

The conduction equation (Eq. \ref{eq:codensity_conduction}), which includes the volumetric sources of buoyancy, allows us to compare the relative influence of thermal and compositional buoyancy sources on convection \citep{davies_2011a} in both the BMO and the core.
The thermal buoyancy sources are the heat sources in the energy budget. This includes radiogenic heating ($q_{R,b}$ and $q_{R,c}$) (Eq. \ref{eq:codensity_radio}), secular cooling ($q_{S,b}$ and $q_{S,c}$) (Eq. \ref{eq:codensity_secular}), viscous or ohmic dissipation of the motion produced by gravitational redistribution of $\mathrm{FeO}$ ($q_{G,b}$) or light elements ($q_{G,c}$) (Eq. \ref{eq:codensity_dissipation}), latent heat released at the ICB ($q_{L,c}$) (Eq. \ref{eq:codensity_latent}), and heat sources at the CMB ($q_{CMB}$) or ICB ($q_{ICB}$) (Eq. \ref{eq:codensity_CMB_ICB}).
The compositional buoyancy sources correspond to the progressive release of dense $\mathrm{FeO}$ in the BMO ($q_{C,b}$), or light elements in the core ($q_{C,c}$) (Eq. \ref{eq:codensity_compositional}). Since the layers are well-mixed, these compositional terms behave as equivalent uniform compositional sinks.

For each buoyancy source, we solve separate conduction equations using the appropriate volumetric source term (Appendix \ref{app:buoyancy_source}). For thermal sources, volumetric source terms are obtained by multiplying the corresponding volumetric heat source by $\bar{\alpha}/\bar{c}_P$, where $\bar{c}_P$ is the average value of the specific heat capacity at constant pressure within the BMO or the core.
For compositional sources, volumetric source terms are obtained by multiplying the rate of release of $\mathrm{FeO}$, or light elements, by $\bar{\rho}\alpha_c$.
This gives a conductive codensity gradient that is functionally equivalent to
the cotemperature gradients of \citet{davies_2011a}, namely
\begin{equation}
    C_0'(r)=-\frac{1}{\kappa_T}\frac{1}{r^2}\int_{r_0}^rx^2\epsilon (x)\mathrm{d}x+\frac{A}{r^2},
    \label{eq:codensity_gradient},
\end{equation}
where $A$ is an integration constant, $\kappa_T$ is a turbulent, not molecular, diffusivity \citep{gubbins_2015}, and $r_0$ is a reference radius.

\sloppy

For each buoyancy source, the integration constant is determined using a boundary condition corresponding to a mass anomaly flux at the appropriate interface (Appendix \ref{app:buoyancy_source}) \citep{gubbins_2015}.
The mass anomaly fluxes, defined as
\begin{equation}
    q(r) = \kappa_T C_0'(r),
    \label{eq:mass_anomaly_flux}
\end{equation}
correspond to the thermal or chemical anomaly that passes through a unit of surface area per unit time \citep{aubert_2009}, and are expressed in $\mathrm{kg.m^{-2}.s^{-1}}$.
The buoyancy sources act as negative, destabilizing codensity gradients and mass anomaly fluxes.
For thermal sources, the mass anomaly fluxes are obtained by multiplying the corresponding heat flux (expressed in $\mathrm{W.m^{-2}}$), which corresponds to a heat flow (expressed in $\mathrm{W}$) per unit of surface at the appropriate interface, by $\bar{\alpha}/\bar{c}_P$.
For compositional sources, the mass anomaly fluxes at the CMB are zero because we neglect exchanges between the BMO and the core, as well as thermodiffusion and barodiffusion, which correspond to negligible buoyancy sinks \citep{davies_2011a}.

\fussy

\subsubsection{Equivalent adiabatic sinks}
\label{sec:model_buoyancy_sinks}

The buoyancy profile also includes sinks that contribute to stabilize the convecting layers.
Thermal conduction corresponds to the main equivalent buoyancy sink in the system. It removes heat from convection and acts as a positive and stabilizing codensity gradient and mass anomaly flux. The adiabatic codensity gradients are obtained by multiplying the adiabatic temperature gradient by $\bar{\rho}\bar{\alpha}$ (Eq. \ref{eq:codensity_definition}). The corresponding mass anomaly fluxes of the BMO and the core are then
\begin{equation}
    \left\{
    \begin{array}{l}
        q_{A,b} = -\kappa_{m_b} \bar{\rho}\bar{\alpha}\frac{\mathrm{d}T}{\mathrm{d}r}, \\
        q_{A,c} = -\kappa_{m_c} \bar{\rho}\bar{\alpha}\frac{\mathrm{d}T}{\mathrm{d}r},
    \end{array}
    \right.
    \label{eq:codensity_adiabatic}
\end{equation}
where $\kappa_{m_b}$ and $\kappa_{m_c}$ are molecular diffusivities and not turbulent ones \citep{gubbins_2015}. These diffusivities can be expressed as a function of the thermal conductivities in the BMO $k_b$ and the core $k_c$, with $\kappa_m=k/(\bar{\rho}\bar{c}_P)$. However, the thermal conductivities, which depend on pressure, temperature and composition, remain uncertain, both in the BMO \citep{holmstrom_2018,soubiran_2018,stixrude_2020} or in the core \citep{ohta_2016,basu_2020,pozzo_2022}. In our model, we therefore set conservative values with $k_{b}=8~\mathrm{W.m^{-1}.K^{-1}}$ and $k_{c}=70~\mathrm{W.m^{-1}.K^{-1}}$.

\subsubsection{Convective stability criterion}
\label{sec:model_buoyancy_stability}

The total mass anomaly flux, which corresponds to the sum of the mass anomaly fluxes associated with the thermocompositional sources and sinks of buoyancy, can be used to determine the local convective stability of the BMO and the core \citep{davies_2011a,pozzo_2012,gubbins_2015}. The reason for using mass anomaly fluxes instead of codensity gradients is that the destabilizing codensity gradients are associated with a turbulent diffusivity whereas the equivalent adiabatic flux uses a molecular diffusivity \citep{gubbins_2015}.

The total mass anomaly flux of the BMO and the core are therefore defined as
\begin{equation}
    \left\{
    \begin{array}{ll}
        q_{b}=q_{R,b}+q_{S,b}+q_{G,b}+q_{L,b}+q_{CMB}+q_{A,b} \\
        q_{c}=q_{R,c}+q_{S,c}+q_{G,c}+q_{L,c}+q_{ICB}+q_{A,c}
    \end{array}
    \right..
    \label{eq:mass_anomaly_flux_total}
\end{equation}
The total mass anomaly flux can be used to determine the local convective stability of the layer and in particular the presence of stably-stratified layers. Convecting and stably-stratified regions are given by an equivalent Schwarzschild criterion
\begin{equation}
    \left\{
    \begin{array}{ll}
        q(r) < 0    & \mathrm{if~convecting} \\
        q(r) \geq 0 & \mathrm{if~stable}
    \end{array}
    \right..
    \label{eq:schwarzschild_criterion}
\end{equation}
The stability boundary is defined where the equivalent adiabatic mass anomaly flux is balanced by the destabilizing mass anomaly fluxes associated with the thermocompositional buoyancy sources. If the sum of the destabilizing mass anomaly fluxes is smaller, in magnitude, than the equivalent adiabatic mass anomaly flux, the BMO and the core may be partly or fully stably-stratified. This corresponds to destabilizing codensity profiles being less steep than the equivalent adiabatic profile.

The stably-stratified layers defined with Eq. \ref{eq:schwarzschild_criterion} are not properly included in our thermal evolution model. In particular, we overestimate the secular cooling of these conductive layers assuming an adiabatic profile throughout the core. Considering such layers would require to compute the buoyancy profiles at every time step when integrating the energy budget.

\subsection{Magnetic evolution}
\label{sec:model_magnetic}

In this model, the magnetic evolution of the planet is obtained by assessing whether the potential dynamo layers, \textit{i.e.} the BMO and the outer core, (i) are convecting, (ii) have sufficient power to balance ohmic dissipation and sustain a dynamo, and (iii) have a magnetic induction larger than magnetic diffusion.
In practice, we assess the layers' convective stability using the criterion in §\ref{sec:model_buoyancy_stability}. Next, we use entropy budgets (§\ref{sec:model_magnetic_entropy}), initially developed for the Earth's core \citep{gubbins_2003,gubbins_2004} and later adapted for the BMO \citep{blanc_2020,orourke_2020}, to assess the dissipation criterion. Then, we estimate the magnetic Reynolds number (§\ref{sec:model_magnetic_reynolds_number}) to assess the induction criterion. Finally, we estimate the intensity of the magnetic field (§\ref{sec:model_magnetic_field_intensity}) powered in each layer.

\subsubsection{Entropy budget}
\label{sec:model_magnetic_entropy}

We write the entropy budget of the BMO and the core as
\begin{small}
    \begin{eqnarray}
        E_{R,b}+E_{S,b}+E_{G,b}+E_{L,b}+E_{CMB}=E_{k,b}+E_{\alpha,b}+E_{\Phi,b}, \\
        \label{eq:entropy_budget_BMO}
        E_{R,c}+E_{S,c}+E_{G,c}+E_{L,c}+E_{ICB}=E_{k,c}+E_{\alpha,c}+E_{\Phi,c},
        \label{eq:entropy_budget_core}
    \end{eqnarray}
\end{small}z
where all terms are rates of entropy production \citep{nimmo_2015}, expressed in $\mathrm{W.K^{-1}}$.
The left-hand side of the equations corresponds to source terms, including the entropy of radiogenic heating $E_R$ (Eq. \ref{eq:entropy_radio}), the entropy of secular cooling $E_S$ (Eq. \ref{eq:entropy_secular}), the entropy of gravitational energy $E_G$ due to the release of $\mathrm{FeO}$ and light elements (Eq. \ref{eq:entropy_gravi}), the entropy of latent heat $E_L$ (Eq. \ref{eq:entropy_latent}), as well as the entropy $E_{CMB}$ and $E_{ICB}$ associated with the heat source at the CMB and the ICB (Eq. \ref{eq:entropy_CMB_ICB}).
The right-hand side corresponds to sink terms, including the entropy of thermal diffusion $E_k$ (Eq. \ref{eq:entropy_thermal}), the entropy of molecular diffusion $E_\alpha$ (Eq. \ref{eq:entropy_compositional}) and the combined entropy of ohmic and viscous dissipation $E_\Phi$.
We neglect entropy terms associated with the heat of reaction, pressure heating and pressure effects on freezing, in both the BMO \citep{blanc_2020,orourke_2020} and the core \citep{gubbins_2003,gubbins_2004}.

\sloppy

Each term can be written as the product of the corresponding heat flow and an efficiency factor, divided by an effective operating temperature (Appendix \ref{app:entropy_budget}). The efficiency factors correspond to the thermodynamic efficiency of the buoyancy sources. Buoyancy sources that are produced at the bottom of the layer are more efficient than buoyancy sources distributed throughout the layer.

\fussy

The dissipation term $E_\Phi$ can be expressed as the sum of ohmic and viscous dissipation. Since $E_\Phi$ is the only unknown in the entropy budgets, we can obtain this term by subtracting the thermal and molecular diffusion terms from the source terms. In addition, we assume that the viscous dissipation term is negligible in comparison with the ohmic dissipation term in both the BMO \citep{blanc_2020,orourke_2020} and the core \citep{aubert_2009}. $E_\Phi$ can thus be related to the gravitational energy that sustains convection \citep{buffett_1996}, which means that it corresponds to the energy fraction that contributes to generate a magnetic field.
In our model, we use $E_\Phi$ as a first criterion for the operation of a dynamo in the BMO \citep{blanc_2020} and the core \citep{davies_2015,greenwood_2021,driscoll_2023}. If $E_\Phi > 0$, the power available is sufficient to sustain a dynamo. Since the total dissipation must exceed the ohmic heating produced in the layer, this criterion actually serves as a lower bound for dynamo operation.

\subsubsection{Magnetic Reynolds number}
\label{sec:model_magnetic_reynolds_number}

\sloppy

In addition to the entropy criterion, the magnetic Reynolds number has to exceed a critical value for magnetic induction to sufficiently overcome magnetic diffusion and a dynamo to operate continuously in the convecting layer. The magnetic Reynolds numbers of the BMO and the core are defined as
\begin{equation}
    \left\{
    \begin{array}{ll}
        R_{m_{b}}=U_{b}L_{b}\mu_0\sigma_{b} \\
        R_{m_{c}}=U_{c}L_{c}\mu_0\sigma_{c}
    \end{array}
    \right.,
    \label{eq:magnetic_reynolds_number}
\end{equation}
where $U_b$, $L_b$, $\sigma_b$ and $U_c$, $L_c$, $\sigma_{c}$ are respectively the characteristic convective velocity, the thickness and the electrical conductivity of the BMO and the core.
Using these definitions, a self-sustained magnetic field might be expected in the respective layers if $R_{m}>40$ \citep{christensen_2006}. This critical value, obtained from magnetohydrodynamic (MHD) simulations in the Earth's core, is not universal but is commonly used as a criterion for dynamo operation, so we adopt it for both the BMO and the core, separately.
Another condition, depending on the approximation of the mean field equations and the interplay between the BMO and the core is the dynamo number \citep{ziegler_2013}. However, we do not consider this criterion here.
Electrical conductivities, like thermal conductivities, remain uncertain, both in the BMO \citep{holmstrom_2018,soubiran_2018,stixrude_2020} and the core \citep{ohta_2016,basu_2020,pozzo_2022}. We therefore set conservative values of $\sigma_{b}=3 \times 10^{4}~\mathrm{S.m^{-1}}$ and $\sigma_{c}=1 \times 10^{6}~\mathrm{S.m^{-1}}$.

The convective velocities are obtained using scaling laws derived from MHD numerical simulations \citep{aubert_2009}. The scaling laws relate the dimensionless convective velocity, \textit{i.e.} a Rossby number $Ro$, to the total dissipation, \textit{i.e.} a dimensionless convective power $\tilde{\Phi}$, determined from the entropy budgets by multiplying the dissipation term by an effective dissipation temperature (Appendix \ref{app:scaling_law}).
These scaling laws, initially developed for the Earth's core, may differ for the BMO since a silicate dynamo would operate in a thinner shell and have a larger Prandtl number. Nevertheless, we use the same scaling law in the BMO and the core, which gives
\begin{equation}
    \left.
    \begin{array}{l}
        Ro_b=\frac{U_{b}}{L_{b}\Omega} \\
        Ro_c=\frac{U_{c}}{L_{c}\Omega}
    \end{array}
    \right\}
    =c_{U}\tilde{\Phi}^{\alpha_{U}},
    \label{eq:U_scaling_law}
\end{equation}
where $c_U$ and $\alpha_U$ are respectively the prefactor and exponent of the scaling law, obtained from a least-square fit on a set of dynamo simulations \citep{aubert_2009}, and $\Omega$ is the rotation rate of the planet. Since the evolution of the Earth's rotation rate over time remains uncertain \citep{touma_1994,daher_2021}, we choose a linear evolution of the length of day from $5~\mathrm{h}$ \citep{canup_2001,cuk_2012} at $t=0$ to one Earth day at $t=4.5~\mathrm{Gyr}$.

\fussy

The exponents of the scaling laws can also be derived analytically from various force balance arguments \citep{christensen_2010}. In the mixing length theory (MLT) regime, the balance between the inertia term and the buoyancy term in the momentum equation implies that the convective velocity scales as $\tilde{\Phi}^{1/3}$. In the Coriolis-Inertia-Archimedean (CIA) regime, the balance between the Coriolis term, the inertia term and the buoyancy term implies that the velocity scales as $\tilde{\Phi}^{2/5}$. Finally, in the Magnetic-Archimedean-Coriolis (MAC) regime, the balance between the Lorentz force, the buoyancy force and the Coriolis force implies a $\tilde{\Phi}^{1/2}$ scaling. The prefactors of the scaling laws are obtained using a least-square fit on the set of dynamo simulations of \citet{aubert_2009} (Appendix \ref{app:scaling_law}). The scaling law parameters are defined in Table S8.

\subsubsection{Magnetic field intensity}
\label{sec:model_magnetic_field_intensity}

The magnetic field of the BMO and the core are also obtained using scaling laws derived from MHD numerical simulations \citep{aubert_2009}. The scaling laws relate the dimensionless magnetic field $\tilde{B}$ to the dimensionless convective power (Appendix \ref{app:scaling_law}). As with the convective velocity, we use the same scaling law in the BMO and the core which gives
\begin{equation}
    \left.
    \begin{array}{l}
        \tilde{B}_{b}=\frac{B_{b}}{\sqrt{\bar{\rho}_{b}\mu_0}\Omega L_{b}} \\
        \tilde{B}_{c}=\frac{B_{c}}{\sqrt{\bar{\rho}_{c}\mu_0}\Omega L_{c}}
    \end{array}
    \right\}
    =c_Bf_\Omega^{1/2}\tilde{\Phi}^{\alpha_{B}},
    \label{eq:B_scaling_law}
\end{equation}
where $c_B$ and $\alpha_B$ are the prefactor and exponent of the scaling law, obtained from a least-square fit of dynamo simulations \citep{aubert_2009}. $B_b$, $\bar{\rho}_{b}$ and $B_c$, $\bar{\rho}_{c}$ are the internal magnetic field and average density of the BMO and the core, respectively. $f_\Omega$ is the ohmic dissipation fraction of the convective power.
Following \citet{aubert_2009} for the core and \citet{orourke_2020} for the BMO, we set $f_\Omega=0.9$ in both layers.
The exponents of the MLT, CIA and MAC balance are respectively $1/3$, $3/10$ and $1/4$ \citep{christensen_2010}. The prefactors are also obtained using a least-square fit on the set of dynamo simulations of \citet{aubert_2009} (Appendix \ref{app:scaling_law}). The scaling law parameters are defined in Table S8.

The surface intensity of the magnetic field produced by the BMO and the core are obtained using a cubic dependence on radius, corresponding to the continuation of the dipole toward the surface.
In the BMO, we use a constant prefactor developed for the core \citep{christensen_2006}, and applied to the BMO \citep{orourke_2020,stixrude_2020}
\begin{equation}
    B_{P,b}=\frac{1}{7}\left(\frac{R_{BMO}}{R_P}\right)^3 B_b.
    \label{eq:surface_magnetic_field_BMO}
\end{equation}
In the core, we first determine the relative strength of the internal field and the dipole field at the CMB using a scaling law \citep{aubert_2009}
\begin{equation}
    b_{dip,c}=c_{dip}\left(1-\frac{R_{ICB}}{R_{CMB}}\right)\left(1+f_{c}\right),
    \label{eq:internal_outer_dipole_core}
\end{equation}
where $f_{c}$ is the fraction of the mass anomaly flow rate produced at the inner boundary of the core, and $c_{dip}=7.3$ is a prefactor obtained from a least-square fit of dynamo simulations \citep{aubert_2009}. A mass anomaly flow rate (expressed in $\mathrm{kg.s^{-1}}$), is the thermal, or chemical, anomaly passing through a surface area per unit time and equals the mass anomaly flux integrated over a surface area. The fraction of the mass anomaly flow rate is defined as
\begin{equation}
    f_{c}=\frac{F_{ICB}}{F_{ICB}+F_{CMB}},
    \label{eq:mass_anomaly_flux_ratio_core}
\end{equation}
where $F_{ICB}$ and $F_{CMB}$ are the mass anomaly flow rates at the ICB and the CMB.
Then, we determine the true dipole moment of the core with
\begin{equation}
    \mathcal{M}_c=\frac{4\pi R_{CMB}^3}{\sqrt{2}\mu_0}\frac{B_{c}}{b_{dip,c}},
    \label{eq:TDM_core}
\end{equation}
and finally obtain the intensity of the magnetic field at the surface with
\begin{equation}
    B_{P,c}=\frac{\mu_0}{4\pi R_P^3}\mathcal{M}_{c}.
    \label{eq:surface_magnetic_field_core}
\end{equation}

\section{Earth-like planet nominal evolution}
\label{sec:results_nominal}

Our nominal model uses the assumptions detailed in §\ref{sec:model}, together with the parameters defined in tables S1-S8, to determine the structural (\ref{sec:results_nominal_structure}), thermal (\ref{sec:results_nominal_thermal}), buoyancy (\ref{sec:results_nominal_buoyancy}) and magnetic (\ref{sec:results_nominal_magnetic}) evolution of the planet.

\subsection{Structural evolution}
\label{sec:results_nominal_structure}

\sloppy

\begin{figure*}[!ht]
    \centering
    \includegraphics[width=\linewidth]{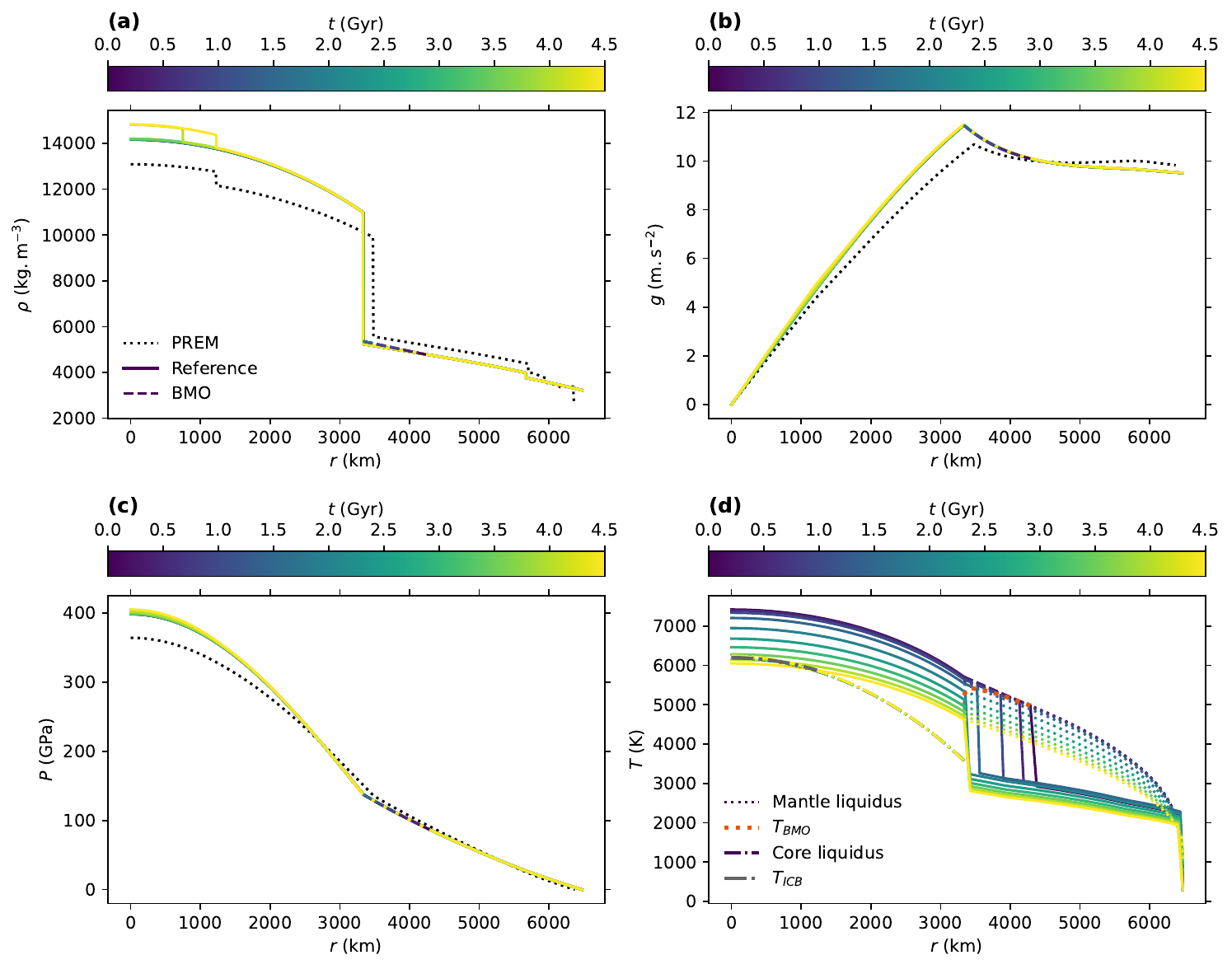}
    \caption{Density $\rho$ (a), gravity $g$ (b), pressure $P$ (c) and temperature $T$ (d) as a function of the radius $r$. Solid lines correspond to profiles obtained using the reference model. Dashed lines correspond to profiles using the BMO model. The colors correspond to the time $t$. Black dotted lines correspond to the Preliminary Reference Earth Model (PREM) defined by \citet{dziewonski_1981}. Panel (d): Colored dotted lines and dash-dotted lines correspond respectively to the mantle and core liquidus. The thick dotted and dash-dotted lines correspond respectively to the temperature at the top of the BMO and at the ICB.}
    \label{fig:structure_nominal}
\end{figure*}

Fig. \ref{fig:structure_nominal} shows the time evolution of the density, gravity, pressure and temperature as a function of radius for the nominal model. The density profile of the BMO corresponds to a linear fit of the reference profile defined for the solid mantle, while the gravity and pressure profiles of the BMO are identical to the reference profile (§\ref{sec:model_structure_BMO}). In the supplementary material, we provide the time evolution of the Grüneisen parameter, thermal expansion coefficient, isothermal bulk modulus, specific heat capacity at constant pressure, mantle viscosity, and mantle thermal conductivity for the nominal model (Fig. S1).

\fussy

The density profile shows a density jump corresponding to the CMB at approximately 3140 km under the surface the planet (Fig. \ref{fig:structure_nominal}a). In the mantle, the density profile also shows two density jumps at approximately 810 km and 2600 km. They correspond respectively to the peridotite to perovskite and to the perovskite to post-perovskite solid phase transitions. In the core, after 3.8 Gyr, the density profile shows density jumps corresponding to the inner core nucleation. These jumps are also visible as slope discontinuities on the gravity (Fig. \ref{fig:structure_nominal}b) and pressure (Fig. \ref{fig:structure_nominal}c) profiles, in particular at the CMB. Since the integration of the internal structure is conducted using an imposed planet mass, the radius of the planet, of the CMB and of the solid phase transitions slightly change over time as the temperature of the planet decreases and upon crystallization of the inner core. For the same reasons, the gravity and pressure profiles also change slowly over time.

In the mantle, the temperature profile (Fig. \ref{fig:structure_nominal}d) is defined by the BMO model (Eq. \ref{eq:adiabat_BMO}) and by the mantle thermal boundary layer model (Eq. \ref{eq:dTdr_UBL}-\ref{eq:dTdr_LBL}). The BMO, defined by the intersection between the adiabatic profile and the mantle liquidus, crystallizes progressively from the top down until it fully solidifies at 2.8 Gyr. The mantle liquidus decreases over time as the BMO is increasingly enriched in $\mathrm{FeO}$ (Eq. \ref{eq:liquidus_BMO}), producing a depression of the melting curve as the BMO crystallizes \citep{boukare_2015,ballmer_2017,blanc_2020}.
In the core, the temperature profile is defined by the reference model (Eq. \ref{eq:reference_adiabat}). The inner core starts to crystallize at 3.8 Gyr, when the temperature profile of the core intersects the core liquidus at the center of the planet. The core liquidus is fixed over time because it is defined for an imposed composition (Eq. \ref{eq:liquidus}).

\subsection{Thermal evolution}
\label{sec:results_nominal_thermal}

\begin{figure*}[!ht]
    \centering
    \includegraphics[width=\linewidth]{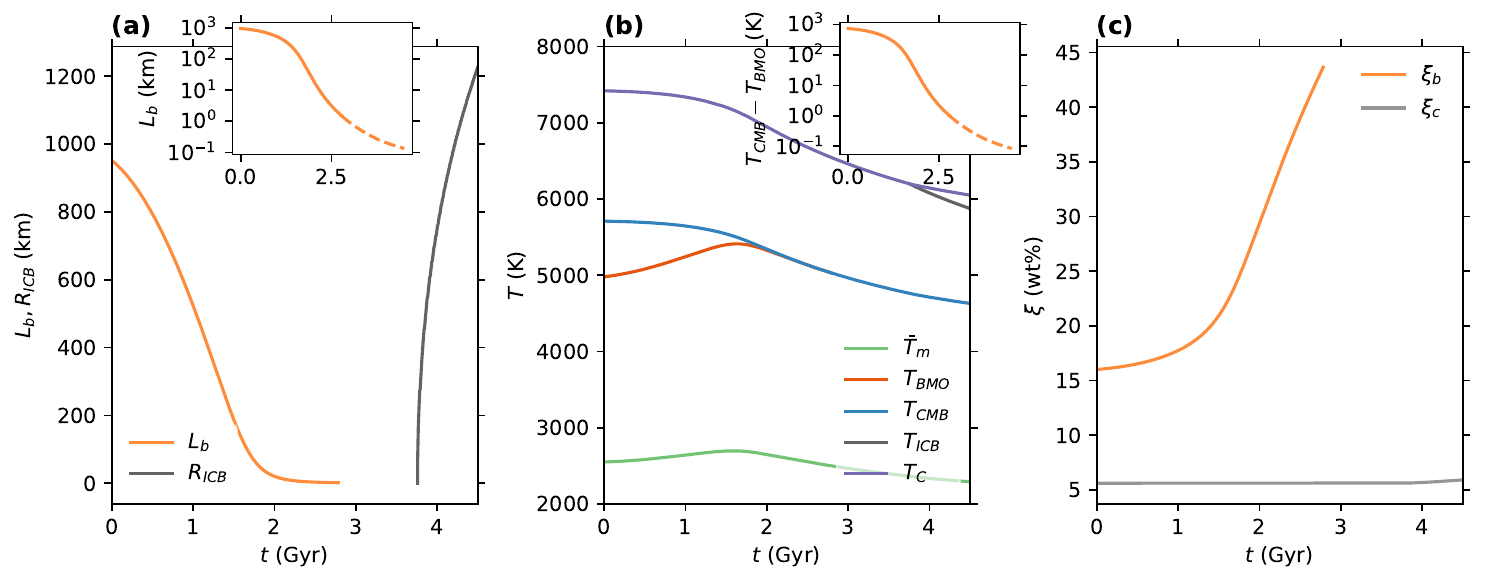}
    \caption{(a) BMO thickness $L_b$ and ICB radius $R_{ICB}$ as a function of time $t$. Inset: $L_b$, in logarithmic scale, as a function of $t$. The dashed line corresponds to thicknesses where the temperature difference across the BMO is less than 1 K. (b) Temperature of the mantle $\bar{T}_m$, at the top of the BMO $T_{BMO}$, at the CMB $T_{CMB}$, at the ICB $T_{ICB}$ and at the center of the planet $T_C$ as a function of time $t$. Inset: Temperature difference across the BMO $T_{CMB}-T_{BMO}$, in logarithmic scale, as a function of $t$. The dashed line corresponds to temperature differences across the BMO that are less than 1 K. (c) Mass fraction of the BMO $\mathrm{FeO}$ $\xi_b$ and of the core light elements $\xi_c$ as a function of time $t$.}
    \label{fig:evolution_nominal}
\end{figure*}

Fig. \ref{fig:evolution_nominal} shows the thickness of the BMO and the radius of the inner core, the temperatures of the mantle, at the top of the BMO, at the CMB, at the ICB and at the center of the planet, as well as the composition of the BMO and the core as a function of time for the nominal model.
The planet initially starts with a fully liquid core and a thick BMO (Fig. \ref{fig:evolution_nominal}a). As the BMO progressively crystallizes, the temperature at the CMB decreases whereas the average temperature of the mantle and the temperature at the top of the BMO increase, as the depth of this upper interface increases (Fig. \ref{fig:evolution_nominal}b).
Before 1.6 Gyr, the average mantle temperature and the temperature at the top of the BMO are dominated by a fast crystallization of the BMO, which explains the temperature increase. After 1.6 Gyr, the BMO crystallization rate significantly diminishes, in relation with observed temperature decrease, caused by overall planet cooling. At 2.8 Gyr, the temperature at the top of the BMO merges with that at the CMB (within 1 K), which means that the BMO is fully crystallized (within 1 km), according to the criterion defined in §\ref{sec:model_thermal_solution}. During the crystallization, the BMO is progressively enriched in $\mathrm{FeO}$ (Fig. \ref{fig:evolution_nominal}c).
Between 2.8 Gyr and 3.8 Gyr, the planet progressively cools, with a fully crystallized BMO and a fully molten core.
At 3.8 Gyr, the central temperature of the planet becomes smaller than the core liquidus (Fig. \ref{fig:evolution_nominal}b), leading to core nucleation. As the size of the inner core increases (Fig. \ref{fig:evolution_nominal}a), the outer core is progressively enriched in light elements (Fig. \ref{fig:evolution_nominal}c).

\begin{figure*}[!ht]
    \centering
    \includegraphics[width=\linewidth]{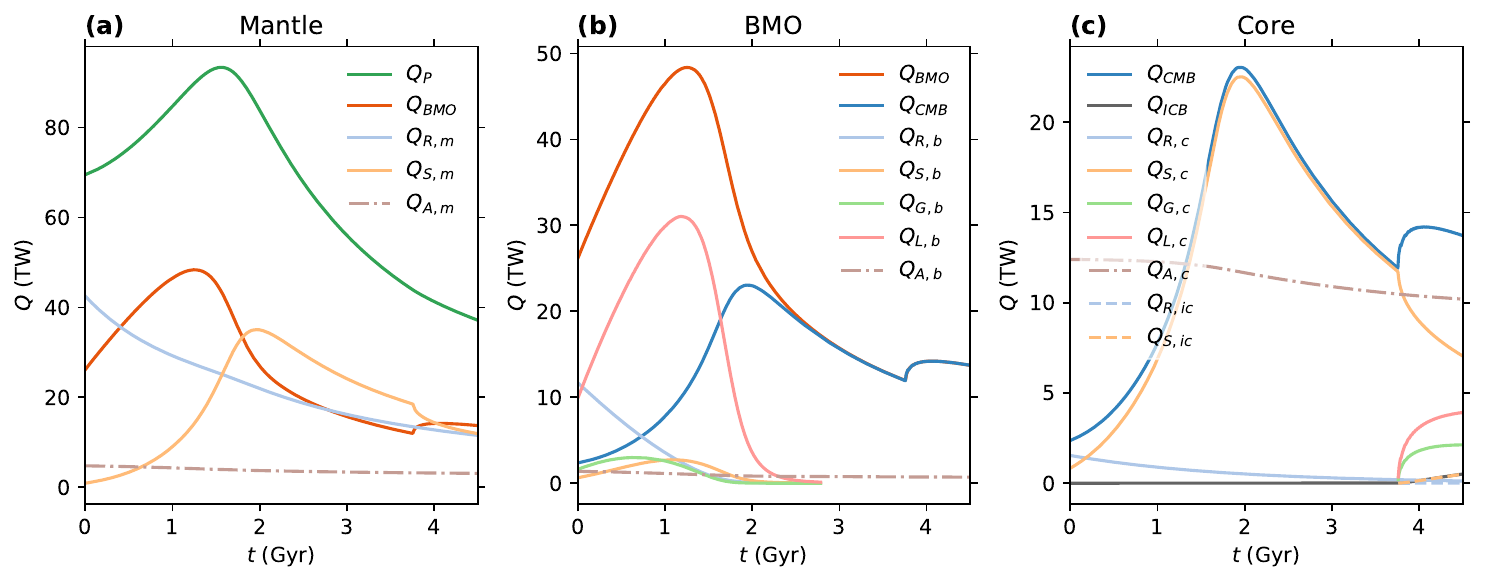}
    \caption{Heat flows $Q$ in the mantle (a), the BMO (b) and the core (c) as a function of time $t$. The radiogenic heating, secular cooling, gravitational energy and latent heat terms are denoted with the subscripts $R$, $S$, $G$ and $L$. The adiabatic heat flows at the top of the mantle, BMO, and core are denoted with the subscript $A$. The heat flows at the surface of the planet, at the top of the BMO, at the CMB and at the ICB are denoted with the subscripts $P$, $BMO$, $CMB$ and $ICB$.}
    \label{fig:energy_nominal}
\end{figure*}

Fig. \ref{fig:energy_nominal} shows the heat flows in the mantle, the BMO and the core as a function of time for the nominal model.
The heat flow at the surface of the planet is mainly driven by the heat flow at the top of the BMO when it exists, \textit{i.e.} by the activity in the BMO and the core (Fig. \ref{fig:energy_nominal}a).
The surface heat flow $Q_P$ (Eq. \ref{eq:heat_flow_surface}) initially increases during 1.6 Gyr, due to an increasing temperature difference across the mantle, explained by the increasing depth of the interface between the BMO and the mantle. After 1.6 Gyr, the surface heat flow decreases as the crystallization rate of the BMO decreases (Fig. \ref{fig:evolution_nominal}a).
Similarly, the secular cooling of the mantle $Q_{S,m}$ (Eq. \ref{eq:secular_mantle}) initially increases for 2.0 Gyr as the cooling rate at the CMB increases (Fig. \ref{fig:evolution_nominal}b) and as the volume of the mantle progressively expands, due to the crystallization of the BMO. After 2.0 Gyr, the secular cooling of the mantle decreases, consistently with a decreasing cooling rate (Fig. \ref{fig:evolution_nominal}b).
Although the volume of the mantle initially increases, the radiogenic heating of the mantle $Q_{R,m}$ (Eq. \ref{eq:radio_mantle}) decreases monotonously due to radioactive decay.
Finally, the adiabatic heat flow at the top of the mantle $Q_{A,m}$ (Eq. \ref{eq:Q_A_mantle}) is lower than the heat flow at the surface, consistent with the operation of thermal convection in the mantle.

The heat flow at the surface of the BMO, $Q_{BMO}$, is mainly driven by the latent heat produced upon crystallization of the BMO (Fig. \ref{fig:energy_nominal}b).
The latent heat flow, and thus the heat flow at the surface of the BMO, initially increase for 1.2 Gyr before decreasing until the BMO is fully crystallized. Since the latent heat $Q_{L,b}$ (Eq. \ref{eq:latent_BMO}) is proportional to the rate of crystallization of the BMO, this change may be explained by the transition from an increasing to a decreasing crystallization rate (Fig. \ref{fig:evolution_nominal}a).
Similarly, the gravitational energy $Q_{G,b}$ (Eq. \ref{eq:gravitational_BMO}) and the secular cooling $Q_{S,b}$ (Eq. \ref{eq:secular_BMO}) initially increase due to increased crystallization (Fig. \ref{fig:evolution_nominal}a) and cooling rates (Fig. \ref{fig:evolution_nominal}b). They eventually decrease as the BMO shrinks and as the crystallization and cooling rates decrease.
The radiogenic heating of the BMO $Q_{R,b}$ (Eq. \ref{eq:radio_BMO}) decreases over time and goes to zero due a BMO decreasing in size, as well as radioactive decay.
Finally, the adiabatic heat flow at the top of the BMO $Q_{A,b}$ (Eq. \ref{eq:Q_A_BMO}) is lower than $Q_{BMO}$, consistent with the operation of thermal convection in the BMO.

Before the nucleation of the inner core, the heat flow at the CMB is mainly driven by the secular cooling of the core (Fig. \ref{fig:energy_nominal}c).
The secular cooling of the core, and thus the heat flow at the CMB, initially increase during 2.0 Gyr before decreasing until the nucleation of the inner core. Since the secular cooling $Q_{S,c}$ (Eq. \ref{eq:secular_core}) is proportional to the cooling rate of the core, this evolution is consistent with the transition from an increasing to a decreasing cooling rate at the CMB (Fig. \ref{fig:evolution_nominal}b). The initial increase of the CMB heat flow is explained by the concomitant crystallisation of the BMO, which increases the surface heat flow, and reduces, after 1.2 Gyr, the cooling rate buffer related to the BMO activity.
After the nucleation of the inner core, the energy budget is significantly modified by contributions from the latent heat $Q_{L,c}$ (Eq. \ref{eq:latent_core}) and gravitational energy $Q_{G,c}$ (Eq. \ref{eq:gravitational_core}) released at the ICB, leading to an increasing heat flow at the CMB. The secular cooling $Q_{S,ic}$ (Eq. \ref{eq:secular_inner_core}) and radiogenic heating $Q_{R,ic}$ (Eq. \ref{eq:radio_inner_core}) of the inner core also contribute to this modification, but to a lesser extent.
In contrast to the mantle and the BMO, the radiogenic heating of the core $Q_{R,c}$ (Eq. \ref{eq:radio_core}) is negligible, except at early times.
Finally, the adiabatic heat flow at the top of the core $Q_{A,c}$ (Eq. \ref{eq:Q_A_core}) is only lower than $Q_{CMB}$ after 1.3 Gyr, indicating the onset of thermal convection in the core.

\subsection{Buoyancy evolution}
\label{sec:results_nominal_buoyancy}

\begin{figure*}[!ht]
    \centering
    \includegraphics[width=\linewidth]{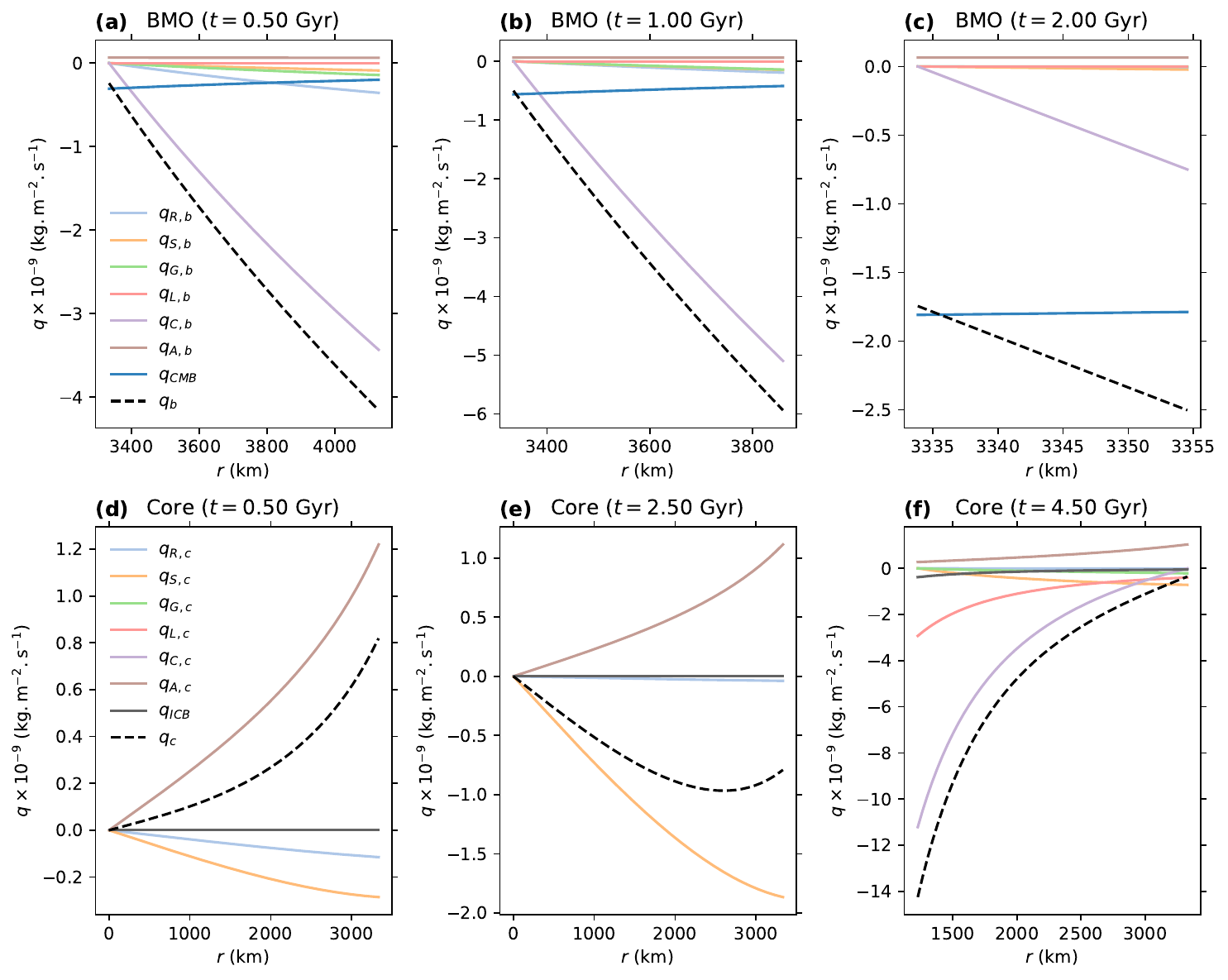}
    \caption{Mass anomaly flux $q$ in the BMO at $t=0.5~\mathrm{Gyr}$ (a), $t=1.0~\mathrm{Gyr}$ (b) and $t=2.0~\mathrm{Gyr}$ (c), and in the outer core at $t=0.5~\mathrm{Gyr}$ (d), $t=2.5~\mathrm{Gyr}$ (e) and $t=4.5~\mathrm{Gyr}$ (f) as a function of radius $r$. Radiogenic heating, secular cooling, dissipation, latent heat, compositional and equivalent adiabatic buoyancy terms are denoted by the subscripts $R$, $S$, $G$, $L$, $C$ and $A$. The contributions from the core at the CMB and from the inner core the ICB are denoted with the subscripts $CMB$ and $ICB$. Dashed lines correspond to the total mass anomaly flux. The equivalent adiabatic terms use respectively $k_{b}=8~\mathrm{W.m^{-1}.K^{-1}}$ and $k_{c}=70~\mathrm{W.m^{-1}.K^{-1}}$.}
    \label{fig:buoyancy_snap_nominal}
\end{figure*}

Fig. \ref{fig:buoyancy_snap_nominal} shows the mass anomaly flux in the BMO and the outer core at different times.
In the BMO, the total mass anomaly flux $q_b$ is negative at all times, which means that the BMO is convective (Fig. \ref{fig:buoyancy_snap_nominal}a-c). This is consistent with the energy budget where $Q_{A,b}<Q_{BMO}$ (Fig. \ref{fig:energy_nominal}b). The positive equivalent adiabatic sink $q_{A,b}$ is negligible in comparison with the negative buoyancy sources. At early times, the convection is mainly driven by the compositional term $q_{C,b}$, related to the release of $\mathrm{FeO}$ upon crystallization of the BMO (Fig. \ref{fig:buoyancy_snap_nominal}a). Close to the CMB, the heat produced by the core $q_{CMB}$ dominates the buoyancy profile. As the BMO crystallizes, the radius fraction dominated by the influence of the core increases (Fig. \ref{fig:buoyancy_snap_nominal}b), until it dominates in the whole BMO (Fig. \ref{fig:buoyancy_snap_nominal}c).
Although the radiogenic heating term $q_{R,b}$ is a major buoyancy source at early times (Fig. \ref{fig:buoyancy_snap_nominal}a), its influence significantly decreases over time due to radioactive decay. The secular cooling $q_{S,b}$ and dissipation $q_{G,b}$ terms are minor sources of buoyancy. The latent heat term $q_{L,b}$ does not contribute to the buoyancy profile since latent heat is released at the top of the BMO (Eq. \ref{eq:codensity_latent}).

In the core, the total mass anomaly flux $q_{c}$ is initially positive before it becomes negative, which means that the core becomes convective after an initial stably-stratified stage (Fig. \ref{fig:buoyancy_snap_nominal}d-f). This is consistent with the energy budget where $Q_{A,c}<Q_{CMB}$ (Fig. \ref{fig:energy_nominal}c) before 1.3 Gyr. At early times, the buoyancy profile is dominated by the positive equivalent adiabatic sink $q_{A,c}$ (Fig. \ref{fig:buoyancy_snap_nominal}d). Subsequently, the negative buoyancy sources overcome the adiabatic sink, facilitating convection (Fig. \ref{fig:buoyancy_snap_nominal}e).
Before nucleation of the inner core, the convection is mainly driven by secular cooling of the core $q_{S,c}$, with a minor and decreasing contribution of radiogenic heating $q_{R,c}$. After nucleation of the inner core, convection is mainly driven by the compositional term $q_{C,c}$, related to the release of light elements, and to a lesser extent by the latent heat term $q_{L,c}$ (Fig. \ref{fig:buoyancy_snap_nominal}f).
The contributions from the inner core $q_{ICB}$, as well as dissipation $q_{G,c}$, are minor sources of buoyancy.

\begin{figure*}[!ht]
    \centering
    \includegraphics[width=\linewidth]{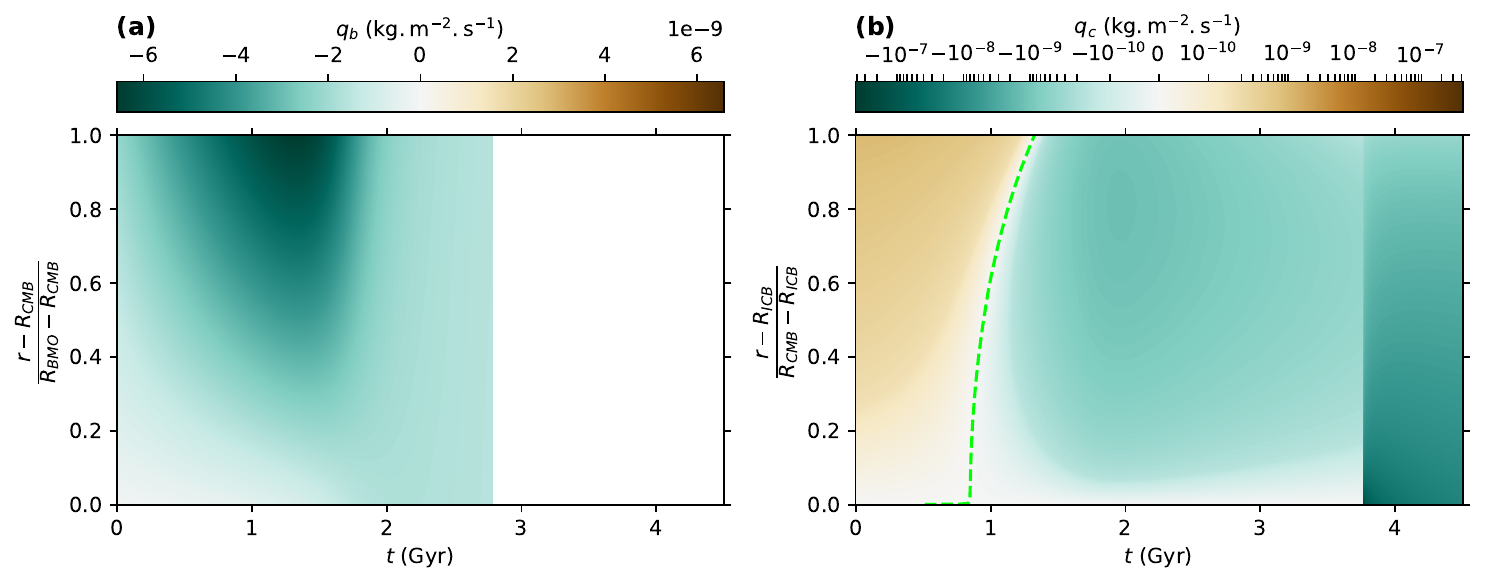}
    \caption{Total mass anomaly flux, \textit{i.e.} buoyancy profiles, in the BMO $q_b$ (a) and the outer core $q_c$ (b) as a function of time $t$. Radii are normalized by the thickness of the corresponding layer. The green line corresponds to a zero total mass anomaly flux.}
    \label{fig:buoyancy_time_nominal}
\end{figure*}

Fig. \ref{fig:buoyancy_time_nominal} shows the time evolution of the total mass anomaly flux in the BMO and the outer core.
In the BMO, the total mass anomaly flux $q_{b}$ is always negative (Fig. \ref{fig:buoyancy_time_nominal}a). Thus, the BMO is fully convective, from the combined action of compositional convection and heating at the CMB (Fig. \ref{fig:buoyancy_snap_nominal}a-c).
In the core, the total mass anomaly flux $q_{c}$ is initially positive, so that the core is fully stably-stratified (Fig. \ref{fig:buoyancy_time_nominal}b). At 0.8 Gyr, the mass anomaly flux becomes progressively negative, starting from the bottom of the core, as the CMB heat flow and the core cooling rate increase. This produces an upper stably-stratified layer overlying a convective core. At 1.3 Gyr, the total mass anomaly flux eventually becomes negative all the way to the top of the core, making it fully convective, due to the increase of the secular cooling term (Fig. \ref{fig:buoyancy_snap_nominal}d-e). At 3.8 Gyr, a discontinuity in the total mass anomaly flux enhances the vigor of convection. The flux magnitude suddenly increases with the release of latent heat at the ICB and the onset of compositional convection due to light elements released from the inner core (Fig. \ref{fig:buoyancy_snap_nominal}f).

\subsection{Magnetic evolution}
\label{sec:results_nominal_magnetic}

\begin{figure*}[!ht]
    \centering
    \includegraphics[width=\linewidth]{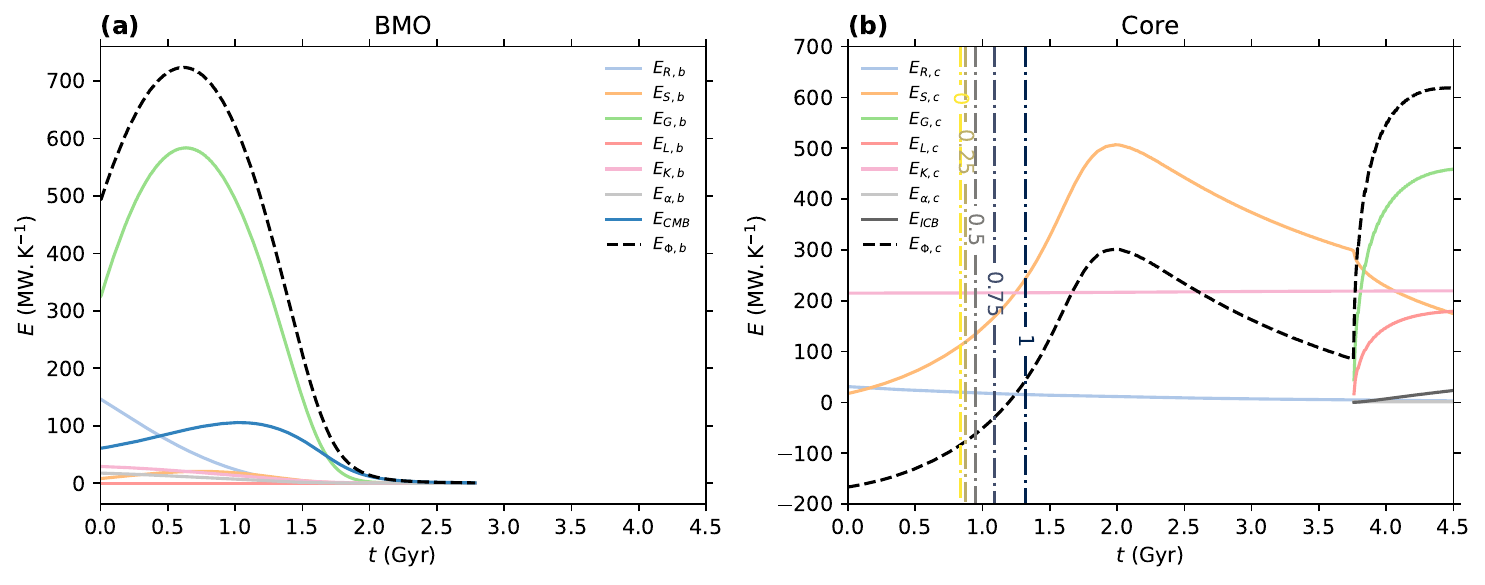}
    \caption{Rates of entropy production $E$ in the BMO (a) and the outer core (b) as a function of time $t$. The radiogenic heating, secular cooling, gravitational energy, latent heat, thermal diffusion, molecular diffusion, and dissipation terms are denoted with the subscripts $R$, $S$, $G$, $L$, $k$, $\alpha$ and $\Phi$. The contributions from the core at the CMB and from the inner core the ICB are denoted with the subscripts $CMB$ and $ICB$. The thermal diffusion terms use respectively $k_{b}=8~\mathrm{W.m^{-1}.K^{-1}}$ and $k_{c}=70~\mathrm{W.m^{-1}.K^{-1}}$. Dash-dotted lines indicate the time when a specific radius fraction of the core is convecting, as determined by the convective stability criterion defined in §\ref{sec:model_buoyancy_stability}.}
    \label{fig:entropy_nominal}
\end{figure*}

Fig. \ref{fig:entropy_nominal} shows the rates of entropy production in the BMO and the core as a function of time for our nominal model.
In the BMO, where convection operates at all times (Fig. \ref{fig:buoyancy_time_nominal}), the dissipation term $E_{\Phi,b}$ is always positive, which means that there is enough power available to sustain a dynamo in this layer (Fig. \ref{fig:entropy_nominal}a).
The entropy sources are dominated by the gravitational energy term $E_{G,b}$ (Eq. \ref{eq:entropy_gravi}), consistently with the high efficiency factor associated with this term \citep{nimmo_2015}, and to a lesser extent by the contribution of the core $E_{CMB}$ (Eq. \ref{eq:entropy_CMB_ICB}), particularly at late times when the BMO has shrunk significantly. The radiogenic heating term $E_{R,b}$ (Eq. \ref{eq:entropy_radio}) also contributes to the entropy budget, in particular at early times. In contrast, the secular cooling term $E_{S,b}$ (Eq. \ref{eq:entropy_secular}) is a minor source of entropy and the latent heat term $E_{L,b}$ (Eq. \ref{eq:entropy_latent}) does not contribute to the entropy budget since latent heat is released at the top the BMO.
The gravitational energy $E_{G,b}$, secular cooling $E_{S,b}$ and core contribution $E_{CMB}$ terms initially increase over time due to the increase of the corresponding heat sources, as a result of enhanced crystallization (Fig. \ref{fig:evolution_nominal}a) and cooling (Fig. \ref{fig:evolution_nominal}b) rates. They eventually decrease as the heat flows diminish, due to a shrinking BMO, reduced cooling and crystallization rates, and as the temperature difference between the top of the BMO and the CMB vanishes. In addition to the decrease in temperature difference, the radiogenic heating term $E_{R,b}$ also decreases due to a diminishing radiogenic heat flow, as a result of a shrinking BMO and radioactive decay.
The entropy sinks are dominated by the thermal diffusion term $E_{k,b}$ (Eq. \ref{eq:entropy_thermal}), with a negligible contribution from the molecular diffusion term $E_{\alpha,b}$ (Eq. \ref{eq:entropy_compositional}). They decrease monotonically over time, mainly because the BMO shrinks, but also because the average temperature of the BMO initially rises as its average depth increases.

Before nucleation of the inner core, the dissipation term $E_{\Phi,c}$ is initially negative for 1.2 Gyr, which means that there is not enough power to sustain an early dynamo in this layer (Fig. \ref{fig:entropy_nominal}b). Although the secular cooling term $E_{S,c}$ (Eq. \ref{eq:entropy_secular}) increases with time, as the CMB heat flow and core cooling increase, the entropy budget is dominated by the thermal diffusion sink $E_{k,c}$ (Eq. \ref{eq:entropy_thermal}).
After 1.2 Gyr, dissipation becomes positive as the secular cooling term continues to increase, and core convection starts, allowing sufficient power to sustain a core dynamo. This transition corresponds to the progressive evolution of the core toward convection, as indicated by the vertical lines that delineate the convective radius fraction of the core (§\ref{sec:model_buoyancy_stability}). Since the core becomes fully convective at 1.3 Gyr, the entropy budget might not be accurate immediately after the entropy of dissipation appears to turn positive \citep{nimmo_2015}. However, at 1.2 Gyr, 82\% of the core is expected to convect, indicating that this inaccuracy is relatively small.
The secular cooling term $E_{S,c}$ initially increases due to the increase of the corresponding heat flow (Fig. \ref{fig:energy_nominal}c) from enhanced cooling at the CMB (Fig. \ref{fig:evolution_nominal}b). At 2.0 Gyr, it eventually decreases as the heat flow diminishes.
The radiogenic heating term $E_{R,c}$ (Eq. \ref{eq:entropy_radio}), which decreases over time due to radioactive decay, is negligible. The thermal diffusion term $E_{k,c}$ is almost constant, due to a decreasing core temperature balanced by a reduced temperature gradient.

After nucleation of the inner core, the dissipation term $E_{\Phi,c}$ increases significantly, mainly due to the gravitational energy $E_{G,c}$ (Eq. \ref{eq:entropy_gravi}) and latent heat $E_{L,c}$ (Eq. \ref{eq:entropy_latent}) released at the ICB. Radiogenic heating and secular cooling of the inner core $E_{ICB}$ (Eq. \ref{eq:entropy_CMB_ICB}) also contribute to this increase.
The gravitational energy $E_{G,c}$, latent heat $E_{L,c}$ and inner core contribution $E_{ICB}$ terms increase over time, until they are diminished by decreasing crystallization and cooling rates, and as the outer core shrinks.

\begin{figure*}[!ht]
    \centering
    \includegraphics[width=\linewidth]{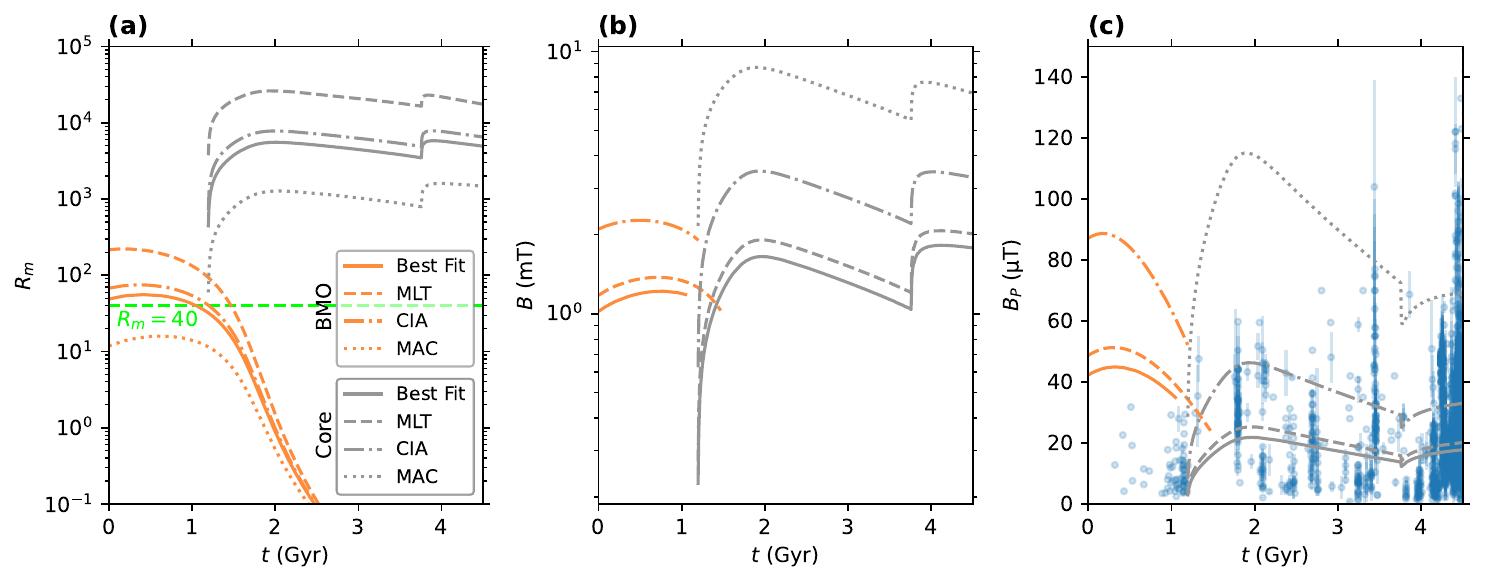}
    \caption{Magnetic Reynolds number $R_m$ (a), internal magnetic field intensity $B$ (b) and surface magnetic field intensity $B_P$ (c) as a function of time $t$.
        Solid lines are obtained using the best-fit scaling law of \citet{aubert_2009}. Dashed, dash-dotted, and dotted lines are obtained using power-law scalings with an imposed exponent derived from the mixing length theory (MLT), the Coriolis-Inertia-Archimedean (CIA), and the Magnetic-Archimedean-Coriolis (MAC) regime, respectively.
        The magnetic Reynolds numbers use respectively $\sigma_{b}=3 \times 10^{4}~\mathrm{S.m^{-1}}$ and $\sigma_{c}=1 \times 10^{6}~\mathrm{S.m^{-1}}$.
        The blue markers correspond to Earth's paleomagnetic measurements of the surface magnetic field from the PINT database \citep{bono_2022}.}
    \label{fig:magnetic_nominal}
\end{figure*}

Fig. \ref{fig:magnetic_nominal} shows the magnetic Reynolds numbers and the internal magnetic field intensities in the BMO and the core, as well as the corresponding surface magnetic field intensities as a function of time for the nominal model. The scaling laws obtained from a fit on dynamo numerical simulations are compared with scaling laws based on force balance arguments (Appendix \ref{app:scaling_law}).

In the BMO, the magnetic Reynolds number obtained using the best-fit scaling law is initially larger than $R_m=40$ during 1.0 Gyr, which means that a self-sustained dynamo is operating in this layer (Fig. \ref{fig:magnetic_nominal}a).
Assuming $\alpha_u = 2/5$ (see Appendix \ref{app:scaling_law}), the magnetic Reynolds number scales as $\Phi^{2/5}\Omega^{-1/5}$ (Eq. \ref{eq:magnetic_reynolds_number}, \ref{eq:U_scaling_law} and \ref{eq:dimensionless_convective_power}). Since the convective power is proportional to the entropy of dissipation (Eq. \ref{eq:convective_power_calculation}), this explains why the magnetic Reynolds number initially increases for 0.6 Gyr before decreasing as dissipation significantly diminishes (Fig. \ref{fig:entropy_nominal}). After 1.0 Gyr, the magnetic Reynolds number decreases below the critical value, thereby suppressing magnetic field growth in the BMO.
The average intensities of the corresponding internal and surface magnetic fields are, respectively, $B=1.2~\mathrm{mT}$ and $B_P=42.3~\mathrm{\mu T}$ (Fig. \ref{fig:magnetic_nominal}b-c).
Assuming $\alpha_B = 1/3$ (see Appendix \ref{app:scaling_law}), the internal magnetic field intensity scales as $\Phi^{1/3}L_b^{-2/3}$ (Eq. \ref{eq:B_scaling_law} and \ref{eq:dimensionless_convective_power}). This explains why the internal intensity initially increases for 0.8 Gyr (Fig. \ref{fig:magnetic_nominal}b), as the entropy of dissipation increases and the BMO thickness decreases. After 0.8 Gyr, the internal intensity decreases as dissipation is reduced, even if the BMO thickness also decreases.
We observe a similar evolution for the surface magnetic field (Fig. \ref{fig:magnetic_nominal}c), but because it is also proportional to $R_{BMO}^3$ (Eq. \ref{eq:surface_magnetic_field_BMO}), it begins to decrease earlier, around 0.3 Gyr, as the BMO depth increases.

In the core, the magnetic Reynolds number obtained using the best-fit scaling law is larger than $R_m=40$ after 1.2 Gyr, before which the power available to sustain a dynamo is insufficient (Fig. \ref{fig:entropy_nominal}b). As with the entropy budget, the core is not fully convective before 1.3 Gyr, which means that the definition of the magnetic Reynolds number might be inaccurate immediately after the entropy of dissipation becomes positive.
The average intensities of the corresponding internal and surface magnetic field are respectively $B=1.4~\mathrm{mT}$ and $B_P=17.3~\mathrm{\mu T}$.

In both the BMO and the core, the estimated magnetic Reynolds number and the magnetic field intensities are sensitive to the scaling law for convective velocity and magnetic field intensity. For the magnetic Reynolds numbers, the CIA scaling law is closer to the best-fit scaling law than the MLT and MAC scaling laws (Fig. \ref{fig:magnetic_nominal}a), whereas for the magnetic field intensities, the MLT scaling law is closer to the best-fit scaling law than the CIA and MAC scaling laws (Fig. \ref{fig:magnetic_nominal}b-c). Depending on the scaling law, the magnetic Reynolds numbers and the magnetic field intensities in the BMO and the core may vary by more than an order of magnitude. In the BMO, the MAC scaling law predicts a magnetic Reynolds number smaller than the critical value. This explains the lack of magnetic field with this scaling law.

\section{Sensitivity to the model parameters}
\label{sec:results_parameters}

In this section, we investigate the sensitivity of the model to several parameters. We focus on the role of the initial temperature at the CMB (§\ref{sec:results_parameters_temperature}), the parameterization of mantle convection (§\ref{sec:results_parameters_convection}), as well as the composition (§\ref{sec:results_parameters_composition}) and radiogenic heating (§\ref{sec:results_parameters_radio}) of the BMO and the core. We also investigate the sensitivity to the thermal and electrical conductivities of these layers (§\ref{sec:results_parameters_conductivity}). However, we do not investigate systematically the thermodynamic parameterization of the mantle, BMO and core, \textit{i.e.} the parameterization of the solid phase transitions, EOS, and liquidus.

As with the nominal model, the entropy of dissipation becomes positive when the core is not fully convective. As a result, the entropy budget and the magnetic Reynolds number might not be accurate immediately after the transition. However, this inaccuracy is likely to remain small since approximately 80\% of the core is consistently expected to convect when dissipation becomes positive.

\subsection{Initial temperature at the CMB}
\label{sec:results_parameters_temperature}

In our model, the initial temperature at the CMB, which corresponds to the temperature at which the BMO adiabat is tangent to the liquidus, is set to $T_{CMB_0}=5710~\mathrm{K}$. This is the maximum value used for our parameter sweep, because a higher temperature would produce a fully molten mantle instead of the intended BMO. The minimum value of the sweep, \textit{i.e.} $T_{CMB_0}=5010~\mathrm{K}$, delineates a transition toward an inconsistent situation whereby the low temperature difference across the mantle and thick upper boundary layer (Eq. \ref{eq:delta_1_beta}) produce insufficient heat flow (\ref{eq:heat_flow_surface}) to accommodate the imposed radiogenic heat flow.

\begin{figure*}[!ht]
    \centering
    \includegraphics[width=\linewidth]{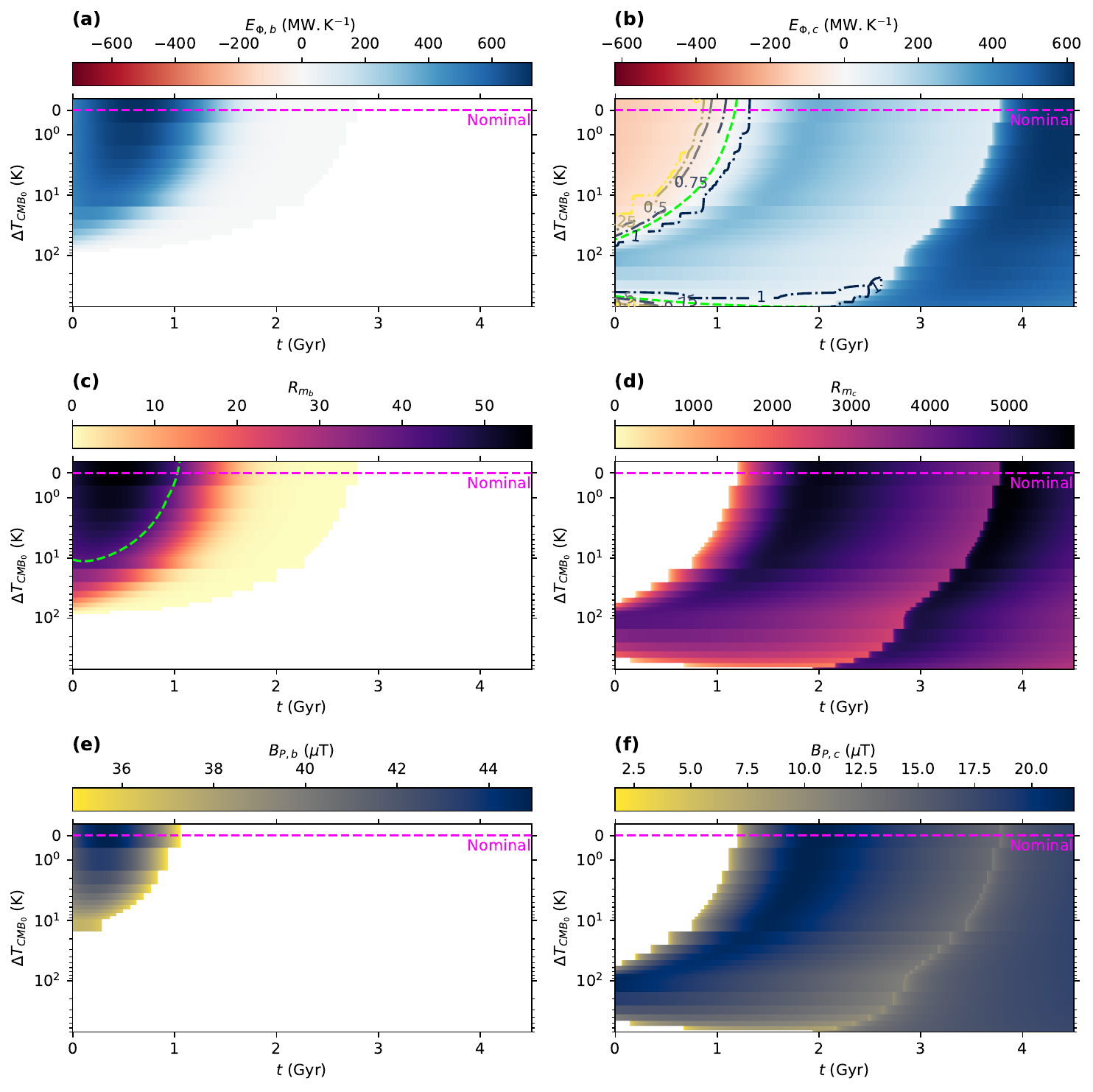}
    \caption{Sensitivity of the entropy of dissipation $E_\Phi$ (a,b), magnetic Reynolds number $R_m$ (c,d) and surface magnetic field intensity $B_P$ (e,f) of the BMO (a,c,e) and the core (b,d,f) to the initial temperature at the CMB $T_{CMB_0}$, expressed as the difference $\Delta T_{CMB_0}$ with the initial temperature used the nominal model, as a function of time $t$. Green dashed lines correspond to $E_\Phi=0$ (a,b) and $R_m=40$ (c,d). Dash-dotted lines indicate the time when a specific radius fraction of the core is convecting, as determined by the convective stability criterion defined in §\ref{sec:model_buoyancy_stability}. Pink dashed lines correspond to the value of the parameter used in the nominal model.}
    \label{fig:T_CMB_0_sweep}
\end{figure*}

Fig. \ref{fig:T_CMB_0_sweep} shows the entropy of dissipation, the magnetic Reynolds number and the intensity of the surface magnetic field as a function of the difference $\Delta T_{CMB_0}$ with the initial temperature used the nominal model, for the BMO and the core.
When $\Delta T_{CMB_0}$ increases, the initial thickness of the BMO decreases, and the surface heat flow increases due to an increased temperature difference between the top of the BMO and the surface. The BMO then crystallizes earlier (Fig. \ref{fig:T_CMB_0_sweep}a) and the silicate dynamo operation window is reduced (Fig. \ref{fig:T_CMB_0_sweep}c,e).
If $T_{CMB_0} \leq 5690~\mathrm{K}$, the silicate dynamo vanishes. If $T_{CMB_0} \leq 5620~\mathrm{K}$, the mantle remains fully solid.
When $\Delta T_{CMB_0}$ increases, the thickness of the BMO diminishes. This increases the temperature at the top of the BMO, and reduces the gravitational energy of the BMO (Eq. \ref{eq:gravitational_BMO}). The entropy of gravitational energy (Eq. \ref{eq:entropy_gravi}), dominant in the BMO entropy budget (Fig. \ref{fig:entropy_nominal}a), thus decreases, reducing the entropy of dissipation (Fig. \ref{fig:T_CMB_0_sweep}a).
Since the magnetic Reynolds number scales as $\Phi^{2/5}\Omega^{-1/5}$ (Eq. \ref{eq:magnetic_reynolds_number}, \ref{eq:U_scaling_law} and \ref{eq:dimensionless_convective_power}) and the convective power is proportional to the entropy of dissipation (Eq. \ref{eq:convective_power_calculation}), the magnetic Reynolds number of the BMO decreases when $\Delta T_{CMB_0}$ increases (Fig. \ref{fig:T_CMB_0_sweep}c).
Similarly, since the magnetic field intensity scales as $\Phi^{1/3}L_b^{-2/3}$ (Eq. \ref{eq:B_scaling_law} and \ref{eq:dimensionless_convective_power}) and the convective power decreases when $\Delta T_{CMB_0}$ increases, the magnetic field intensity of the BMO is reduced, even if the BMO thickness also decreases. Given that the surface magnetic field is also proportional to $R_{BMO}^3$ (Eq. \ref{eq:surface_magnetic_field_BMO}), this explains why the surface magnetic field decreases when $\Delta T_{CMB_0}$ increases (Fig. \ref{fig:T_CMB_0_sweep}e).

If the BMO exists, increasing $\Delta T_{CMB_0}$ enhances the heat flow at the CMB, due to an increased surface heat flow and an earlier BMO crystallization. Sufficient power is then available to allow an earlier core dynamo onset (Fig. \ref{fig:T_CMB_0_sweep}b,d,f). Furthermore, the resulting colder core crystallizes earlier.
In contrast, if the mantle remains solid, increasing $\Delta T_{CMB_0}$ reduces the temperature difference through the mantle and therefore the heat flow at the surface (Eq. \ref{eq:heat_flow_surface}). If $T_{CMB_0} \leq 5210~\mathrm{K}$, the CMB heat flow then becomes insufficient to power a core dynamo at early times (Fig. \ref{fig:T_CMB_0_sweep}b,d,f). Inner core nucleation then becomes necessary to sustain a magnetic field.
When corrected for the time shift related to inner core nucleation, the entropy of dissipation (Fig. \ref{fig:T_CMB_0_sweep}b), magnetic Reynolds number (Fig. \ref{fig:T_CMB_0_sweep}d), and magnetic field intensity (Fig. \ref{fig:T_CMB_0_sweep}f) of the core are relatively independent of $T_{CMB_0}$.

\subsection{Mantle convection}
\label{sec:results_parameters_convection}

In our model, mantle convection is controlled by the thermal boundary layer model, which depends on the $Nu \propto Ra^\beta$ relationship, as well as the mantle viscosity and thermal conductivity. Here, we focus on the reference viscosity of the mantle $\mu_{m_0}$ (Eq. \ref{eq:viscosity_mantle}), but provide a parameter sweep of $\beta$ (Fig. S2) and the mantle thermal conductivity (Fig. S3) in the supplementary material.
In order to obtain an Earth-like evolution, the nominal value of $\mu_{m_0}$ is set to $1.1 \times 10^{15}~\mathrm{Pa.s}$. The maximum value of the sweep, \textit{i.e.} $\mu_{m_0}=2.1 \times 10^{15}~\mathrm{Pa.s}$, corresponds to the value above which the thickness of the BMO, and the temperature at the CMB, would initially increase, leading to a fully molten mantle instead of the intended BMO. Such a large $\mu_{m_0}$ would produce a thick upper boundary layer (Eq. \ref{eq:delta_1_beta}), leading to an insufficient release of energy (Eq. \ref{eq:heat_flow_surface}) to obtain a BMO that decreases in size, and reduces the CMB temperature. The minimum value of the sweep is arbitrarily set to $\mu_{m_0}=1.0 \times 10^{12}~\mathrm{Pa.s}$.

\begin{figure*}[!ht]
    \centering
    \includegraphics[width=\linewidth]{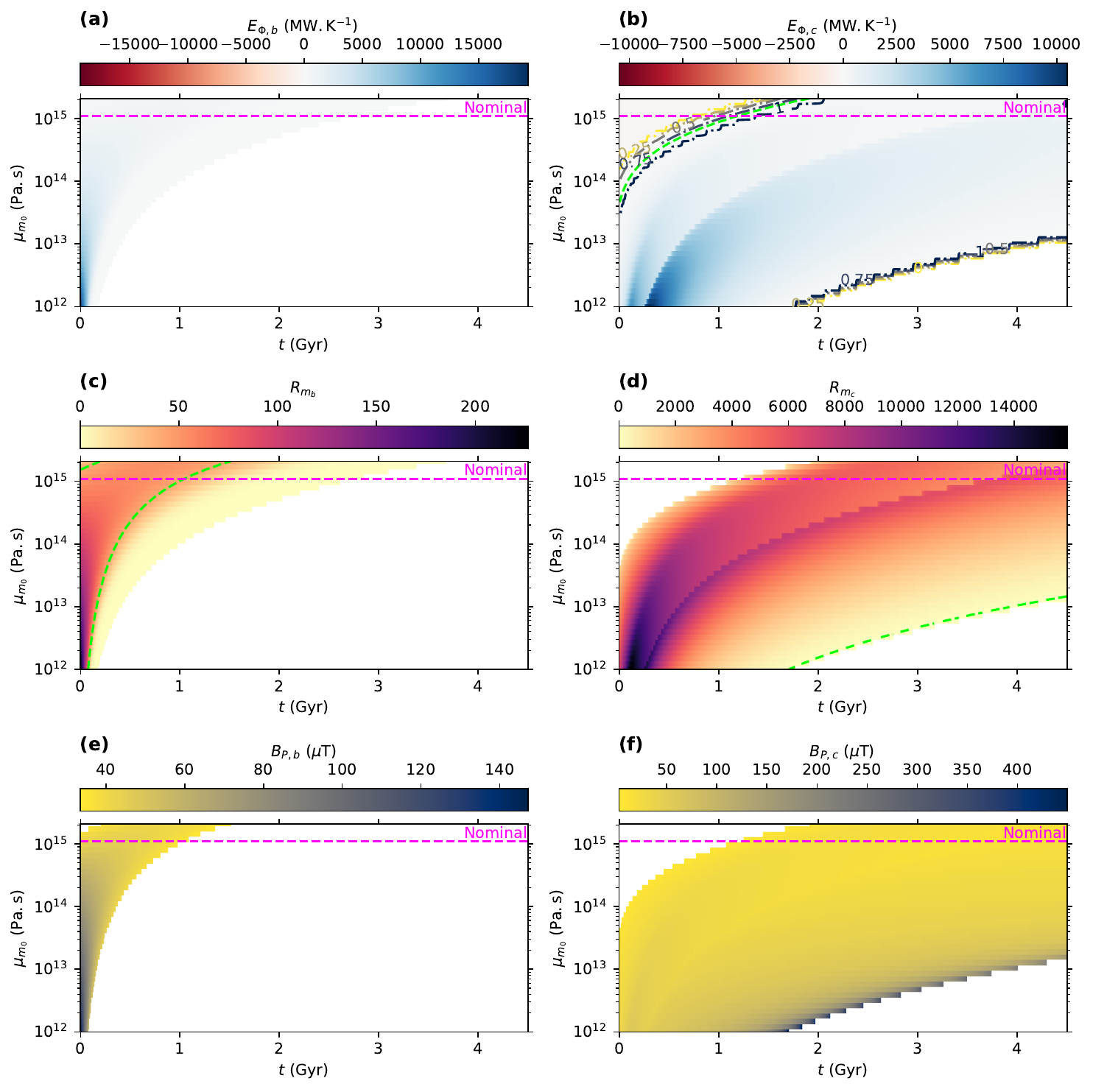}
    \caption{Sensitivity of the entropy of dissipation $E_\Phi$ (a,b), magnetic Reynolds number $R_m$ (c,d) and surface magnetic field intensity $B_P$ (e,f) of the BMO (a,c,e) and the core (b,d,f) to the reference viscosity of the mantle $\mu_{m_0}$ (Eq. \ref{eq:viscosity_mantle}), as a function of time $t$. Green dashed lines correspond to $E_\Phi=0$ (a,b) and $R_m=40$ (c,d). Dash-dotted lines indicate the time when a specific radius fraction of the core is convecting, as determined by the convective stability criterion defined in §\ref{sec:model_buoyancy_stability}. Pink dashed lines correspond to the value of the parameter used in the nominal model.}
    \label{fig:eta_earth_sweep}
\end{figure*}

Fig. \ref{fig:eta_earth_sweep} shows the entropy of dissipation, the magnetic Reynolds number and the intensity of the surface magnetic field as a function of $\mu_{m_0}$ for the BMO and the core.
When $\mu_{m_0}$ decreases, the upper boundary layer becomes smaller (Eq. \ref{eq:delta_1_beta}), which produces an enhanced surface heat flow (Eq. \ref{eq:heat_flow_surface}). This intensification of mantle convection leads to an earlier solidification of the BMO (Fig. \ref{fig:eta_earth_sweep}a), associated with a shorter silicate dynamo operation window (Fig. \ref{fig:eta_earth_sweep}c).
When $\mu_{m_0}$ decreases, the BMO crystallizes faster, increasing the dominant entropy of gravitational energy (Eq. \ref{eq:entropy_gravi}), and the associated entropy of dissipation (Fig. \ref{fig:eta_earth_sweep}a).
The magnetic Reynolds number also increases (Fig. \ref{fig:eta_earth_sweep}c), since the convective power increases with the entropy of dissipation (Eq. \ref{eq:convective_power_calculation}), and because the magnetic Reynolds number scales as $\Phi^{2/5}\Omega^{-1/5}$ (Eq. \ref{eq:magnetic_reynolds_number}, \ref{eq:U_scaling_law} and \ref{eq:dimensionless_convective_power}).
Similarly, the magnetic field intensity increases, since the convective power and the thickness of the BMO respectively increases and decreases when $\mu_{m_0}$ decreases, and because the magnetic field scales as $\Phi^{1/3}L_b^{-2/3}$ (Eq. \ref{eq:B_scaling_law} and \ref{eq:dimensionless_convective_power}). This explains why the surface magnetic field increases when $\mu_{m_0}$ decreases (Fig. \ref{fig:eta_earth_sweep}e), even if a reduced BMO thickness contributes to decrease the surface magnetic field (Eq. \ref{eq:surface_magnetic_field_BMO}).

In the core, the decrease of $\mu_{m_0}$ allows for an earlier dynamo onset, starting at initial time when $\mu_{m_0} \leq 4 \times 10^{13}~\mathrm{Pa.s}$ (Fig. \ref{fig:eta_earth_sweep}b,d,f). This may be explained by an increased surface heat flow and an earlier solidification of the BMO, increasing the heat flow at the CMB and providing sufficient power to sustain an earlier magnetic field. The enhanced cooling of the core is also responsible for an earlier inner core nucleation, leading to a full crystallization when $\mu_{m_0} \leq 1 \times 10^{13}~\mathrm{Pa.s}$ (Fig. \ref{fig:eta_earth_sweep}b,d,f).
Since the cooling rate of the core increases when $\mu_{m_0}$ decreases, the entropy source terms increase along with the entropy of dissipation (Fig. \ref{fig:eta_earth_sweep}b). The magnetic Reynolds number (Fig. \ref{fig:eta_earth_sweep}d) and the magnetic field intensity (Fig. \ref{fig:eta_earth_sweep}f) also increase from the enhanced convective power (Eq. \ref{eq:U_scaling_law} and \ref{eq:B_scaling_law}).

\subsection{Composition}
\label{sec:results_parameters_composition}

In our model, the composition of the BMO strongly influences the planet evolution, and is governed by the the iron partition coefficient $D_{BMO}$ and the initial mass fraction $\xi_{b_0}$ of $\mathrm{FeO}$. However, our model is independent of the core composition because we define the EOS and the liquidus of the core with a fixed light element content. In addition, the planet evolution is also influenced by the compositional density jump at the top of the BMO $\Delta_\xi\rho_{BMO}$ and at the ICB $\Delta_\xi\rho_{ICB}$. Here, we focus on $D_{BMO}$, but we provide a parameter sweep of $\xi_{b_0}$ (Fig. S4), $\Delta_\xi\rho_{BMO}$ (Fig. S5) and $\Delta_\xi\rho_{ICB}$ (Fig. S6) in the supplementary material.
To obtain an Earth-like evolution, we set the nominal value of the iron partition coefficient to $D_{BMO}=0.85$. We conduct the parameter sweep in the range $0 < D_{BMO} < 1$, where iron is preferentially partitioned into the melt.

\sloppy

\begin{figure*}[!ht]
    \centering
    \includegraphics[width=\linewidth]{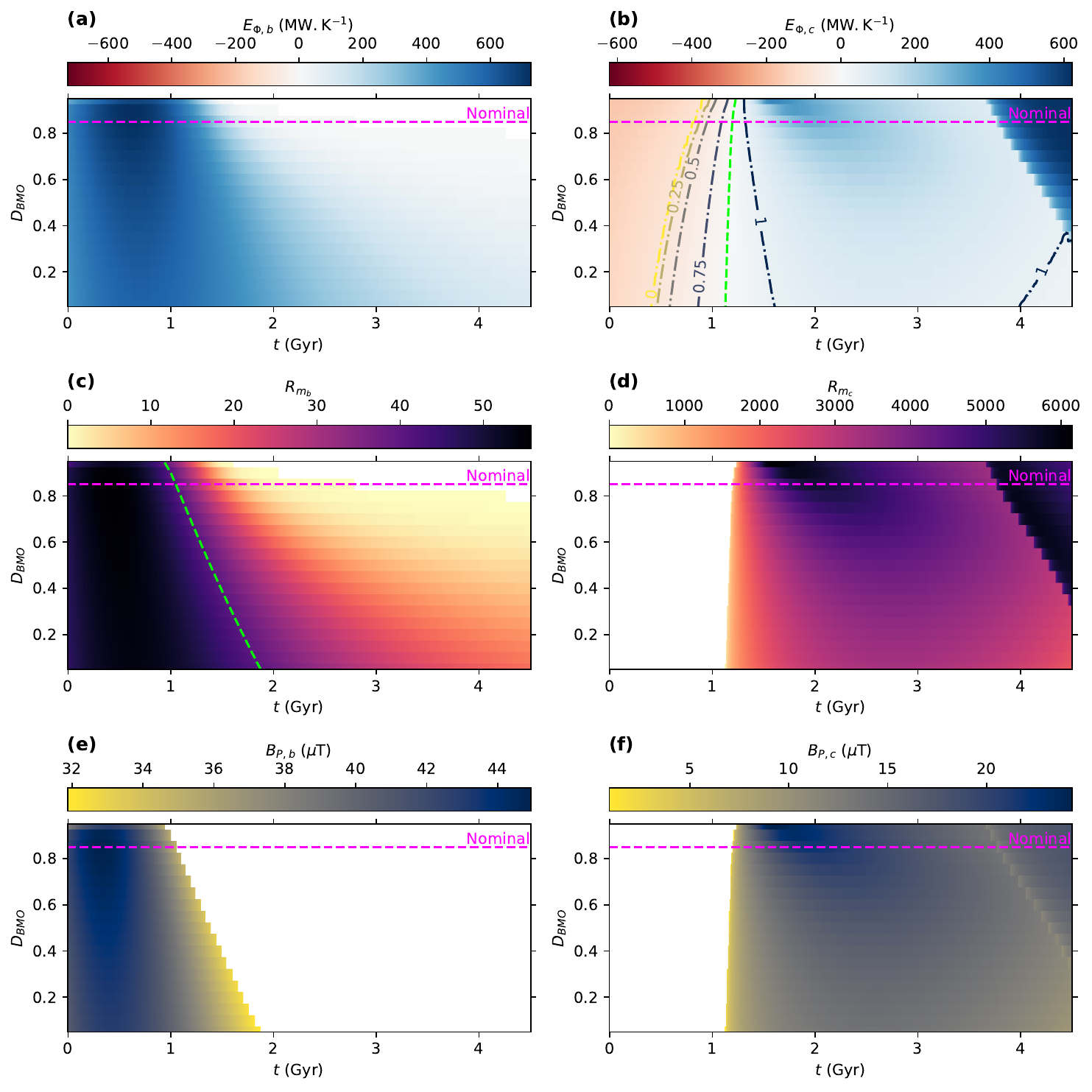}
    \caption{Sensitivity of the entropy of dissipation $E_\Phi$ (a,b), magnetic Reynolds number $R_m$ (c,d) and surface magnetic field intensity $B_P$ (e,f) of the BMO (a,c,e) and the core (b,d,f) to the iron partition coefficient $D_{BMO}$, as a function of time $t$. Green dashed lines correspond to $E_\Phi=0$ (a,b) and $R_m=40$ (c,d). Dash-dotted lines indicate the time when a specific radius fraction of the core is convecting, as determined by the convective stability criterion defined in §\ref{sec:model_buoyancy_stability}. Pink dashed lines correspond to the value of the parameter used in the nominal model.}
    \label{fig:D_BMO_0_sweep}
\end{figure*}

Fig. \ref{fig:D_BMO_0_sweep} shows the entropy of dissipation, the magnetic Reynolds number and the intensity of the surface magnetic field as a function of $D_{BMO}$ for the BMO and the core.
When $D_{BMO}$ decreases, the rate of release of $\mathrm{FeO}$ increases, raising the $\mathrm{FeO}$ content and depressing the liquidus. Furthermore, the temperature difference through the mantle decreases, reducing the surface heat flow (Eq. \ref{eq:heat_flow_surface}). This reduces the BMO crystallization rate (Fig. \ref{fig:D_BMO_0_sweep}a) and extends the BMO-hosted dynamo operation window (Fig. \ref{fig:D_BMO_0_sweep}c).
When $D_{BMO}$ decreases, the change in composition across the interface between the BMO and the mantle increases (Eq. \ref{eq:composition_change_BMO}), reducing the compositional expansion coefficient (Eq. \ref{eq:compositional_expansivity_BMO}, and decreasing the gravitational energy of the BMO (Eq. \ref{eq:gravitational_BMO}). The dominant entropy of gravitational energy (Eq. \ref{eq:entropy_gravi}) therefore decreases, reducing the entropy of dissipation (Fig. \ref{fig:D_BMO_0_sweep}a).
Since the magnetic Reynolds number scales as $\Phi^{2/5}\Omega^{-1/5}$ (Eq. \ref{eq:magnetic_reynolds_number}, \ref{eq:U_scaling_law} and \ref{eq:dimensionless_convective_power}), it slightly decreases with $D_{BMO}$ (Fig. \ref{fig:D_BMO_0_sweep}c) due to the reduced convective power (Eq. \ref{eq:convective_power_calculation}).
Similarly, the magnetic field intensity, which scales as $\Phi^{1/3}L_b^{-2/3}$ (Eq. \ref{eq:B_scaling_law} and \ref{eq:dimensionless_convective_power}), decreases with $D_{BMO}$, due to the reduced convective power and thicker BMO. This explains why the surface magnetic field decreases with $D_{BMO}$ (Fig. \ref{fig:D_BMO_0_sweep}e), even if a thicker BMO contributes to increase the surface magnetic field (Eq. \ref{eq:surface_magnetic_field_BMO}).

In the core, dynamo onset is largely independent of $D_{BMO}$ (Fig. \ref{fig:D_BMO_0_sweep}b,d,f), although it occurs slightly earlier for lower values of $D_{BMO}$. Lower $D_{BMO}$ means an increase in the CMB cooling rate and heat flow at early time ($t \lesssim 1.3~\mathrm{Gyr}$), compensating for reduced BMO heat flow and crystallization rate, from the depression of the liquidus.
In contrast, lowering $D_{BMO}$ means that the inner core nucleation is significantly delayed, leading to a fully liquid core at 4.5 Gyr, for $D_{BMO} \leq 0.35$ (Fig. \ref{fig:D_BMO_0_sweep}b,d,f). This is explained by a lower CMB cooling rate from a lower surface heat flow for most of the later planet evolution ($t \gtrsim 1.3~\mathrm{Gyr}$).

If $t \lesssim 1.3~\mathrm{Gyr}$, the CMB cooling rate slightly increases when $D_{BMO}$ decreases. The entropy of secular cooling, which dominates the entropy budget of the core before inner core nucleation, thus increases, leading to a slight increase of the entropy of dissipation (Fig. \ref{fig:D_BMO_0_sweep}b). In contrast, if $t \gtrsim 1.3~\mathrm{Gyr}$, lower CMB cooling rates lead to decreasing entropy of secular cooling and entropy of dissipation.
Similarly, since the onset of core dynamo occurs when $t \gtrsim 1.3~\mathrm{Gyr}$, the magnetic Reynolds number (Fig. \ref{fig:D_BMO_0_sweep}d) and the magnetic field intensity (Fig. \ref{fig:D_BMO_0_sweep}f) decrease due to reduced convective power (Eq. \ref{eq:U_scaling_law} and \ref{eq:B_scaling_law}).

\fussy

\subsection{Radiogenic heating}
\label{sec:results_parameters_radio}

In our model, radiogenic heating in the BMO and the mantle is controlled by the mantle fraction $f_h$ of BSE radiogenic production. In the core, heat production is driven by the $\mathrm{^{40}K}$ core concentration $c_c$. Here, we focus on $f_h$, but we provide a parameter sweep of $c_c$ in the supplementary material (Fig. S9).
The nominal value of the mantle fraction of the BSE radiogenic production is $f_h=0.58$, and the sweep is conducted for $f_h > 0$. As with previous parameters, the maximum value of this parameter sweep, \textit{i.e.} $f_h=0.7$, corresponds to the value above which the thickness of the BMO, and the temperature at the CMB, would initially increase, leading to a fully molten mantle instead of the intended BMO. Such a large $f_h$ would be responsible for low heat flows at the top of the BMO and at the CMB, leading to an insufficient release of energy to obtain a obtain a BMO decreasing in size, and reduce the CMB temperature.

\begin{figure*}[!ht]
    \centering
    \includegraphics[width=\linewidth]{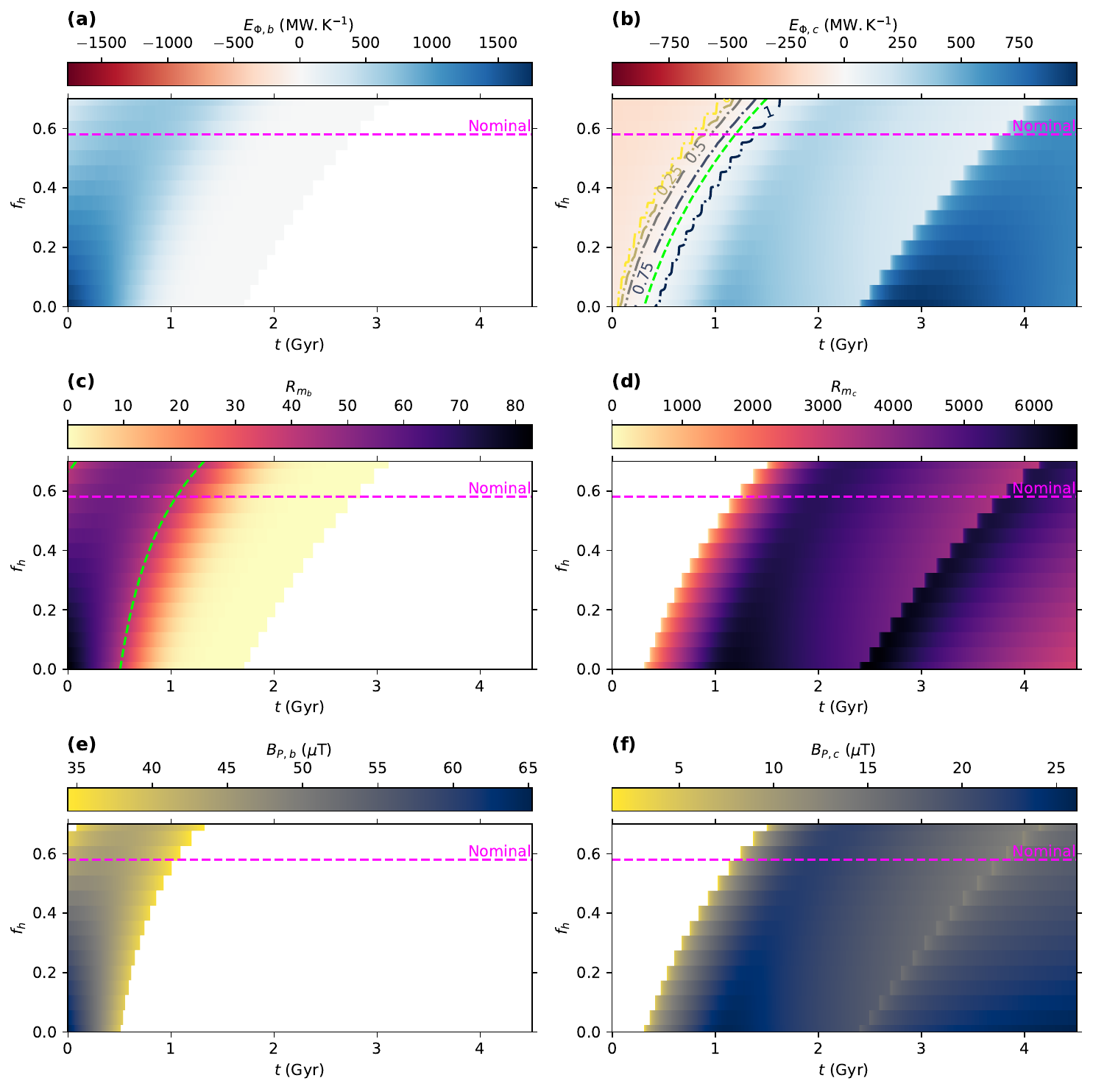}
    \caption{Sensitivity of the entropy of dissipation $E_\Phi$ (a,b), magnetic Reynolds number $R_m$ (c,d) and surface magnetic field intensity $B_P$ (e,f) of the BMO (a,c,e) and the core (b,d,f) to the mantle fraction of the BSE radiogenic heating $f_h$, as a function of time $t$. Green dashed lines correspond to $E_\Phi=0$ (a,b) and $R_m=40$ (c,d). Dash-dotted lines indicate the time when a specific radius fraction of the core is convecting, as determined by the convective stability criterion defined in §\ref{sec:model_buoyancy_stability}. Pink dashed lines correspond to the value of the parameter used in the nominal model.}
    \label{fig:f_R_mantle_sweep}
\end{figure*}

Fig. \ref{fig:f_R_mantle_sweep} shows the entropy of dissipation, the magnetic Reynolds number and the intensity of the surface magnetic field as a function of $f_h$ for the BMO and the core.
When $f_h$ increases, the internal heat production of the BMO and the solid mantle increases, leading to a reduced cooling rate of the BMO (Eq. \ref{eq:energy_budget}), and therefore protracted crystallization (Fig. \ref{fig:f_R_mantle_sweep}a) and silicate dynamo operation (Fig. \ref{fig:f_R_mantle_sweep}c).
The entropy of dissipation decreases (Fig. \ref{fig:f_R_mantle_sweep}a), as a larger $f_h$ reduces the $\mathrm{FeO}$ release rate, reducing the dominant entropy source from gravitational energy (Eq. \ref{eq:entropy_gravi}).
The magnetic Reynolds number also decreases (Fig. \ref{fig:f_R_mantle_sweep}c), since the convective power diminishes with decreasing entropy of dissipation, and because the magnetic Reynolds number scales as $\Phi^{2/5}\Omega^{-1/5}$.
Similarly, the magnetic field intensity decreases, since the convective power and the thickness of the BMO respectively decrease and increase as $f_h$ increases, and because the magnetic field scales as $\Phi^{1/3}L_b^{-2/3}$ (Eq. \ref{eq:B_scaling_law} and \ref{eq:dimensionless_convective_power}). This explains why the surface magnetic field decreases when $f_h$ increases (Fig. \ref{fig:f_R_mantle_sweep}e), even if a thicker BMO contributes to increase the surface magnetic field (Eq. \ref{eq:surface_magnetic_field_BMO}).

In the core, the increase of $f_h$ promotes a lower heat flow at the CMB, explaining a delayed dynamo onset and inner core nucleation (Fig. \ref{fig:f_R_mantle_sweep}b,d,f).
Since the cooling rate of the core decreases when $f_h$ increases, the source terms of the entropy budget decrease, leading to a decreasing entropy of dissipation (Fig. \ref{fig:eta_earth_sweep}b). The magnetic Reynolds number (Fig. \ref{fig:eta_earth_sweep}d) and the magnetic field intensity (Fig. \ref{fig:eta_earth_sweep}f) also decrease due to reduced convective power (Eq. \ref{eq:U_scaling_law} and \ref{eq:B_scaling_law}).

\subsection{Thermal and electrical conductivity}
\label{sec:results_parameters_conductivity}

\sloppy

In the BMO, dynamo operation is mostly limited by the relatively low electrical conductivity $\sigma_b$ of the silicates. In our model, we set the nominal value of $\sigma_b$ to $3 \times 10^{4}~\mathrm{S.m^{-1}}$ \citep{holmstrom_2018,soubiran_2018,stixrude_2020}. To obtain a consistent parameter sweep, the thermal conductivity of the BMO $k_b$ is defined concomitantly with $\sigma_b$, using the Wiedemann-Franz law. We define the total thermal conductivity as the sum of ionic $k_{b,l}$ and electronic $k_{b,e}$ contributions
\begin{equation}
    k_{b}=k_{b,l}+k_{b,e}.
    \label{eq:thermal_conductivity_BMO}
\end{equation}
Following \citet{deng_2021}, the ionic contribution is
\begin{equation}
    k_{b,l}=k_{b,l_0}\left(\frac{T_{k}}{T}\right)^{a_{k,b}}\left(\frac{\rho}{\rho_{k}}\right)^{b_{k,b}},
    \label{eq:thermal_conductivity_phonon_BMO}
\end{equation}
where the prefactor $k_{b,l_0}$, the exponents $a_{k,b}$ and $b_{k,b}$, as well as the reference temperature $T_k$ and density $\rho_k$ are defined in Table S6, and where temperature and density are averaged over the BMO at any given time.
The electronic contribution is related to the electrical conductivity of the BMO by the Wiedemann-Franz law
\begin{equation}
    k_{b,e}=Lo T \sigma_{b},
    \label{eq:thermal_conductivity_electron_BMO}
\end{equation}
where the temperature is averaged over the BMO at any given time and $Lo$ is the Lorenz number. Since $Lo$ remains uncertain in the BMO \citep{soubiran_2018}, we use the theoretical value $Lo=2.443 \times 10^{-8}~\mathrm{V^2.K^{-2}}$.
The time-averaged thermal conductivity of the BMO, which varies only marginally over time (see Fig. S10), is obtained by using the nominal value of $\sigma_b$ in Eq. \ref{eq:thermal_conductivity_electron_BMO}. This gives $k_b=8.03~\mathrm{W.m^{-1}.K^{-1}}$, where $k_{b,l}=4.26~\mathrm{W.m^{-1}.K^{-1}}$ and $k_{b,e}=3.77~\mathrm{W.m^{-1}.K^{-1}}$, consistently with the imposed $k_b=8~\mathrm{W.m^{-1}.K^{-1}}$ in the nominal model.

In the core, dynamo operation is mostly limited by the thermal conductivity $k_c$ of the liquid metal, the core being an efficient electrical conductor \citep{williams_2018}.
In our model, we set the nominal value of $k_c$ to $70~\mathrm{W.m^{-1}.K^{-1}}$ \citep{ohta_2016,basu_2020,hsieh_2020,pozzo_2022}. As with the BMO, we define the electrical conductivity of the core $\sigma_c$ concomitantly with $k_c$, using the Wiedemann-Franz law. However, since the electronic contribution is the only factor contributing to $k_c$, we have
\begin{equation}
    k_c=Lo T \sigma_c,
    \label{eq:thermal_conductivity_core}
\end{equation}
where temperature is averaged over the outer core at any given time. Since $Lo$ remains uncertain in the core \citep{williams_2018}, we also use the theoretical value $Lo=2.443 \times 10^{-8}~\mathrm{V^2.K^{-2}}$.
The time-averaged electrical conductivity of the core, which also marginally varies over time (see Fig. S10), is obtained by using the nominal value of $k_c$ in Eq. \ref{eq:thermal_conductivity_core}. This gives $\sigma_c=4.88 \times 10^{5}~\mathrm{S.m^{-1}}$, consistently with the imposed $\sigma_c = 1 \times 10^{6}~\mathrm{S.m^{-1}}$ in the nominal model.

\fussy

\begin{figure*}[!ht]
    \centering
    \includegraphics[width=\linewidth]{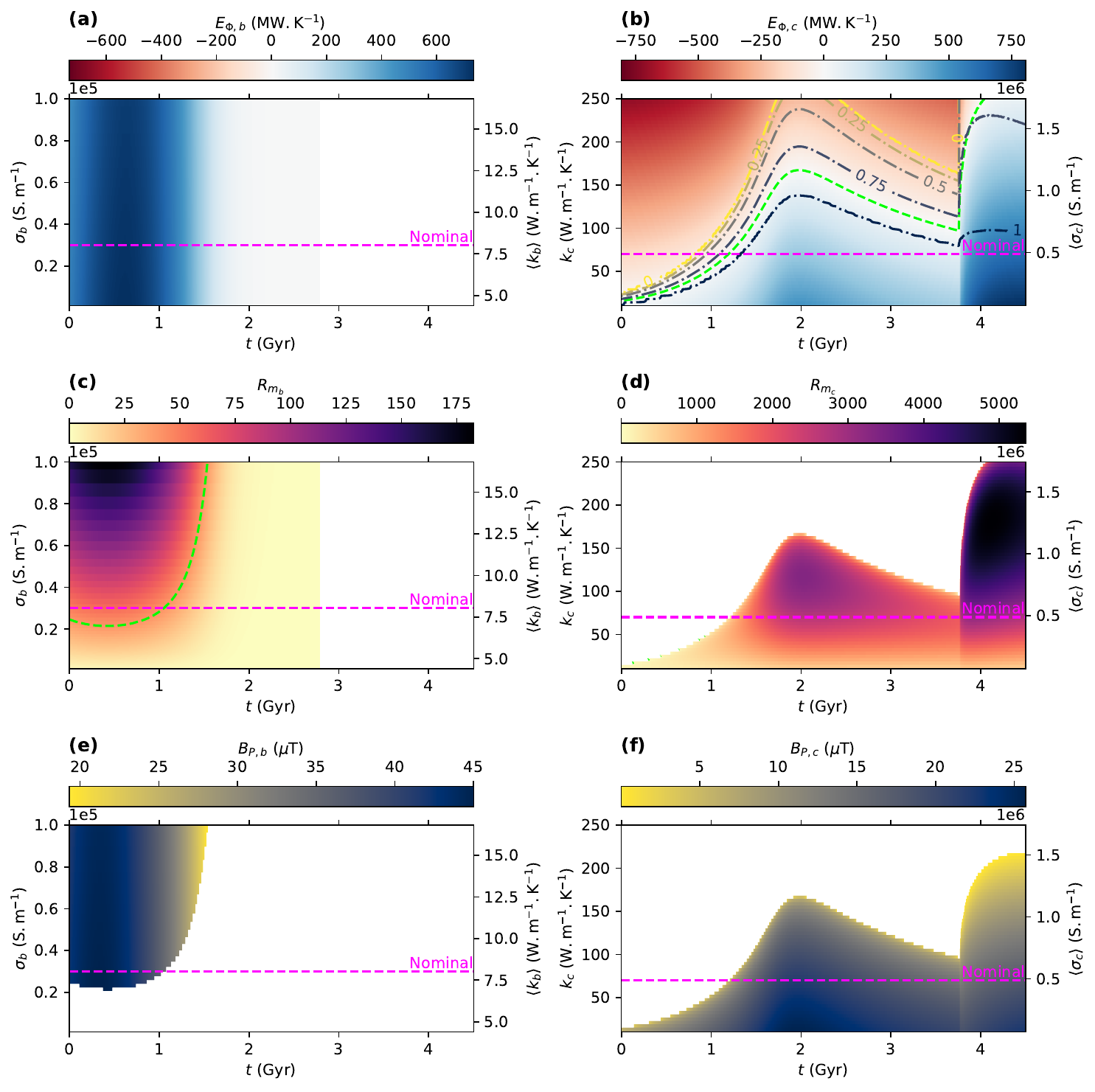}
    \caption{Influence of the BMO electrical conductivity $\sigma_b$ (a,c,e) and of the core thermal conductivity $k_c$ (b,d,f) on the entropy of dissipation $E_\Phi$ (a,b), magnetic Reynolds number $R_m$ (c,d) and surface magnetic field intensity $B_P$ (e,f), as a function of time $t$. The right y-axis shows the time-averaged thermal (a,c,e) and electrical (b,d,f) conductivities calculated from the imposed left y-axis conductivity using the Wiedemann-Franz law. Green dashed lines correspond to $E_\Phi=0$ (a,b) and $R_m=40$ (c,d). Dash-dotted lines indicate the time when a specific radius fraction of the core is convecting, as determined by the convective stability criterion defined in §\ref{sec:model_buoyancy_stability}. Pink dashed lines correspond to the value of the parameter used in the nominal model.}
    \label{fig:k_sigma_sweep}
\end{figure*}

Fig. \ref{fig:k_sigma_sweep} shows the entropy of dissipation, the magnetic Reynolds number, and the intensity of the surface magnetic field as a function of $\sigma_b$ and $k_c$. Because the temperature and the density of the BMO and the core vary over time, the conductivities shown on the right y-axis, calculated using the Wiedemann-Franz law, correspond to time-averaged values.

In the BMO, the entropy of dissipation is always positive and varies by less than 10\% over the range of thermal and electrical conductivity investigated (Fig. \ref{fig:k_sigma_sweep}a). If the thermal conductivity increases, the entropy of dissipation, and therefore the power available to the dynamo, slightly decreases as the entropy of thermal diffusion increases (Eq. \ref{eq:entropy_thermal}). In contrast, the magnetic Reynolds number increases significantly with electrical conductivity (Eq. \ref{eq:magnetic_reynolds_number}), making a silicate dynamo possible if $\sigma_b \geq 2.15 \times 10^{4}~\mathrm{S.m^{-1}}$ (Fig. \ref{fig:k_sigma_sweep}c). However, the intensity of the magnetic field produced in the BMO is relatively independent of the electrical conductivity (Fig. \ref{fig:k_sigma_sweep}e), consistently with the relatively constant entropy of dissipation (Eq. \ref{eq:B_scaling_law} and \ref{eq:convective_power_calculation}).

In the core, the entropy of dissipation decreases when the thermal conductivity increases (Fig. \ref{fig:k_sigma_sweep}b), due to an increasing entropy of thermal diffusion (Eq. \ref{eq:entropy_thermal}). As a result, the time at which enough power is available to power a dynamo is delayed. In particular, if $k_c \geq 15.8~\mathrm{W.m^{-1}.K^{-1}}$, a core dynamo is impossible at early times. Furthermore, if $k_c \geq 168~\mathrm{W.m^{-1}.K^{-1}}$, the dynamo is impossible before the additional supply of entropy from inner core nucleation. If sufficient power is available, the magnetic Reynolds number is always greater than $R_m=40$ (Fig. \ref{fig:k_sigma_sweep}d), which means that a core dynamo is possible. Although an increasing thermal conductivity contributes to reduce the magnetic Reynolds number through reduced convective velocities, the evolution of the magnetic Reynolds number is dominated by the electrical conductivity (Eq. \ref{eq:magnetic_reynolds_number}). Thus, the magnetic Reynolds number increases with the transport properties of the core. Finally, the intensity of the magnetic field decreases as the thermal conductivity increases (Fig. \ref{fig:k_sigma_sweep}f), consistently with a reduced entropy of dissipation (Eq. \ref{eq:B_scaling_law} and \ref{eq:convective_power_calculation}).

\section{Discussion}
\label{sec:discussion}

\subsection{Comparison with Earth}
\label{sec:discussion_Earth}

Our nominal model converges toward the present-day Earth structure. The density, gravity and pressure profiles of the planet (§\ref{sec:results_nominal_structure}) are consistent with the Preliminary Reference Earth Model (PREM) \citep{dziewonski_1981}. In detail, our model CMB is 4.3\% deeper than PREM because Earth's mantle includes iron and is thus denser than pure $\mathrm{MgSiO_3}$ \citep{boujibar_2020}. In addition, the core density is 11.1\% larger in our model since our core EOS uses pure $\mathrm{Fe}$. As such, the gravity at the CMB and the pressure at the center are respectively 11.1\% and 11.2\% larger than in PREM.

Temperatures at the CMB and at the ICB (§\ref{sec:results_nominal_thermal}) are respectively $T_{CMB}=4630~\mathrm{K}$ and $T_{ICB}=5874~\mathrm{K}$ at 4.5 Gyr, which is within, or close to, present-day estimates in the Earth's core, \textit{i.e.} $T_{CMB}=3400-4600~\mathrm{K}$ \citep{hirose_2021} and $T_{ICB}=5050-6250~\mathrm{K}$ \citep{nimmo_2015a}. The average temperature of the mantle is $\bar{T}_{m}=2291~\mathrm{K}$ at 4.5 Gyr, which is larger than present-day estimates for the Earth \citep{jaupart_2015}. The final radius of the model inner core is 1226 km, close to the 1221 km of actual Earth's inner core \citep{dziewonski_1981}.
At $t=0$, the model temperature of the CMB is $T_{CMB_0}=5710~\mathrm{K}$. This is significantly higher than the upper limit $T_{CMB_0}=5375~\mathrm{K}$ obtained from metal-silicate equilibration \citep{fischer_2017}, but is compatible with previous models that give initial temperatures in the range $T_{CMB_0}=4570-6400~\mathrm{K}$ \citep{driscoll_2023}.

Heat flows values at the CMB and the surface (§\ref{sec:results_nominal_thermal}) are respectively $Q_{CMB}=13.7~\mathrm{TW}$ and $Q_{P}=37.2~\mathrm{TW}$ at 4.5 Gyr. These are within the range estimated for the Earth, namely $Q_{CMB}=5-17~\mathrm{TW}$ at the CMB and $Q_{P}=35-41~\mathrm{TW}$ at the surface of the mantle \citep{jaupart_2015}.

\sloppy

Paleomagnetic data \citep{bono_2022} are generally compatible with the magnetic field produced by a core dynamo (§\ref{sec:results_nominal_magnetic}). However, our nominal model present-day surface magnetic field $B_P=17.8~\mathrm{\mu T}$ is slightly smaller than the typical range estimated for the Earth, \textit{i.e.} $B_P=25-65~\mathrm{\mu T}$ \citep{alken_2021}. The nominal model predicts a BMO silicate dynamo lifetime consistent with previous models, with typical cessation times in the range $1-2.5~\mathrm{Gyr}$ \citep{ziegler_2013,blanc_2020,orourke_2020,stixrude_2020}. The BMO magnetic field predicted before 1.0 Gyr is only slightly larger than paleomagnetic measurements, and compatible with previous models \citep{orourke_2020,stixrude_2020}.
The model sensitivity to the magnetic scaling laws (Appendix \ref{app:scaling_law}), initially developed for the Earth's core, as well as various parameters (§\ref{sec:results_parameters}), may contribute to explaining these differences with the nominal model. Uncertainties in paleomagnetic data must also be considered \citep{tarduno_2010,tarduno_2015,borlina_2020,tarduno_2020,taylor_2023}.

\fussy

\subsection{Comparison to previous works}
\label{sec:discussion_previous}

In this section, we highlight the main methodological extensions proposed by our model. We therefore propose a comparison to previous models that compute planetary evolution with energy budgets, leaving aside higher-dimension models \citep[\textit{e.g.}][]{zhang_2022}.

In our model, we partially couple the radially integrated internal structure of an Earth-like planet with a BMO to its thermal evolution. This differs from earlier works \citep{labrosse_2007,ziegler_2013,blanc_2020,orourke_2020} where the internal structure of the planet is simplified by using thermodynamic properties independent of time, a radially-averaged BMO, and where the core is either radially-averaged or parameterized using radially-dependent polynomials.
We also invoke a simple parameterization of the time-dependent compositional evolution of the BMO and consider its feedback on the EOS and melting curve. This differs from previous work for which the thermodynamic properties of the BMO are independent of composition \citep{labrosse_2007,ziegler_2013,blanc_2020,orourke_2020}, and the melting curve is either parameterized with a heuristic temperature depression \citep{blanc_2020}, or defined using a linear phase diagram fixed over time \citep{labrosse_2007,ziegler_2013,orourke_2020}.

\sloppy

We couple the core, BMO, and mantle in the planet energy budget, and prescribe the surface heat flow with a thermal boundary layer model of the mantle. This augments previous work where the core energy budget is simplified \citep{labrosse_2007,ziegler_2013,blanc_2020} and the cooling history of the BMO is imposed with a mantle boundary layer model  \citep{labrosse_2007,ziegler_2013}, an exponential decay based on radiogenic heating \citep{blanc_2020}, and a linear decay \citep{orourke_2020}.
Furthermore, we determine the thermocompositional stability of the BMO and the core, highlighting the dominant sources of convection and revealing potential stably-stratified layers. This builds upon previous work where this method was only applied to the core \citep{davies_2011a,pozzo_2012,gubbins_2015}.
Finally, we determined the conditions of dynamo operation in the BMO and the core by using entropy budgets to assessing if sufficient power is available to sustain a magnetic field, and if magnetic induction exceeds magnetic diffusion, using magnetic Reynolds numbers. This extends previous work where the dynamo activity is only assessed through magnetic Reynolds numbers \citep{ziegler_2013,scheinberg_2018,orourke_2020,stixrude_2020}, or entropy budgets \citep{blanc_2020}.

\fussy

Although these methodological extensions introduce new physical ingredients to previous planetary evolution models, we obtain a nominal model relatively similar to previous works targeted at Earth. However, a large diversity of planetary evolution can arise from these new ingredients, as shown by the sensitivity of the model to multiple key parameters. As a result, these new ingredients should be included in models of other planets in the Solar System and beyond.

\subsection{Sensitivity of the model}
\label{sec:discussion_sensitivity}

The planet evolution is sensitive to several parameters, including the initial temperature at the CMB (§\ref{sec:results_parameters_temperature}), the parameterization of mantle convection (§\ref{sec:results_parameters_convection}), the composition of the BMO (§\ref{sec:results_parameters_composition}), and the radiogenic content of the planet (§\ref{sec:results_parameters_radio}).
Since the model is sensitive to these parameters, this highlights the need to better constrain these quantities with experiments and simulations at planetary interior conditions.

The initial temperature at the CMB, $T_{CMB_0}$, depends on the accretion history of the planet, and the efficiency of heat partitioning between the mantle and the core \citep{rubie_2015}. This requires understanding the conditions of magma ocean formation following giant impacts at the end of accretion \citep{tonks_1993,nakajima_2015,nakajima_2021}, the processes that lead to the formation of long-lived BMO \citep{labrosse_2015a}, and heat partitioning between metal and silicates during \citep{landeau_2021,lherm_2022} and after \citep{deguen_2014} the impacts.
Furthermore, given the definition of $T_{CMB_0}$ (§\ref{sec:model_thermal_solution}), variations of this parameter may also be interpreted as uncertainties in the EOS, through the adiabatic temperature profile, and the liquidus of the BMO.

Mantle convection involves several parameters, including a critical Rayleigh number and the exponent of the $Nu \propto Ra^\beta$ relationship \citep{ricard_2015}. These parameters depend on the proportion of internal heating, the rheology of the mantle and partial melting \citep{driscoll_2014}, as well as the dynamics of plate tectonics. Mantle convection is also influenced by the mantle viscosity, which depends nonlinearly on temperature, pressure, stress and mineralogical variables \citep{ricard_2015}, the thermal conductivity, which depends on pressure and temperature \citep{vandenberg_2010}, and the EOS, which depends on the composition of the solid mantle \citep{fumagalli_2015,irifune_2015}.

\sloppy

The evolution of the planet depends on parameters related to composition, including in particular the iron partition coefficient $D_{BMO}$, the initial mass fraction $\xi_{b_0}$ of $\mathrm{FeO}$, the compositional density jump at the top of the BMO $\Delta_\xi\rho_{BMO}$, and at the compositional density jump at the ICB $\Delta_\xi\rho_{ICB}$.
$D_{BMO}$ has been determined experimentally for moderate iron concentrations, giving values in the range $0.1-0.6$ \citep{nomura_2011,andrault_2012,tateno_2014}. Experimental dispersion might be explained by differences in the sample analyses, variations in the iron oxidation and spin states, and the presence of aluminium \citep{boukare_2015}. $D_{BMO}$ may also change at high iron concentrations \citep{blanc_2020}.
$\xi_{b_0}$ is expected to depend on the formation mechanism of the BMO \citep{labrosse_2015a}, and metal-silicate equilibration during core formation \citep{rubie_2015}.
In addition, the value of $\Delta_\xi\rho_{ICB}$, which depends on the light elements content of the core, is uncertain \citep{hirose_2013}, as is $\Delta_\xi\rho_{BMO}$ \citep{caracas_2019}.
Finally, uncertainties in the EOS and liquidus of $\mathrm{MgSiO_3}$ \citep{fratanduono_2018,wolf_2018} and $\mathrm{FeO}$ \citep{morard_2022,dobrosavljevic_2023} are likely to influence the value of these parameters. Our model is also expected to be sensitive to the crystallization sequence of the BMO \citep{boukare_2015,nabiei_2021}.

\fussy

The model depends on parameters related to radiogenic heating, including the mantle fraction $f_h$ of BSE radiogenic production and the $\mathrm{^{40}K}$ core concentration $c_c$.
$f_h$ depends on heat production in the convective mantle and the lithosphere. However, these radioactive heat sources remain uncertain, since the range of mantle and lithosphere heat production are respectively $9-17~\mathrm{TW}$ and $7-8~\mathrm{TW}$ \citep{jaupart_2015}. Similarly, the concentration of $\mathrm{^{40}K}$ in the core is also uncertain, with proposed concentration exceeding $130~\mathrm{ppm}$ \citep{hirose_2013}.

The magnetic evolution of the planet is also sensitive to the electrical conductivity $\sigma_b$ of the silicates and the thermal conductivity $k_c$ of the liquid metal (§\ref{sec:results_parameters_conductivity}).
In the BMO, \textit{ab initio} calculations show that the thermal conductivity is expected to increase by several order of magnitude at high pressure and temperature conditions \citep{scipioni_2017}, and with the addition of metal \citep{holmstrom_2018,soubiran_2018,stixrude_2020}. These calculations typically yield electrical conductivities in the range $10^4-10^5~\mathrm{Pa.s}$, which cover our silicate dynamo threshold of $\sigma_b \geq 2.15 \times 10^{4}~\mathrm{S.m^{-1}}$. As a result, the existence of a silicate dynamo on an Earth-like planet remains open to debate.
In the core, recent experiments and simulations provide a relatively broad range of thermal conductivity \citep{ohta_2016,basu_2020,hsieh_2020,pozzo_2022}, falling within $20-250~\mathrm{W.m^{-1}.K^{-1}}$. This raises questions about the age of the dynamo onset and inner core nucleation, as well as the need for an alternative mechanism to explain paleomagnetic data.
In any case, the sensitivity of the model to these transport properties highlights the need to better constrain these quantities.

\subsection{Model limitations and future opportunities}
\label{sec:discussion_limitations}

\sloppy

Our model has a number of limitations that offer opportunities for further work.
The internal structure of the planet should be integrated self-consistently \citep{zhang_2022} using an EOS that accurately parameterizes the effects of thermal boundary layers, melting, and evolving compositions. The sensitivity of the model to the EOS and melting curves should also be explicitly evaluated.
In the solid mantle, the parameterization of the EOS should be improved by including iron \citep{irifune_2015} and additional phase transitions \citep{tsuchiya_2011,umemoto_2017}.
In the BMO, an evolving multi-component composition in realistic pressure and temperature conditions should be included, by using for example, a crystallization sequence in the $\mathrm{MgO-FeO-SiO_2}$ system \citep{boukare_2015,nabiei_2021}. This could be used to parameterize the mantle liquidus more accurately than a linear phase diagram. The EOS of the BMO should also be parameterized using self-consistent integration, rather than simply the reference density profile of the solid mantle.
In the core, light elements should influence the parameterization of both the EOS and the liquidus \citep{hirose_2021}.

\fussy

The planet evolution should include conductive layers \citep{gomi_2013,laneuville_2018,greenwood_2021} with conductive temperature profiles in the solid inner core, the stably-stratified layers, and the thermal boundary layers, where we presently overestimate the cooling rates by using adiabatic profiles.
The dynamics of multi-phase flows beyond the low melt fraction rheological transition \citep{boukare_2015,boukare_2017} should be parameterized to allow for situations where the BMO is not necessarily well-mixed and adiabatic. The role of the density contrast between melt and crystals on the formation and evolution of the BMO should also be considered \citep{boukare_2015,caracas_2019,nabiei_2021}.
Radiogenic production should account for the fractionation of radioisotopes between the BMO and the mantle, as well as between between the outer and the inner core, during the crystallization of the liquid phases.

\sloppy

The magnetic evolution of the planet relies on scaling laws extrapolated over several orders of magnitude to reach the conditions of planetary interiors, potentially leading to dynamics invisible in the dynamo simulations \citep{christensen_2010}. These scaling laws, originally developed for the Earth's core, may be modified in the BMO, with a dynamo operating in a thinner shell \citep{heimpel_2005} at a larger Prandtl number \citep{simitev_2005}.
Furthermore, a dynamo produced in the BMO may interact with the underlying core. For example, even if the core is stably-stratified, vigorous horizontal flows are likely to develop, converting the BMO poloidal field into a toroidal field and leading to a reinforcement of the silicate dynamo \citep{ziegler_2013}. Such interaction may require a criterion for dynamo operation that accounts for the magnetic Reynolds number in both layers, \textit{e.g.} a dynamo number \citep{ziegler_2013}.

\fussy

Beyond an Earth-like planet investigated in this paper, the model can be used to analyse the evolution of other objects likely to have a BMO. In addition to Venus \citep{orourke_2020} and the Moon \citep{hamid_2023}, Mars has had a BMO that still exists today \citep{khan_2023,samuel_2023}. Beyond the Solar System, terrestrial exoplanets may host a BMO, in particular Super-Earths \citep{blaske_2021,zhang_2022}.

\section{Conclusion}
\label{sec:conclusion}

\sloppy

In this paper, we have investigated the structural, thermal, buoyancy, and magnetic evolution of an Earth-like planet, with a coupled mantle, long-lived BMO, and solid and/or liquid core. Using conservative parameters, we have produced an Earth-like nominal model that includes an early silicate dynamo. Our parameter sensitivity sweep of the model provides insight into the range of BMO electrical conductivity and core thermal conductivity required to sustain a dynamo.

\fussy

In our nominal model, the Earth-like planet starts with high temperature, thick iron-rich BMO, and a fully liquid core. Vigorous convection in the BMO, primarily compositionally driven by release of $\mathrm{FeO}$ from above, provides a supercritical magnetic Reynolds number and sufficient power to sustain a silicate dynamo. The intense activity of the BMO limits the heat flow at the CMB, producing a stably-stratified core with no core dynamo at early times.

As the planet cools, the solidifying BMO is progressively enriched in iron, while the release rate of $\mathrm{FeO}$, the crystallization rate of the BMO and the cooling rate at the CMB all progressively increase. Core convection then develops from the bottom up at 0.8 Gyr. Eventually, the convective power available to sustain a core dynamo increases, initiating a core dynamo at 1.2 Gyr.

At 1.0 Gyr, the silicate dynamo vanishes as the magnetic Reynolds number becomes subcritical in a progressively thinner BMO. The BMO solidifies at 2.8 Gyr, leaving a solid mantle overlying a liquid core.
At 3.8 Gyr, the central core temperature drops below the core liquidus, nucleating the inner core. The onset of compositional convection from the release of light elements at the ICB suddenly increases the convective power, enhancing the core dynamo and magnetic field.

Since our model is sensitive to the initial temperature at the CMB, the parameterization of mantle convection, the composition of the BMO, and the radiogenic content of the planet, the actual evolution of Earth might have differed from our Earth-like nominal model. Nonetheless, our nominal evolution is broadly consistent with the present-day Earth structure, as well as previous evolution models involving a long-lived BMO. The inner core radius, temperatures at the CMB and ICB, and heat flows at the CMB and surface are close to Earth's values. Our model is also largely consistent with paleomagnetic data, except before 1.0 Gyr where the model predicts a silicate dynamo with a relatively strong magnetic field intensity. However, its value is particularly sensitive to the magnetic scaling laws being used.

\section*{Declaration of competing interest}

The authors declare that they have no known competing financial interests or personal relationships that could have appeared to influence the work reported in this paper.

\section*{Data availability}

\sloppy

The source code required to reproduce the above findings is available to download from \href{https://doi.org/10.17632/3gj9r8rzx9.1}{https://doi.org/10.17632/3gj9r8rzx9.1}.

\fussy

\section*{Funding}

\sloppy

Funding for this research was provided by the Center for Matter at Atomic Pressure (CMAP), a National Science Foundation (NSF) Physics Frontier Center, under award PHY-2020249. Any opinions, findings, conclusions or recommendations expressed in this material are those of the author(s) and do not necessarily reflect those of the National Science Foundation.
E.B. acknowledges working in part at the Aspen Center for Physics, which is supported by National Science Foundation grant PHY-2210452.
M.N. was supported in part by the National Aeronautics and Space Administration (NASA) grant numbers 80NSSC19K0514, 80NSSC21K1184, and 80NSSC22K0107. Partial funding for M.N. was also provided by NSF EAR-2237730 and by the Alfred P. Sloan Foundation under grant G202114194.

\fussy

\appendix

\section{Equation of state of the mantle}
\label{app:EOS_mantle}

The isothermal bulk modulus $K_T$ of the solid mantle is defined using a third order Vinet-Rydberg EOS
\begin{equation}
    K_T=K_0X^{-2}\exp\left[\eta (1-X)\right]\left[1+(1-X)(\eta X+1)\right]
    \label{eq:reference_isothermal_bulk_modulus}
\end{equation}
where $X=(\hat{V}/\hat{V}_0)^{1/3}$ and $\eta=\frac{3}{2}\left(K_0'-1\right)$ \citep{vinet_1989}. The molar volume $\hat{V}$ is related to the density through $\rho \hat{V}=\rho_0 \hat{V}_0=M$, where $M$ is the molar mass. Furthermore, $\hat{V}_0$, $\rho_0$, $K_0$ and $K_0'$ are the reference molar volume, density, bulk modulus and derivative of the bulk modulus with respect to pressure. These parameters, identical to those of \citet{boujibar_2020}, are defined for each phase of the reference model (Table S2).
The volume dependence of the thermal expansivity and the Grüneisen parameter are respectively
\begin{eqnarray}
    \label{eq:reference_thermal_expansivity}
    \alpha&=&\alpha_0x^{3},\\
    \label{eq:reference_gruneisen}
    \gamma&=&\gamma_0x^{\gamma_1},
\end{eqnarray}
where $x=\hat{V}/\hat{V}_0$. $\alpha_0$, $\gamma_0$ and $\gamma_1$ are parameters defined in Table S2.

\section{Equation of state of the basal magma ocean}
\label{app:EOS_BMO}

\sloppy

For $\mathrm{MgSiO_3}$, we use a generalized Rosenfeld-Tarazona EOS \citep{rosenfeld_1998,wolf_2018}
\begin{equation}
    \begin{array}{r}
        \hat{F}_\mathrm{MgSiO_3}(\hat{V},T)=\hat{U}_0+\hat{E}_0(\hat{V})+\Delta \hat{E}(\hat{V},T_0 \rightarrow T) \\
        +\hat{S}_0(T_0-T)+T_0 \Delta \hat{S}(\hat{V},T_{0_S} \rightarrow T_0)                                      \\
        -T\Delta \hat{S}(\hat{V},T_{0_S} \rightarrow T)+\hat{F}_e(\hat{V},T)
    \end{array}
    \label{eq:free_energy_MgSiO3}
\end{equation}
where $\hat{U}_0$ is the reference energy at reference temperature $T_0$ and volume $\hat{V}_0$, $\hat{E}_0$ is the potential contribution of the free energy at $T_0$, $\Delta \hat{E}$ is the perturbation in internal energy from $T_0$, $\Delta \hat{S}$ is the perturbation in entropy from $T_{0_S}$, with $T_{0_S}$ the reference adiabatic temperature profile at reference entropy $\hat{S}_0$, and $\hat{F}_e$ is the electronic contribution of the free energy.
The potential contribution is expressed as a third order Vinet-Rydberg EOS \citep{vinet_1989}
\begin{equation}
    \begin{array}{r}
        \hat{E}_0=9K_0\hat{V}_0\eta^{-2}\left\{1-\left[1-\eta(1-X)\right]\right. \\
        \left.\times\exp\left[(1-X)\eta\right]\right\}.
    \end{array}
    \label{eq:energy_vinet}
\end{equation}
The perturbation in internal energy is
\begin{equation}
    \begin{array}{rcl}
        \Delta \hat{E}(\hat{V}, T_0 \rightarrow T) & = & b(\hat{V})\left[\left(\frac{T}{T_0}\right)^m-1\right] \\
                                                   & + & \frac{3}{2} n R(T-T_0),
    \end{array}
    \label{eq:energy_perturbation_MgSiO3}
\end{equation}
where $m$ is the power-law exponent of the generalized Rosenfeld-Tarazona model, $n$ is the number of atoms in the chemical formula of the compound and $R$ is the gas constant. Furthermore, $b(\hat{V})$ is a polynomial representation of the thermal coefficients $b_n$, \textit{i.e.}
\begin{equation}
    b(\hat{V})=\sum_{n=0}^{4} b_n(x-1)^n.
    \label{eq:thermal_coefficients_MgSiO3}
\end{equation}
The perturbation in entropy is
\begin{equation}
    \begin{array}{r}
        \Delta \hat{S}(\hat{V}, T_{0_S} \rightarrow T)=b(\hat{V})\frac{m}{m-1}T_0^{-m}\left[T^{m-1}-T_{0_S}^{m-1}\right] \\
        +\frac{3}{2} n R\ln\left(\frac{T}{T_{0_S}}\right).
    \end{array}
    \label{eq:entropy_perturbation_MgSiO3}
\end{equation}
The reference adiabatic temperature profile $T_{0_S}$ is obtained with
\begin{equation}
    T_{0_S}=T_0\exp\left(-\int_{\hat{V}_{0}}^{\hat{V}} \frac{\gamma}{\hat{V}'}\mathrm{d}\hat{V}'\right),
    \label{eq:adiabat_MgSiO3}
\end{equation}
where we use a modified Altshuler form of the Grüneisen parameter of $\mathrm{MgSiO_3}$ \citep{fratanduono_2018} as the reference Grüneisen parameter
\begin{equation}
    \gamma
    =\gamma_\infty+(\gamma_0-\gamma_\infty)\left(\frac{\rho}{\rho_{0_\gamma}}\right)^{-\gamma_1}
    +\gamma_{01}\exp\left[-\frac{(\rho-\rho_e)^2}{\rho_\sigma^2}\right].
    \label{eq:gruneisen_MgSiO3}
\end{equation}
The electronic contribution is expressed as
\begin{small}
    \begin{equation}
        \hat{V}_e = \left\{
        \begin{array}{lcl}
            \zeta\left[T T_e\ln\left(\frac{T}{T_e}\right)-\frac{1}{2}(T^2-T_e^2)\right] & \mathrm{if} & T > T_e    \\
            0                                                                           & \mathrm{if} & T \leq T_e
        \end{array}
        \right.,
        \label{eq:electronic_contribution_MgSiO3}
    \end{equation}
\end{small}
The thermo-electronic heat capacity coefficient $\zeta$ and the electronic temperature $T_e$ both depend on volume as
\begin{eqnarray}
    \label{eq:elec_heat_capacity_MgSiO3}
    \zeta&=&\zeta_0 x^{\zeta_1}, \\
    \label{eq:elec_temperature_MgSiO3}
    T_e&=&T_{e_0} x^{T_{e_1}}.
\end{eqnarray}
The parameters used for the EOS of $\mathrm{MgSiO_3}$ are defined in Table S5.

\fussy

For $\mathrm{FeO}$ \citep{dorogokupets_2017,morard_2022}, we use
\begin{equation}
    \begin{array}{r}
        \hat{F}_\mathrm{FeO}(\hat{V},T)=\hat{U}_0+\hat{E}_0(\hat{V})+\hat{F}_{th}(\hat{V},T)+\hat{F}_e(\hat{V},T) \\
        -\hat{F}_{th}(\hat{V},T_0)-\hat{F}_e(\hat{V},T_0)-a_SR(T-T_0),
    \end{array}
    \label{eq:free_energy_FeO}
\end{equation}
where $\hat{F}_{th}$ is the thermal contribution of the free energy and where the part proportional to $a_S$ corresponds to the residual entropy of the liquid at 0 K.
The potential contribution is expressed as a third order Vinet-Rydberg EOS (Eq. \ref{eq:energy_vinet}).
The thermal contribution is expressed with the Einstein model, \textit{i.e.} the Debye model in the high temperature limit
\begin{equation}
    \hat{F}_{th}=3nRT\ln\left[1-\exp\left(-\frac{\Theta}{T}\right)\right].
    \label{eq:thermal_contribution_FeO}
\end{equation}
The Einstein temperature $\Theta$ is
\begin{equation}
    \Theta=\Theta_0 x^{-\gamma_\infty}\exp\left[\frac{\gamma_0-\gamma_\infty}{\gamma_1}\left(1-x^{\gamma_1}\right)\right],
    \label{eq:einstein_temperature}
\end{equation}
where $\Theta_0$ is the Einstein temperature prefactor. $\gamma_0$, $\gamma_1$ and $\gamma_\infty$ are defined in the Altshuler form of the Grüneisen parameter
\begin{equation}
    \gamma=\gamma_\infty+(\gamma_0-\gamma_\infty)\left(\frac{\rho}{\rho_{0}}\right)^{-\gamma_1}.
    \label{eq:gruneisen_FeO}
\end{equation}
The electronic contribution is expressed as
\begin{equation}
    \hat{F}_e=-\frac{3}{2}nRe_0x^{e_1}T^2
    \label{eq:electronic_contribution_FeO}
\end{equation}
The parameters used for the EOS of $\mathrm{FeO}$ are defined in Table S5.

\section{Thermal conductivity of the mantle}
\label{app:k_mantle}

In this model \citep{vandenberg_2010}, the thermal conductivity is the sum of the phonon $k_l$, photon $k_r$ and electronic $k_e$ contributions, \textit{i.e.}
\begin{equation}
    k=k_l+k_r+k_e.
    \label{eq:conductivity_mantle}
\end{equation}
The phonon contribution corresponds to a truncated version of the model of \citet{hofmeister_1999}
\begin{equation}
    k_l=\left({k_l^*}^{-2}+k_\mathrm{trunc}^{-2}\right)^{-1/2},
    \label{eq:conductivity_mantle_phonon_1}
\end{equation}
with
\begin{equation}
    \begin{array}{r}
        k_l^*=k_{l_0}\left(\frac{T_{P}}{T}\right)^{a_k}\exp\left[-\left(4\gamma+\frac{1}{3}\right)\alpha(T-T_{P})\right] \\
        \times \left(1+\frac{P}{K_T}\frac{\mathrm{d}K_T}{\mathrm{d}P}\right),
    \end{array}
    \label{eq:conductivity_mantle_phonon_2}
\end{equation}
where $k_{l_0}$ and $a_k$ are the phonon conductivity prefactor and exponent and $k_\mathrm{trunc}$ is the truncation value.
The photon contribution is set to $k_{r_0}$ in an operational temperature window defined by $T_{r_1}$, $T_{r_2}$, $T_{r_3}$ and $T_{r_4}$, and tapered off to zero outside of this window using sine functions
\begin{small}
    \begin{equation}
        k_r=\left\{
        \begin{array}{lll}
            0                                                                                                                & \mathrm{if} & T < T_{r_1}              \\
            \frac{k_{r_0}}{2}\left\{1+\sin\left[\pi\left(\frac{T-T_{r_1}}{T_{r_2}-T_{r_1}}-\frac{1}{2}\right)\right]\right\} & \mathrm{if} & T_{r_1} \leq T < T_{r_2} \\
            k_{r_0}                                                                                                          & \mathrm{if} & T_{r_2} \leq T < T_{r_3} \\
            \frac{k_{r_0}}{2}\left\{1+\sin\left[\pi\left(\frac{T-T_{r_3}}{T_{r_4}-T_{r_3}}+\frac{1}{2}\right)\right]\right\} & \mathrm{if} & T_{r_3} \leq T < T_{r_4} \\
            0                                                                                                                & \mathrm{if} & T \geq T_{r_4}           \\
        \end{array}
        \right..
        \label{eq:conductivity_mantle_photon}
    \end{equation}
\end{small}
The electronic contribution is defined as
\begin{equation}
    k_e=k_{e_0}\exp\left(-\frac{E_k}{k_B T}\right),
    \label{eq:conductivity_mantle_electron}
\end{equation}
where $k_{e_0}$ and $E_k$ are the electron conductivity prefactor and an electron transport activation energy and $k_B$ is the Boltzmann constant. The parameters of the model are defined in Table S6.

\section{Expression of the crystallization rates}
\label{app:crystallization_rate}

\begin{figure*}[!ht]
    \centering
    \includegraphics[width=\linewidth]{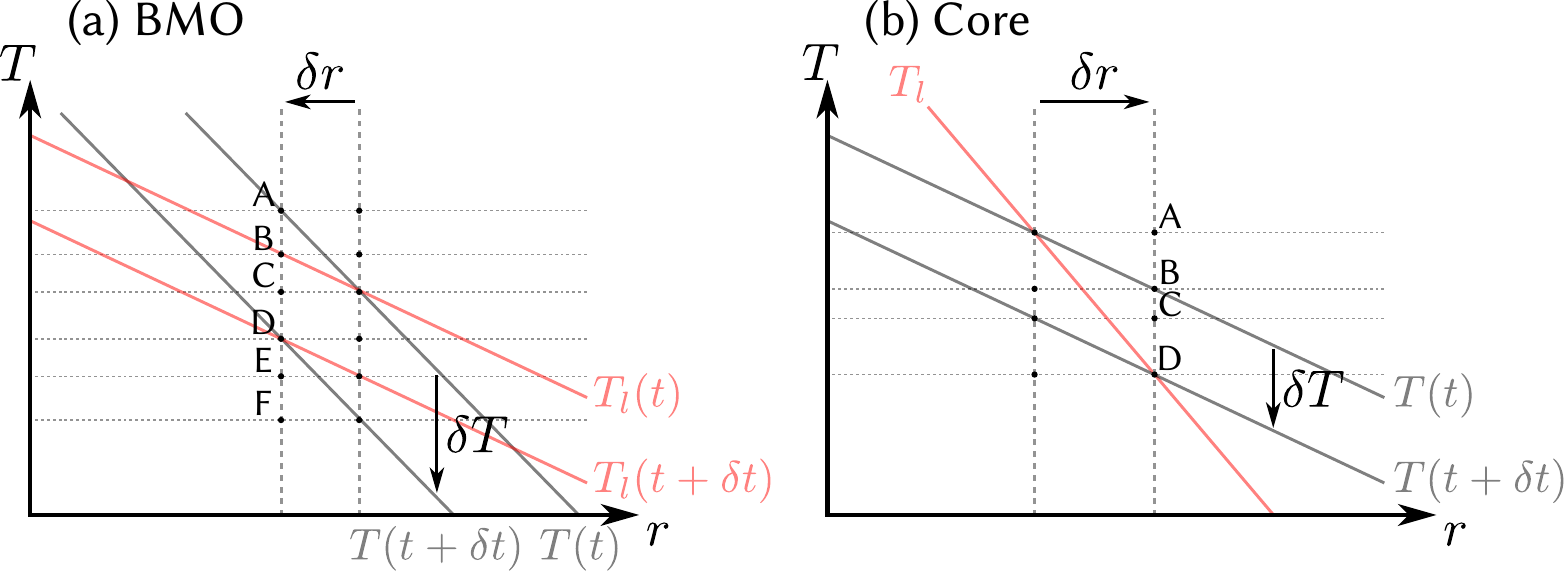}
    \caption{(a) Variation $\delta r$ of the interface between the BMO and the mantle caused by a temperature decrease $\delta T$ during a time $\delta t$. The adiabatic temperature $T$ varies in response to the cooling of the BMO and changes in the composition of the BMO influence the mantle liquidus $T_l$. (b) Variation $\delta r$ of the ICB radius caused by a temperature decrease $\delta T$ during a time $\delta t$. The adiabatic temperature $T$ varies in response to the cooling of the core but the mantle liquidus $T_l$ is independent of the core composition.}
    \label{fig:radius_sketch}
\end{figure*}

In the BMO (Fig. \ref{fig:radius_sketch}a), the infinitesimal temperature drop that occurs during $\delta t$ is $\delta T=AD=AC+BD-BC$, where $AC=-(\mathrm{d}T/\mathrm{d}r)\delta r$, $BC=-(\mathrm{d}T_l/\mathrm{d}r)\delta r$ and $BD=(\mathrm{d}T_l/\mathrm{d}t)\delta t$. $\delta r$ is the corresponding radius decrease of the interface between the BMO and the mantle.
Since we assume an adiabatic temperature profile, the cooling rate at the top of the BMO is proportional to the cooling rate at the CMB (Eq. \ref{eq:adiabatic_simplification}). Thus $\delta T=(T_{BMO}/T_{CMB})\delta T_{CMB}$, where $\delta T_{CMB}$ is the infinitesimal temperature drop at the CMB.
Using Eq. \ref{eq:composition_rate_BMO}, we have $\mathrm{d}T_l/\mathrm{d}t=(\mathrm{d}T_l/\mathrm{d}\xi)C_{\xi_{b}}(\mathrm{d}R_{BMO}/\mathrm{d}t)$, where $\mathrm{d}T_l/\mathrm{d}\xi$ is the slope of the linear phase diagram (Eq. \ref{eq:liquidus_BMO}) at a given pressure.
We finally write the crystallization rate as
\begin{equation}
    \begin{array}{rcl}
        \frac{\mathrm{d}R_{BMO}}{\mathrm{d}t}
        &=&\frac{T_{BMO}}{T_{CMB}}
        \left[\left(\frac{\mathrm{d}T_l}{\mathrm{d}r}-\frac{\mathrm{d}T}{\mathrm{d}r}\right)_{R_{BMO}}\right. \\
        &+&\left.\left(\frac{\mathrm{d}T_l}{\mathrm{d}\xi}\right)_{R_{BMO}}C_{\xi_{b}}\right]^{-1}\frac{\mathrm{d}T_{CMB}}{\mathrm{d}t} \\
        &=&C_{R_{b}}\frac{\mathrm{d}T_{CMB}}{\mathrm{d}t}.
    \end{array}
    \label{eq:crystallization_rate_BMO_full}
\end{equation}

In the core (Fig. \ref{fig:radius_sketch}b), the infinitesimal temperature drop is $\delta T=AC=AD-CD$, where $AD=(\mathrm{d}T_l/\mathrm{d}r)\delta r$ and $CD=(\mathrm{d}T/\mathrm{d}r)\delta r$. $\delta r$ is the corresponding radius increase of the ICB.
Since we assume an adiabatic temperature profile, the cooling rate at the ICB is in particular proportional to the cooling rate at the CMB (Eq. \ref{eq:adiabatic_simplification}), which means that $\delta T=(T_{ICB}/T_{CMB})\delta T_{CMB}$.
We can then write the crystallization rate as
\begin{equation}
    \begin{array}{rcl}
        \frac{\mathrm{d}R_{ICB}}{\mathrm{d}t}
        &=&\frac{1}{\left(\frac{\mathrm{d}T_l}{\mathrm{d}r}-\frac{\mathrm{d}T}{\mathrm{d}r}\right)_{R_{ICB}}}\frac{T_{ICB}}{T_{CMB}}\frac{\mathrm{d}T_{CMB}}{\mathrm{d}t} \\
        &=&C_{R_{c}}\frac{\mathrm{d}T_{CMB}}{\mathrm{d}t}.
    \end{array}
    \label{eq:crystallization_rate_core_full}
\end{equation}

\section{Energy budget terms}
\label{app:energy_budget}

The radiogenic heating in the mantle is integrated over the volume $V_m$ of the mantle
\begin{equation}
    Q_{R,m}=\int_{V_m} \rho h_m f_h \mathrm{d}V
    \label{eq:radio_mantle}
\end{equation}
where $h_m$ is the specific heat production rate of the BSE and $f_h$ is the mantle fraction of the BSE production.
We write the specific heat production rate as
\begin{equation}
    h_m=\sum_{i}c_{m_i}a_ie_i\frac{N_A\lambda_i}{M_i}\exp\left[-\lambda_i(t-t_P)\right]
    \label{eq:heat_production_BSE}
\end{equation}
where $c_{m_i}$, $a_i$, $e_i$, $M_i$ and $\lambda_i$ are respectively the BSE concentration at $t=t_P=4.5~\mathrm{Gyr}$, the natural abundance, the energy per atom, the molar mass and the decay constant of the radioactive isotopes $i$.
The radiogenic heating parameters are defined in Table S7.

The secular cooling of the mantle is
\begin{equation}
    Q_{S,m}=-\int_{V_m} \rho c_P\frac{\mathrm{d}T_m}{\mathrm{d}t}\mathrm{d}V.
    \label{eq:secular_mantle}
\end{equation}
Using Eq. \ref{eq:adiabatic_simplification}, the adiabatic temperature profile in the BMO allows to write \citep{gubbins_2003}
\begin{equation}
    \frac{1}{T_{BMO}}\frac{\mathrm{d}T_{BMO}}{\mathrm{d}t}=\frac{1}{T_{CMB}}\frac{\mathrm{d}T_{CMB}}{\mathrm{d}t},
    \label{eq:secular_mantle_adiabatic}
\end{equation}
which using Eq. \ref{eq:T_m_R_m} gives
\begin{equation}
    \frac{\mathrm{d}T_m}{\mathrm{d}t}=\frac{1}{2}\frac{T_{BMO}}{T_{CMB}}\frac{\mathrm{d}T_{CMB}}{\mathrm{d}t}.
    \label{eq:secular_mantle_simplification}
\end{equation}

The adiabatic heat flow at the top of the mantle is
\begin{equation}
    Q_{A,m}=-4\pi R_P^2 \bar{k}_m \frac{T_P-T_{BMO}}{L_m},
    \label{eq:Q_A_mantle}
\end{equation}
where $\bar{k}_m$ is the average thermal conductivity of the mantle (Appendix \ref{app:k_mantle}).

The radiogenic heating in the BMO is integrated over the volume $V_{b}$ of the BMO
\begin{equation}
    Q_{R,b}=\int_{V_{b}} \rho h_m f_h \mathrm{d}V.
    \label{eq:radio_BMO}
\end{equation}

The secular cooling of the BMO is
\begin{equation}
    Q_{S,b}=-\int_{V_{b}} \rho c_P \frac{T}{T_{CMB}}\frac{\mathrm{d}T_{CMB}}{\mathrm{d}t}\mathrm{d}V.
    \label{eq:secular_BMO}
\end{equation}
In Eq. \ref{eq:secular_BMO}, we assume an adiabatic temperature profile in the BMO (Eq. \ref{eq:adiabatic_simplification}).

The gravitational energy of the BMO is
\begin{equation}
    Q_{G,b}=\int_{V_{b}} \alpha_{c_{b}}\rho\psi\frac{\mathrm{D}\xi_{b}}{\mathrm{D}t}\mathrm{d}V,
    \label{eq:gravitational_BMO}
\end{equation}
where $\psi$ is the gravitational potential. In Eq. \ref{eq:gravitational_BMO}, the compositional expansion coefficient $\alpha_{c_{b}}$ \citep{gubbins_2004} is
\begin{equation}
    \alpha_{c_{b}}=-\frac{1}{\rho}\left(\frac{\partial\rho}{\partial\xi}\right)_{P,T} \simeq \frac{\Delta_\xi\rho_{BMO}}{\rho_{BMO}^+\Delta\xi_{BMO}},
    \label{eq:compositional_expansivity_BMO}
\end{equation}
where $\rho_{BMO}^+$ is the density in the solid mantle at the BMO interface and $\Delta_\xi\rho_{BMO}$ is the compositional contribution to the density jump due to a change in composition $\Delta\xi_{BMO}$ across the interface. Following \citet{blanc_2020}, we use data from \citet{caracas_2019} to set $\Delta_\xi\rho_{BMO}=250~\mathrm{kg.m^{-3}}$.

The latent heat of the BMO is
\begin{equation}
    Q_{L,b}=-4\pi R_{BMO}^2\Delta S_{m} T_{BMO}\rho_{BMO}^-\frac{\mathrm{d}R_{BMO}}{\mathrm{d}t},
    \label{eq:latent_BMO}
\end{equation}
where $\Delta S_{m}=300~\mathrm{J.kg^{-1}.K^{-1}}$ is the entropy of fusion in the mantle \citep{labrosse_2007}. The sensitivity of the model to $\Delta S_{m}$ is investigated in Fig. S7.

The adiabatic heat flow at the top of the BMO is
\begin{equation}
    Q_{A,b}=-4\pi R_{BMO}^2 k_b \left(\frac{\mathrm{d}T}{\mathrm{d}r}\right)_{BMO},
    \label{eq:Q_A_BMO}
\end{equation}
where the adiabatic gradient is obtained from Eq. \ref{eq:adiabat_BMO}.

The radiogenic heating in the core is integrated over the volume $V_{oc}$ of the outer core
\begin{equation}
    Q_{R,c}=\int_{V_{oc}} \rho h_c \mathrm{d}V,
    \label{eq:radio_core}
\end{equation}
where $h_c$ is the specific heat production rate of the core.
The heat production rate is
\begin{equation}
    h_c=c_{c}a_{\mathrm{^{40}K}}e_{\mathrm{^{40}K}}\frac{N_A\lambda_{\mathrm{^{40}K}}}{M_{\mathrm{^{40}K}}}\exp\left[-\lambda_{\mathrm{^{40}K}}(t-t_P)\right],
    \label{eq:heat_production_core}
\end{equation}
where $c_c$, $a_\mathrm{^{40}K}$, $e_\mathrm{^{40}K}$, $M_\mathrm{^{40}K}$ and $\lambda_\mathrm{^{40}K}$ are respectively the core concentration at $t=t_P$, the natural abundance, the energy per atom, the molar mass and the decay constant of $\mathrm{^{40}K}$.
The radiogenic heating parameters are defined in Table S7.

The secular cooling of the outer core is
\begin{equation}
    Q_{S,c}=-\int_{V_{oc}} \rho c_P \frac{T}{T_{CMB}}\frac{\mathrm{d}T_{CMB}}{\mathrm{d}t}\mathrm{d}V.
    \label{eq:secular_core}
\end{equation}
In Eq. \ref{eq:secular_core}, we assume an adiabatic temperature profile in the outer core (Eq. \ref{eq:adiabatic_simplification}).

The gravitational energy of the core is
\begin{equation}
    Q_{G,c}=\int_{V_{oc}} \alpha_{c_{c}}\rho\psi\frac{\mathrm{D}\xi_{c}}{\mathrm{D}t}\mathrm{d}V.
    \label{eq:gravitational_core}
\end{equation}
As with the BMO, the compositional expansion coefficient $\alpha_{c_{c}}$ \citep{gubbins_2004} is
\begin{equation}
    \alpha_{c_{c}}=-\frac{1}{\rho}\left(\frac{\partial\rho}{\partial\xi}\right)_{P,T} \simeq \frac{\Delta_\xi\rho_{ICB}}{\rho_{ICB}^-\Delta\xi_{ICB}},
    \label{eq:compositional_expansivity_core}
\end{equation}
where $\rho_{ICB}^-$ is the density of the solid inner core at the ICB and $\Delta_\xi\rho_{ICB}$ is the compositional contribution to the density jump due to a change in composition $\Delta\xi_{ICB}$ across the interface. Following \citet{labrosse_2015}, we use data from \citet{alfe_1999} and \citet{masters_2003} to set $\Delta_\xi\rho_{ICB}=580~\mathrm{kg.m^{-3}}$.

The latent heat of the core is
\begin{equation}
    Q_{L,c}=4\pi R_{ICB}^2\Delta S_c T_{ICB}\rho_{ICB}^+\frac{\mathrm{d}R_{ICB}}{\mathrm{d}t},
    \label{eq:latent_core}
\end{equation}
where $\Delta S_{c}=127~\mathrm{J.kg^{-1}.K^{-1}}$ is the entropy of fusion in the core \citep{hirose_2013,labrosse_2015}. The sensitivity of the model to $\Delta S_{c}$ is investigated in Fig. S8.

The adiabatic heat flow at the top of the core is
\begin{equation}
    Q_{A,c}=-4\pi R_{CMB}^2 k_c \left(\frac{\mathrm{d}T}{\mathrm{d}r}\right)_{CMB},
    \label{eq:Q_A_core}
\end{equation}
where the adiabatic gradient is obtained from the reference temperature profile (Eq. \ref{eq:reference_adiabat}).

The radiogenic heating of the inner core is integrated over the volume $V_{ic}$ of the inner core
\begin{equation}
    Q_{R,ic}=\int_{V_{ic}} \rho h_c \mathrm{d}V.
    \label{eq:radio_inner_core}
\end{equation}

The secular cooling of the inner core is
\begin{equation}
    Q_{S,ic}=-\int_{V_{ic}} \rho c_P \frac{T_{ICB}}{T_{CMB}} \frac{\mathrm{d}T_{CMB}}{\mathrm{d}t}\mathrm{d}V.
    \label{eq:secular_inner_core}
\end{equation}
In Eq. \ref{eq:secular_inner_core}, we assume an adiabatic temperature profile in the outer core (Eq. \ref{eq:adiabatic_simplification}).

\section{Buoyancy source terms}
\label{app:buoyancy_source}

The volumetric buoyancy sources associated with radiogenic heating are
\begin{equation}
    \left\{
    \begin{array}{l}
        \epsilon_{R_b} = \frac{\bar{\alpha}}{\bar{c}_P}\left(\rho h_m f_h\right) \\
        \epsilon_{R_c} = \frac{\bar{\alpha}}{\bar{c}_P}\left(\rho h_c\right)
    \end{array}
    \right..
    \label{eq:epsilon_radio}
\end{equation}
The boundary conditions, corresponding to the radiogenic heat flow at the top of the layer, are
\begin{equation}
    \left\{
    \begin{array}{rcl}
        \kappa_T C_{0_{R,b}}'(r=R_{BMO}) & = & -Q_{R,b}\frac{\bar{\alpha}}{\bar{c}_P}\frac{1}{4\pi R_{BMO}^2} \\
        \kappa_T C_{0_{R,c}}'(r=R_{CMB}) & = & -Q_{R,c}\frac{\bar{\alpha}}{\bar{c}_P}\frac{1}{4\pi R_{CMB}^2}
    \end{array}
    \right..
    \label{eq:BC_radio}
\end{equation}
The radiogenic mass anomaly fluxes are then
\begin{equation}
    \left\{
    \begin{array}{l}
        q_{R,b} = -\frac{1}{4\pi r^2}\frac{\bar{\alpha}}{\bar{c}_P}Q_{R,b}\left(\frac{r^3-R_{CMB}^3}{R_{BMO}^3-R_{CMB}^3}\right) \\
        q_{R,c} = -\frac{1}{4\pi r^2}\frac{\bar{\alpha}}{\bar{c}_P}Q_{R,c}\left(\frac{r^3-R_{ICB}^3}{R_{CMB}^3-R_{ICB}^3}\right)
    \end{array}
    \right..
    \label{eq:codensity_radio}
\end{equation}

The volumetric buoyancy sources associated with secular cooling are
\begin{equation}
    \epsilon_{S_b} = \epsilon_{S_c} = \frac{\bar{\alpha}}{\bar{c}_P}\left(-\rho c_P \frac{T}{T_{CMB}}\frac{\mathrm{d}T_{CMB}}{\mathrm{d}t}\right).
    \label{eq:epsilon_secular}
\end{equation}
The boundary conditions, corresponding to the secular heat flow at the top of the layer, are
\begin{equation}
    \left\{
    \begin{array}{rcl}
        \kappa_T C_{0_{S,b}}'(r=R_{BMO}) & = & -Q_{S,b}\frac{\bar{\alpha}}{\bar{c}_P}\frac{1}{4\pi R_{BMO}^2} \\
        \kappa_T C_{0_{S,c}}'(r=R_{CMB}) & = & -Q_{S,c}\frac{\bar{\alpha}}{\bar{c}_P}\frac{1}{4\pi R_{CMB}^2}
    \end{array}
    \right..
    \label{eq:BC_secular}
\end{equation}
The secular mass anomaly fluxes are then
\begin{equation}
    \left\{
    \begin{array}{l}
        q_{S,b} = -\frac{1}{ r^2}\int_{R_{CMB}}^r x^2\epsilon_{S_b}\mathrm{d}x \\
        q_{S,c} = -\frac{1}{ r^2}\int_{R_{ICB}}^r x^2\epsilon_{S_c}\mathrm{d}x \\
    \end{array}
    \right..
    \label{eq:codensity_secular}
\end{equation}

\sloppy

The volumetric buoyancy sources associated with dissipation are
\begin{equation}
    \left\{
    \begin{array}{l}
        \epsilon_{G_b} = \frac{\bar{\alpha}}{\bar{c}_P}\left(\frac{Q_{G,b}}{V_{b}}\right) \\
        \epsilon_{G_c} = \frac{\bar{\alpha}}{\bar{c}_P}\left(\frac{Q_{G,c}}{V_{oc}}\right)
    \end{array}
    \right..
    \label{eq:epsilon_dissipation}
\end{equation}
This term is not involved in the heat equation under the Boussinesq approximation, but it contributes to the total mass anomaly flux as a heat source for the convecting system \citep{gubbins_2015}. The distribution of the dissipation term depends on the actual velocity and the magnetic fields and is therefore unknown. Here, we thus assume a uniform dissipation throughout the layer.
The boundary conditions, corresponding to the heat flow associated with the gravitational energy at the top of the layer, are
\begin{equation}
    \left\{
    \begin{array}{rcl}
        \kappa_T C_{0_{G,b}}'(r=R_{BMO}) & = & -Q_{G,b}\frac{\bar{\alpha}}{\bar{c}_P}\frac{1}{4\pi R_{BMO}^2} \\
        \kappa_T C_{0_{G,c}}'(r=R_{CMB}) & = & -Q_{G,c}\frac{\bar{\alpha}}{\bar{c}_P}\frac{1}{4\pi R_{CMB}^2}
    \end{array}
    \right..
    \label{eq:BC_dissipation}
\end{equation}
The dissipation mass anomaly fluxes are then
\begin{equation}
    \left\{
    \begin{array}{l}
        q_{G,b} = -\frac{1}{4\pi r^2}\frac{\bar{\alpha}}{\bar{c}_P}Q_{G,b}\left(\frac{r^3-R_{CMB}^3}{R_{BMO}^3-R_{CMB}^3}\right) \\
        q_{G,c} = -\frac{1}{4\pi r^2}\frac{\bar{\alpha}}{\bar{c}_P}Q_{G,c}\left(\frac{r^3-R_{ICB}^3}{R_{CMB}^3-R_{ICB}^3}\right)
    \end{array}
    \right..
    \label{eq:codensity_dissipation}
\end{equation}

\fussy

The volumetric buoyancy sources associated with the latent heat are
\begin{equation}
    \epsilon_{L_b} = \epsilon_{L_c} = 0.
    \label{eq:epsilon_latent}
\end{equation}
In the BMO, the latent heat is released at the top of the layer, it does not contribute to the buoyancy profile. The corresponding mass anomaly flux is therefore zero. In the core, the latent heat is released at the ICB, it passes through the outer core. Since there is no internal buoyancy source, the mass anomaly flux is determined by the boundary condition, obtained by converting the heat flow associated with latent heat at the ICB. This gives
\begin{equation}
    \left\{
    \begin{array}{rcl}
        \kappa_T C_{0_{L,b}}'(r=R_{BMO}) & = & 0                                                               \\
        \kappa_T C_{0_{L,c}}'(r=R_{ICB}) & = & -Q_{L,c}\frac{\bar{\alpha}}{\bar{c}_P}\frac{1}{4\pi R_{ICB}^2}
    \end{array}
    \right..
    \label{eq:BC_latent}
\end{equation}
The latent heat mass anomaly fluxes are then
\begin{equation}
    \left\{
    \begin{array}{l}
        q_{L,b} = 0 \\
        q_{L,c} = -\frac{1}{4\pi r^2}\frac{\bar{\alpha}}{\bar{c}_P}Q_{L,c}
    \end{array}
    \right..
    \label{eq:codensity_latent}
\end{equation}

\sloppy

The volumetric buoyancy sources associated with the heat source at the CMB and the ICB are
\begin{equation}
    \epsilon_{CMB} = \epsilon_{ICB} = 0.
    \label{eq:epsilon_CMB_ICB}
\end{equation}
Since the heat is released at the bottom of the layer, the internal buoyancy source is zero. The boundary conditions, corresponding to the heat flow at the CMB and the ICB, are
\begin{equation}
    \left\{
    \begin{array}{rcl}
        \kappa_T C_{0_{CMB}}'(r=R_{CMB}) & = & -Q_{CMB}\frac{\bar{\alpha}}{\bar{c}_P}\frac{1}{4\pi R_{CMB}^2} \\
        \kappa_T C_{0_{ICB}}'(r=R_{ICB}) & = & -Q_{ICB}\frac{\bar{\alpha}}{\bar{c}_P}\frac{1}{4\pi R_{ICB}^2}
    \end{array}
    \right..
    \label{eq:BC_CMB_ICB}
\end{equation}
The mass anomaly fluxes associated with the heat flow at the CMB and the ICB are then
\begin{equation}
    \left\{
    \begin{array}{l}
        q_{CMB} = -\frac{1}{4\pi r^2}\frac{\bar{\alpha}}{\bar{c}_P}Q_{CMB} \\
        q_{ICB} = -\frac{1}{4\pi r^2}\frac{\bar{\alpha}}{\bar{c}_P}Q_{ICB}
    \end{array}
    \right..
    \label{eq:codensity_CMB_ICB}
\end{equation}

\fussy

The volumetric buoyancy sources associated with compositional convection are
\begin{equation}
    \left\{
    \begin{array}{ll}
        \epsilon_{C_b} = \bar{\rho}\alpha_{c_b}\frac{\mathrm{D\xi_b}}{\mathrm{D}t} \\
        \epsilon_{C_c} = \bar{\rho}\alpha_{c_c}\frac{\mathrm{D\xi_c}}{\mathrm{D}t}
    \end{array}
    \right..
    \label{eq:epsilon_compositional}
\end{equation}
If we neglect exchanges between the BMO and the core, as well as thermodiffusion and barodiffusion, the flux of $\mathrm{FeO}$ and light elements at the CMB are zero. The boundary conditions, corresponding to a zero mass anomaly flux at the CMB, are
\begin{equation}
    \kappa_T C_{0_{C,b}}'(r=R_{CMB}) = \kappa_T C_{0_{C,c}}'(r=R_{CMB}) = 0.
    \label{eq:BC_compositional}
\end{equation}
The compositional mass anomaly fluxes, are then
\begin{equation}
    \left\{
    \begin{array}{rcl}
        q_{C,b} = -\frac{1}{3 r^2}\epsilon_{C_b}\left(r^3-R_{CMB}^3\right) \\
        q_{C,c} = -\frac{1}{3 r^2}\epsilon_{C_c}\left(R_{CMB}^3-r^3\right)
    \end{array}
    \right..
    \label{eq:codensity_compositional}
\end{equation}

\section{Entropy budget terms}
\label{app:entropy_budget}

The radiogenic heating terms are respectively integrated over the volume of the BMO and the outer core
\begin{equation}
    \left\{
    \begin{array}{l}
        E_{R,b}=\int_{V_{b}} \rho h_m f_h \left(\frac{1}{T_{BMO}}-\frac{1}{T}\right)\mathrm{d}V \\
        E_{R,c}=\int_{V_{oc}} \rho h_c \left(\frac{1}{T_{CMB}}-\frac{1}{T}\right)\mathrm{d}V
    \end{array}
    \right..
    \label{eq:entropy_radio}
\end{equation}

The secular cooling terms, where we assume an adiabatic temperature profile (Eq. \ref{eq:adiabatic_simplification}), are
\begin{equation}
    \left\{
    \begin{array}{l}
        E_{S,b}=-\int_{V_{b}} \rho c_P \frac{1}{T_{CMB}}\left(\frac{T}{T_{BMO}}-1\right)\frac{\mathrm{d}T_{CMB}}{\mathrm{d}t}\mathrm{d}V \\
        E_{S,c}=-\int_{V_{oc}} \rho c_P \frac{1}{T_{CMB}}\left(\frac{T}{T_{CMB}}-1\right)\frac{\mathrm{d}T_{CMB}}{\mathrm{d}t}\mathrm{d}V
    \end{array}
    \right..
    \label{eq:entropy_secular}
\end{equation}

The entropy terms, associated with the gravitational energy of the BMO and the core, are
\begin{equation}
    \left\{
    \begin{array}{l}
        E_{G,b}=\frac{Q_{G,b}}{T_{BMO}} \\
        E_{G,c}=\frac{Q_{G,c}}{T_{CMB}}
    \end{array}
    \right..
    \label{eq:entropy_gravi}
\end{equation}
This terms are expected to have a higher efficiency than the other buoyancy sources due to the intrinsic definition of their efficiency factor \citep{nimmo_2015}.

The latent heat terms in the BMO and the core are
\begin{equation}
    \left\{
    \begin{array}{l}
        E_{L,b}=0 \\
        E_{L,c}=Q_{L,c}\left(\frac{1}{T_{CMB}}-\frac{1}{T_{ICB}}\right)
    \end{array}
    \right..
    \label{eq:entropy_latent}
\end{equation}
Since the latent heat of the BMO is released at the top of the layer, this term does not contribute to the convection in the BMO and the corresponding rate of entropy production is zero \citep{blanc_2020}.

The entropy terms associated with the heat source at the CMB and the ICB are respectively
\begin{equation}
    \left\{
    \begin{array}{l}
        E_{CMB}=Q_{CMB}\left(\frac{1}{T_{BMO}}-\frac{1}{T_{CMB}}\right) \\
        E_{ICB}=Q_{ICB}\left(\frac{1}{T_{CMB}}-\frac{1}{T_{ICB}}\right)
    \end{array}
    \right..
    \label{eq:entropy_CMB_ICB}
\end{equation}

The thermal diffusion terms, obtained using adiabatic temperature profiles, are
\begin{equation}
    \left\{
    \begin{array}{l}
        E_{k,b}=\int_{V_{b}} k_{b} \left(\frac{\nabla T}{T}\right)^2\mathrm{d}V \\
        E_{k,c}=\int_{V_{oc}} k_c \left(\frac{\nabla T}{T}\right)^2\mathrm{d}V
    \end{array}
    \right..
    \label{eq:entropy_thermal}
\end{equation}

The molecular diffusion terms, corresponding to a flux of $\mathrm{FeO}$, or light elements, in a hydrostatic pressure gradient \citep{gubbins_2004}, are
\begin{equation}
    \left\{
    \begin{array}{l}
        E_{\alpha,b}=\alpha_{c_{b}}^2\alpha_{D}\int_{V_{b}} \frac{g^2}{T}\mathrm{d}V \\
        E_{\alpha,c}=\alpha_{c_{c}}^2\alpha_{D}\int_{V_{oc}} \frac{g^2}{T}\mathrm{d}V
    \end{array}
    \right.,
    \label{eq:entropy_compositional}
\end{equation}
where $\alpha_{D}=1.0 \times 10^{12}~\mathrm{kg.m^{-3}.s}$ \citep{gubbins_2004} is a material constant assumed to be identical in the BMO and the core.
In these barodiffusion terms, we neglect thermodiffusion and assume a well-mixed liquid associated with zero concentration diffusion.

\section{Velocity and magnetic scaling laws}
\label{app:scaling_law}

The convective power, or total dissipation \citep{buffett_1996,lister_2003}, available to the dynamo in the BMO or the outer core is approximated by
\begin{equation}
    \Phi=F_{i}\left(\bar{\psi}-\psi_{i}\right)+F_{o}\left(\psi_{o}-\bar{\psi}\right),
    \label{eq:convective_power_definition}
\end{equation}
where $F_i$ and $F_o$ are the mass anomaly flow rates at the inner and outer boundary of the layer, $\psi_i$ and $\psi_o$ are the gravitational potential at the corresponding boundaries and $\bar{\psi}$ is the mass-averaged gravitational potential in the layer.
Eq. \ref{eq:convective_power_definition} shows that the convective power is produced by taking the mass anomaly flow rate at the gravitational potential of the inner and outer boundaries, and redistributing it in the layer at the mean gravitational potential, which implies in particular that the layer is well-mixed.

\sloppy

In practice, we obtain the convective power from the dissipation term of the entropy budget with
\begin{equation}
    \Phi=E_\Phi T_D=
    \left\{
    \begin{array}{l}
        E_\Phi\left(\frac{1}{M_{b}}\int_{V_{b}}\frac{\rho}{T}\mathrm{d}V\right)^{-1} \\
        E_\Phi\left(\frac{1}{M_{oc}}\int_{V_{oc}}\frac{\rho}{T}\mathrm{d}V\right)^{-1}
    \end{array}
    \right.,
    \label{eq:convective_power_calculation}
\end{equation}
where $T_D$ is an effective dissipation temperature calculated using the mass-averaged adiabatic temperature profile \citep{labrosse_2003}.
Following \citet{aubert_2009}, the convective power is made dimensionless with
\begin{equation}
    \tilde{\Phi}=
    \left\{
    \begin{array}{l}
        \frac{\Phi}{M_{b}\Omega^3L_{b}^2} \\
        \frac{\Phi}{M_{oc}\Omega^3L_{c}^2}
    \end{array}
    \right..
    \label{eq:dimensionless_convective_power}
\end{equation}

\fussy

\begin{figure*}[!ht]
    \centering
    \includegraphics[width=\linewidth]{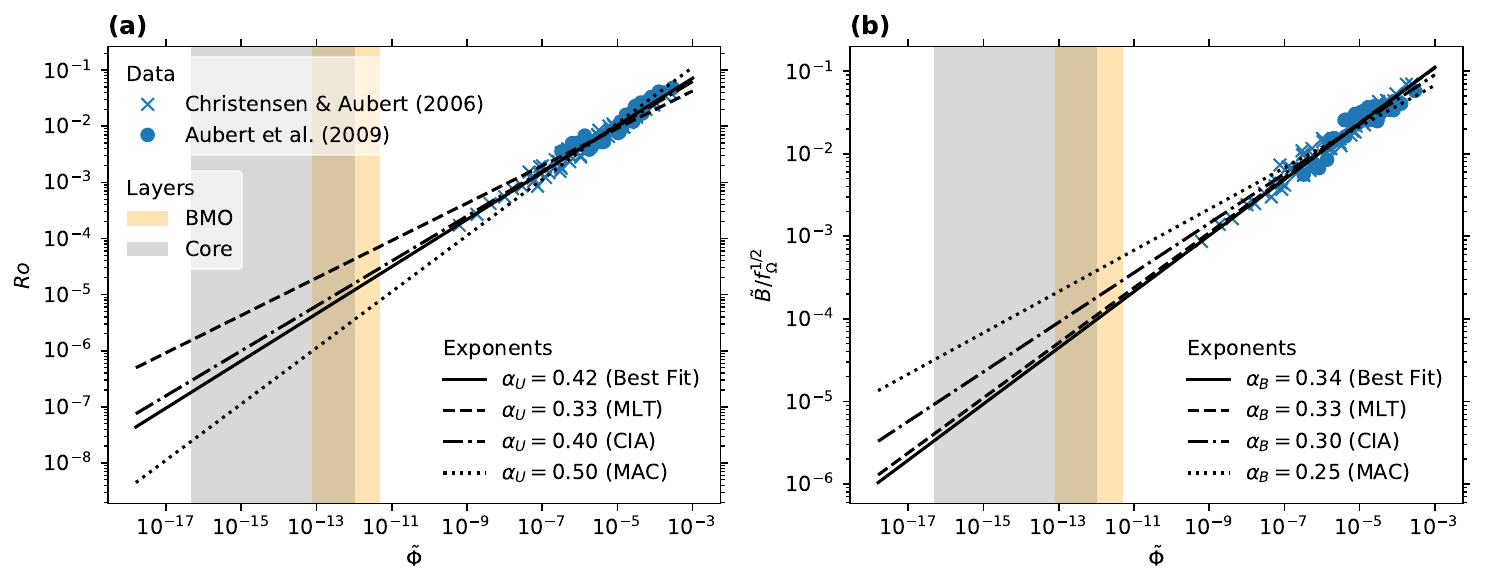}
    \caption{(a) Scaling of the Rossby number $Ro$ as a function of the dimensionless convective power $\tilde{\Phi}$.
        (b) Scaling of the dimensionless internal magnetic field intensity, normalized by $f_\Omega^{1/2}$, where $f_\Omega$ is the ohmic dissipation fraction of the convective power, as a function of the dimensionless convective power $\tilde{\Phi}$.
        The blue markers correspond to the set of MHD simulations extracted from \citet{aubert_2009}, but respectively computed by \citet{christensen_2006} (crosses) and \citet{aubert_2009} (disks).
        Solid lines correspond to the best-fit power-law scaling obtained by \citet{aubert_2009} using a least-square fit in the set of MHD simulations.
        Dashed, dash-dotted, and dotted lines correspond to power-law scalings, obtained using a least-square fit on the set of MHD simulations, with an imposed exponent derived from the mixing length theory (MLT), the Coriolis-Inertia-Archimedean (CIA), and the Magnetic-Archimedean-Coriolis (MAC) regime, respectively \citep{christensen_2010}.
        The orange and gray shaded areas correspond respectively to the dimensionless convective power derived from the entropy budget in the BMO and the core, with $R_m>40$.}
    \label{fig:scaling_laws}.
\end{figure*}

Fig. \ref{fig:scaling_laws} shows the Rossby number and the dimensionless internal magnetic field intensity as a function of the dimensionless convective power. Scaling laws, fitted on a set of numerical dynamo simulations \citep{aubert_2009}, are compared with scaling laws based on force balance arguments \citep{christensen_2010}, \textit{i.e.} the MLT, CIA and MAC regimes. The scaling law parameters are defined in Table S8.
In both the BMO and the core, the range of dimensionless convective power predicted by the entropy budget (shaded areas) is several orders of magnitude smaller than the convective power obtained in the dynamo simulations (blue markers). This implies in particular that even a moderate variation of the scaling law exponent may be responsible for significant changes in the convective velocities and magnetic field intensities. This explains the magnetic evolution variability observed in the nominal model, depending on the exponent of the scaling law (Fig. \ref{fig:magnetic_nominal}).

\sloppy

\printcredits

\fussy

\bibliographystyle{cas-model2-names}

\bibliography{biblio}











\end{document}


\begin{titlepage}

\begin{center}
    
{\LARGE\noindent Supplementary Material for}

\vspace*{1cm}

{\bfseries\Large\noindent "Thermal and magnetic evolution of an Earth-like planet with a basal magma ocean"}

\vspace*{2cm}

{\Large\noindent Victor Lherm$^{a,b,*}$, Miki Nakajima$^{a,b}$, Eric G. Blackman$^{b,c}$}

\vspace{2cm}

{\footnotesize\itshape\noindent
\begin{tabular}{c}
    $^a$Department of Earth and Environmental Sciences, University of Rochester, 227 Hutchison Hall, Rochester, NY 14627, USA\\
    $^b$Department of Physics and Astronomy, University of Rochester, 206 Bausch \& Lomb Hall, Rochester, NY 14627, USA\\
    $^c$Laboratory for Laser Energetics, University of Rochester, 250 E River Rd, Rochester, NY 14623, USA
\end{tabular}
}  

\end{center}

\vfill

{\small\noindent $^*$Corresponding author: V. Lherm (vlherm@ur.rochester.edu)}

\end{titlepage}

\setcounter{page}{2}

\begin{table}[!ht]
    \caption{General parameters of the model. References: L07=\citet{labrosse_2007}; N11=\citet{nomura_2011}; A12=\citet{andrault_2012}; D14=\citet{driscoll_2014}; L15=\citet{labrosse_2015}; C19=\citet{caracas_2019}.}
    \centering
    \begin{tabular*}{\linewidth}{@{\extracolsep{\stretch{1}}}*{4}{l}@{}}
        \toprule
        Symbol & Value & Units & Definition \\
        \midrule
        \multicolumn{4}{@{}l@{}}{Planet parameters} \\
        $M_P$ & $5.972 \times 10^{24}$ & $\mathrm{kg}$ & Planet mass \\
        $\mathrm{CMF}$ & 0.32 & - & Core mass fraction \\
        $t_P$ & 4.5 & $\mathrm{Gyr}$ & Age of the planet \\
        $T_{CMB_0}$ & 5710 & $\mathrm{K}$ & Initial CMB temperature \\
        \multicolumn{4}{@{}l@{}}{Composition parameters} \\
        $\Delta_\xi\rho_{BMO}$ & 250 & $\mathrm{kg.m^{-3}}$ & Compositional density jump of the BMO (C19) \\
        $\Delta_\xi\rho_{ICB}$ & 580 & $\mathrm{kg.m^{-3}}$ & Compositional density jump of the core (L15) \\
        $\xi_{b_0}$ & 16 & $\mathrm{wt\%}$ & Initial $\mathrm{FeO}$ mass fraction in the BMO (C19) \\
        $\xi_{c_0}$ & 5.6 & $\mathrm{wt\%}$ & Initial light elements mass fraction in the core (L15) \\
        $D_{BMO}$ & 0.85 & - & $\mathrm{FeO}$ partition coefficient (N11 \& A12) \\
        \multicolumn{4}{@{}l@{}}{Mantle convection parameters} \\
        $\beta$ & $1/3$ & - & Exponent of the $Nu \propto Ra^\beta$ scaling law (D14) \\
        $Ra_c$ & 660 & - & Critical Rayleigh number (D14)\\
        \multicolumn{4}{@{}l@{}}{Entropy of fusion} \\
        $\Delta S_m$ & 300 & $\mathrm{J.kg^{-1}.K^{-1}}$ & Entropy of fusion of the mantle (L07) \\
        $\Delta S_c$ & 127 & $\mathrm{J.kg^{-1}.K^{-1}}$ & Entropy of fusion of the core (L15) \\
        \multicolumn{4}{@{}l@{}}{Universal constants} \\
        $G$ & $6.674 \times 10^{-11}$ & $\mathrm{m^3.kg^{-1}.s^{-2}}$ & Gravitational constant \\
        $k_B$ & $1.381 \times 10^{-23}$ & $\mathrm{J.K^{-1}}$ & Boltzmann constant \\
        $Lo$ & $2.443 \times 10^{-8}$ & $\mathrm{V^2.K^{-2}}$ & Theoretical Lorenz number \\
        $\mu_0$ & $1.257 \times 10^{-6}$ & $\mathrm{N.A^{-2}}$ & Magnetic constant \\
        $N_A$ & $6.022 \times 10^{23}$ & $\mathrm{mol^{-1}}$ & Avogadro constant \\
        $R$ & $8.314$ & $\mathrm{J.K^{-1}.mol^{-1}}$ & Gas constant \\
        \bottomrule
    \end{tabular*}
    \label{tab:parameters_main}
\end{table}

\begin{table}[!ht]
    \caption{Thermodynamic parameters of peridotite (Pd), perovskite (Pv), post-perovskite (pPv), liquid iron ($\mathrm{Fe_l}$) and solid iron ($\mathrm{Fe_s}$) in the reference model \citep{boujibar_2020}. Data are derived from \citet{stixrude_2005} for Pd, from \citet{dorfman_2013} and \citet{lundin_2008} for Pv, from \citet{sakai_2016} for pPv, from \citet{smith_2018} for $\mathrm{Fe_l}$ and from \citet{wicks_2018} for $\mathrm{Fe_s}$.}
    \centering
    \begin{tabular*}{\linewidth}{@{\extracolsep{\stretch{1}}}*{6}{l}@{}}
        \toprule
        & Pd & Pv & pPv & $\mathrm{Fe_l}$ & $\mathrm{Fe_s}$ \\
        \midrule
        $M~(\mathrm{g.mol^{-1}})$ & 100.39 & 100.39 & 100.39 & 55.845 & 55.845 \\
        $\rho_0~(\mathrm{kg.m^{-3}})$ & 3226 & 4109 & 4260 & 7700 & 8160 \\
        $\hat{V}_0~(\mathrm{cm^{3}.mol^{-1}})$ & 31.12 & 24.43 & 23.57 & 7.25 & 6.84 \\
        $K_0~(\mathrm{GPa})$ & 128 & 261 & 324 & 125 & 165 \\
        $K_0'$ & 4.2 & 4.0 & 3.3 & 4.5 & 4.9 \\
        $\gamma_0$ & 0.99 & 1.0 & 1.48 & 1.6 & 1.6 \\
        $\gamma_1$ & 2.1 & 1.4 & 1.4 & 0.92 & 0.92 \\
        $\alpha_0~(10^{-5}~\mathrm{K^{-1}})$ & 2 & 2 & 2 & 4 & 4 \\
        \bottomrule
    \end{tabular*}
    \label{tab:parameters_reference}
\end{table}

\begin{table}[!ht]
    \caption{Parameters used in the definition of the solid phase transitions of the reference model \citep{boujibar_2020}. Data are derived from \citet{ito_1989} and \citet{hernlund_2007}.}
    \centering
    \begin{tabular*}{0.5\linewidth}{@{\extracolsep{\stretch{1}}}*{3}{l}@{}}
        \toprule
        & Pd - Pv & Pv - pPv \\
        \midrule
        $P_c~(\mathrm{GPa})$ & 28.3 & 124 \\
        $T_c~(\mathrm{K})$ & 0 & 2500 \\
        $\gamma_c~(\mathrm{MPa.K^{-1}})$ & -2.8 & 8 \\
        \bottomrule
    \end{tabular*}
    \label{tab:parameters_transitions}
\end{table}

\begin{table}[!ht]
    \caption{Parameters used in the definition of the liquidus. Data are derived from \citet{fei_2021} for $\mathrm{MgSiO_3}$ (lower estimate), \citet{morard_2022} for $\mathrm{FeO}$ and \citet{stixrude_2014} for $\mathrm{Fe}$.}
    \centering
    \begin{tabular*}{0.7\linewidth}{@{\extracolsep{\stretch{1}}}*{4}{l}@{}}
        \toprule
        & $\mathrm{MgSiO_3}$ & $\mathrm{FeO}$ & $\mathrm{Fe}$\\
        \midrule
        $T_{l_0}~(\mathrm{K})$ & 6000 & 1650 & 6500 \\
        $P_{l_0}~(\mathrm{GPa})$ & 140 & 0 & 340 \\
        $a_{l}~(\mathrm{GPa})$ & 140 & 14.89 & 340 \\
        $b_{l}$ & 0.26 & 0.36 & 0.515 \\
        \bottomrule
    \end{tabular*}
    \label{tab:liquidus}
\end{table}

\begin{table}[!ht]
    \caption{Thermodynamic parameters of the BMO end-members. Data are derived from \citet{fratanduono_2018} and \citet{wolf_2018} for $\mathrm{MgSiO_3}$ and from \citet{morard_2022} for $\mathrm{FeO}$.}
    \centering
    \begin{tabular*}{0.7\linewidth}{@{\extracolsep{\stretch{1}}}*{3}{l}@{}}
        \toprule
        & $\mathrm{MgSiO_3}$ & $\mathrm{FeO}$ \\
        \midrule
        $n$ & 5 & 2 \\
        $M~(\mathrm{g.mol^{-1}})$ & 100.39 & 71.84 \\
        $T_0~(\mathrm{K})$ & 3000 & 1650 \\
        $\hat{V}_0~(\mathrm{cm^{3}.mol^{-1}})$ & 38.99 & 15.97 \\
        $\rho_0~(\mathrm{kg.m^{-3}})$ & 2574.7 & 4498.7 \\
        $\hat{U}_0~(\mathrm{kJ.mol^{-1}})$ & -9935.73 & -442.22 \\
        $K_0~(\mathrm{GPa})$ & 13.20 & 40.72 \\
        $K_0'$ & 8.238 & 5.33 \\
        $\gamma_0$ & 0.3675 & 2.499 \\
        $\gamma_1$ & 1.0 & -0.021 \\
        $\gamma_\infty$ & 0.5 & 0 \\
        $\gamma_{01}$ & 0.65 & - \\
        $\rho_{0_\gamma}~(\mathrm{kg.m^{-3}})$ & 2743.4 & - \\
        $\rho_{e}~(\mathrm{kg.m^{-3}})$ & 5195 & - \\
        $\rho_\sigma~(\mathrm{kg.m^{-3}})$ & $1707 \times 10^{3}$ & - \\
        $m$ & 0.6 & - \\
        $b_0~(\mathrm{kJ.mol^{-1}})$ & 473.84 & - \\
        $b_1~(\mathrm{kJ.mol^{-1}})$ & 296.98 & - \\
        $b_2~(\mathrm{kJ.mol^{-1}})$ & 633.28 & - \\
        $b_3~(\mathrm{kJ.mol^{-1}})$ & -1455.48 & - \\
        $b_4~(\mathrm{kJ.mol^{-1}})$ & -1991.94 & - \\
        $T_{e_0}~(\mathrm{K})$ & 2466.6 & - \\
        $T_{e_1}$ & -0.4578 & - \\
        $\zeta_0~(\mathrm{mJ.mol^{-1}.K^{-2}})$ & 1.88849 & - \\
        $\zeta_1$ & 0.67774 & - \\
        $\Theta_0~(\mathrm{K})$ & - & 1247.8 \\
        $a_S$ & - & 18.81 \\
        $e_0~(\mathrm{10^{-5}~K^{-1}})$ & - & -4.972 \\
        $e_1$ & - & 4.96 \\
        \bottomrule
    \end{tabular*}
    \label{tab:parameters_BMO}
\end{table}

\begin{table}[!ht]
    \caption{Transport properties parameters. Data are derived from \citet{stixrude_2020} for the nominal conductivities of the BMO, \citet{driscoll_2023} for the nominal conductivities of the core, \citet{deng_2021} for the thermal conductivity of the BMO, \citet{vandenberg_2010} for the thermal conductivity of the mantle, \citet{driscoll_2014} for the mantle viscosity and \citet{gubbins_2004} for the material constant involved in the entropy of molecular diffusion.}
    \centering
    \begin{tabular*}{0.7\linewidth}{@{\extracolsep{\stretch{1}}}*{3}{l}@{}}
        \toprule
        Symbol & Value & Units \\
        \midrule
        \multicolumn{3}{l}{Nominal conductivities of the BMO} \\
        $k_b$ & 8 & $\mathrm{W.m^{-1}.K^{-1}}$ \\
        $\sigma_b$ & $3 \times 10^4$ & $\mathrm{S.m^{-1}}$ \\
        \multicolumn{3}{l}{Nominal conductivities of the core} \\
        $k_c$ & 70 & $\mathrm{W.m^{-1}.K^{-1}}$ \\
        $\sigma_c$ & $1 \times 10^6$ & $\mathrm{S.m^{-1}}$ \\
        \multicolumn{3}{l}{Thermal conductivity model of the BMO}\\
        $k_{b,l_0}$ & 1.16 & $\mathrm{W.m^{-1}.K^{-1}}$ \\
        $T_k$ & 3000 & $\mathrm{K}$ \\
        $a_{k,b}$ & -0.09 & - \\
        $\rho_k$ & 2580.7 & $\mathrm{kg.m^{-3}}$ \\
        $b_{k,b}$ & 1.75 & - \\
        \multicolumn{3}{l}{Thermal conductivity model of the mantle}\\
        $k_{l_0}$ & 4.7 & $\mathrm{W.m^{-1}.K^{-1}}$ \\
        $a_k$ & 0.3 & - \\
        $k_\mathrm{trunc}$ & 15 & $\mathrm{W.m^{-1}.K^{-1}}$ \\
        $k_{r_0}$ & 2 & $\mathrm{W.m^{-1}.K^{-1}}$ \\
        $T_{r_1},T_{r_2},T_{r_3},T_{r_4}$ & $573,4000,4500,7000$ & $\mathrm{K}$ \\
        $k_{e_0}$ & 487.54 & $\mathrm{W.m^{-1}.K^{-1}}$ \\
        $E_k$ & 3 & $\mathrm{eV}$ \\
        \multicolumn{3}{l}{Viscosity model of the mantle}\\
        $\mu_{m_0}$ & $1.1 \times 10^{15}$ & $\mathrm{Pa.s}$ \\
        $A_\mu$ & $3 \times 10^5$ & $\mathrm{J.mol^{-1}}$ \\
        $f_\mu$ & 10 & - \\
        \multicolumn{3}{l}{Molecular diffusion}\\
        $\alpha_{D}$ & $1.0 \times 10^{-12}$ & $\mathrm{kg.m^{-3}.s}$ \\
        \bottomrule
    \end{tabular*}
    \label{tab:parameters_transport_properties}
\end{table}

\begin{table}[!ht]
    \caption{Radiogenic heating parameters. Concentrations are derived from \citet{jaupart_2015} for the mantle and from \citet{hirose_2013} for the core. The radioisotope properties are derived from \citet{dye_2012} and \citet{jaupart_2015}. The mantle fraction of the BSE radiogenic heating is $f_h=0.58$ \citep{jaupart_2015}.}
    \centering
    \begin{tabular*}{\linewidth}{@{\extracolsep{\stretch{1}}}*{5}{l}@{}}
        \toprule
        & $\mathrm{^{238}U}$ & $\mathrm{^{235}U}$ & $\mathrm{^{232}Th}$ & $\mathrm{^{40}K}$ \\
        \midrule
        $c_m~(\mathrm{ppm})$ & 0.02 & 0.02 & 0.08 & 240 \\
        $c_c~(\mathrm{ppm})$ & 0 & 0 & 0 & 20 \\
        $a$ & 0.9927 & 0.0072 & 0.9998 & 0.000117 \\
        $e~(\mathrm{pJ})$ & 7.65 & 7.11 & 6.48 & 0.11 \\
        $\lambda~(\mathrm{s^{-1}})$ & $4.92 \times 10^{-18}$ & $3.12 \times 10^{-17}$ & $1.57 \times 10^{-18}$ & $1.74 \times 10^{-17}$ \\
        $M~(\mathrm{g.mol^{-1}})$ & 238.05 & 235.04 & 232.04 & 39.96 \\ 
        \bottomrule
    \end{tabular*}
    \label{tab:parameters_radio}
\end{table}

\begin{table}[!ht]
    \caption{Convective velocity and magnetic field intensity scaling laws. Data are derived from \citet{christensen_2006} and \citet{aubert_2009}.}
    \centering
    \begin{tabular*}{\linewidth}{@{\extracolsep{\stretch{1}}}*{7}{l}@{}}
        \toprule
        & $c_U$ & $\alpha_U$ & $c_B$ & $\alpha_B$ & $c_{dip}$ & $f_\Omega$ \\
        \midrule
        Best-fit & 1.31 & 0.42 & 1.17 & 0.34 & 7.3 & 0.9 \\
        MLT & 0.43 & $1/3$ & 1.10 & $1/3$ & 7.3 & 0.9 \\
        CIA & 0.99 & $2/5$ & 0.72 & $3/10$ & 7.3 & 0.9 \\
        MAC & 3.55 & $1/2$ & 0.38 & $1/4$ & 7.3 & 0.9 \\
        \bottomrule
    \end{tabular*}
    \label{tab:scaling_law}
\end{table}

\begin{figure}[!ht]
    \centering
    \includegraphics[width=\linewidth]{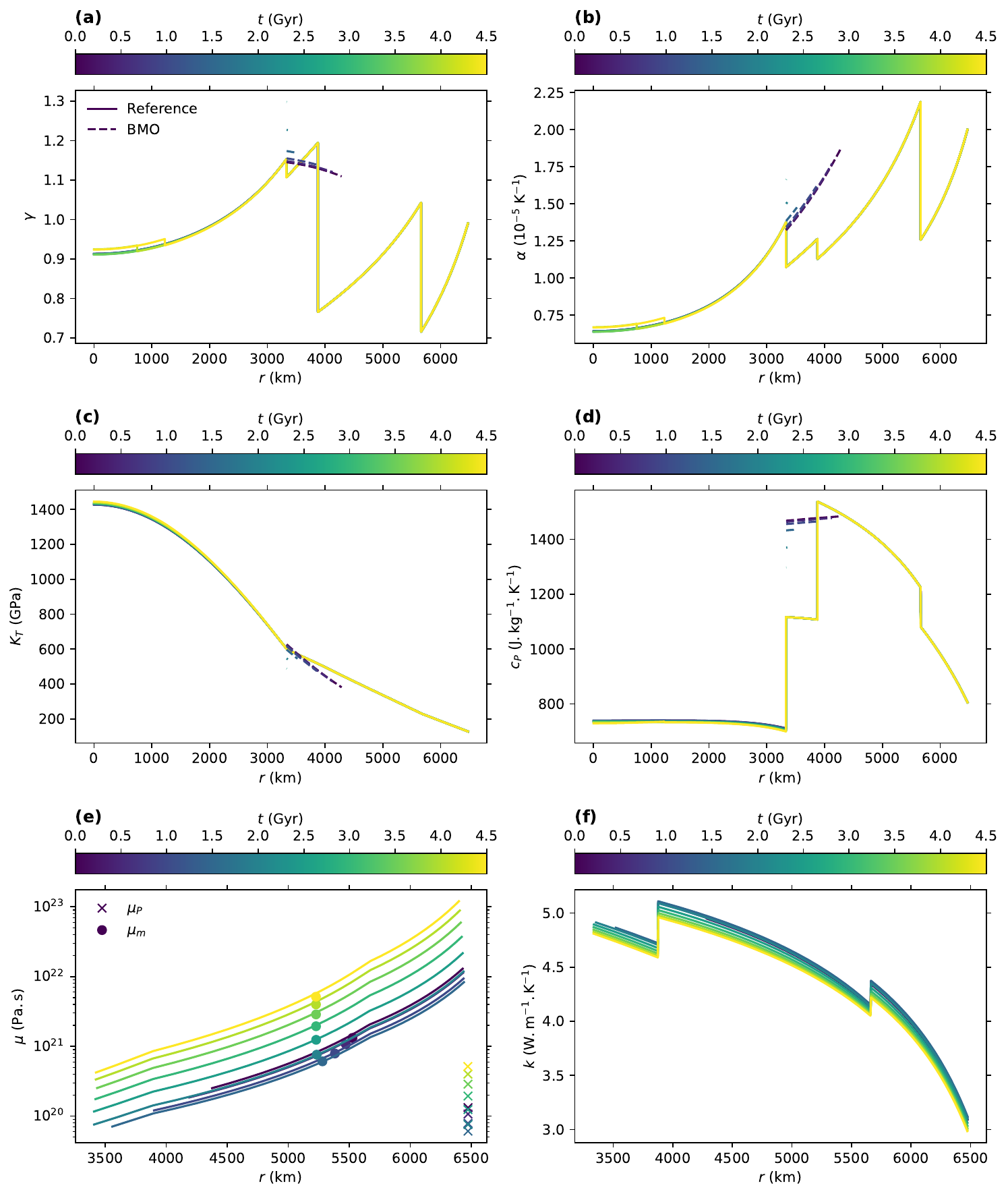}
    \caption{Grüneisen parameter $\gamma$ (a), thermal expansion coefficient $\alpha$ (b), isothermal bulk modulus $K_T$ (c), and specific heat capacity at constant pressure $c_P$ (d), as a function of the radius $r$. Solid lines correspond to profiles obtained using the reference model. Dashed lines correspond to profiles using the BMO model.
    (e): Mantle viscosity as a function of the radius $r$. Solid lines are calculated with Eq. 18, using the adiabatic temperature profile of the mantle outside of the thermal boundary layers. X markers show the mantle viscosity at the surface $\eta_P$ as a function of the radius of the planet. Circle markers show the average viscosity of the mantle $\eta_m$ (Eq. 18) as a function of its average radius.
    (f): Mantle thermal conductivity as a function of the radius $r$ (Appendix C).
    The colors correspond to the time $t$.}
    \label{fig:thermodynamic_nominal}
\end{figure}

\begin{figure}[!ht]
    \centering
    \includegraphics[width=\linewidth]{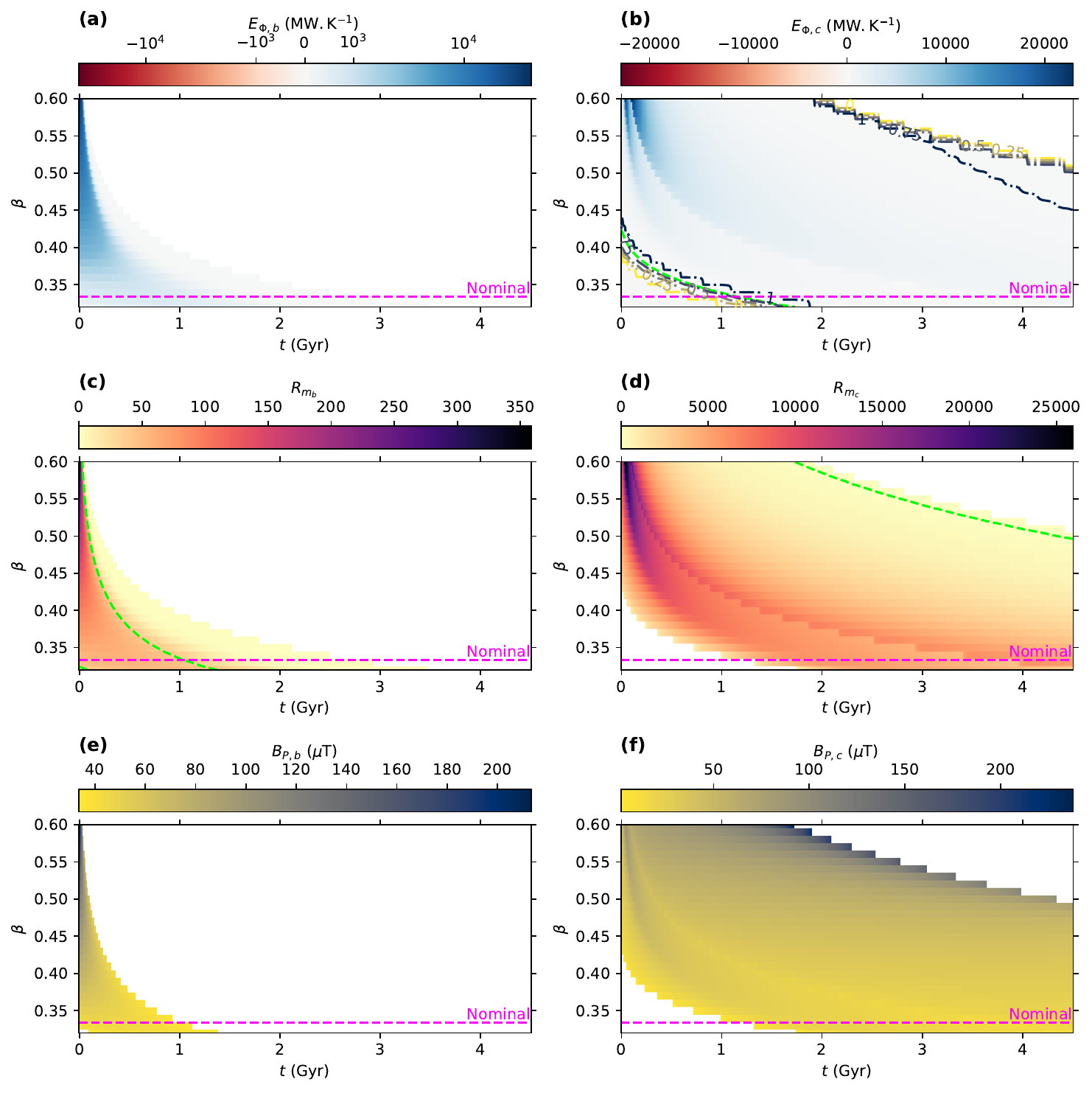}
    \caption{Sensitivity of the entropy of dissipation $E_\Phi$ (a,b), magnetic Reynolds number $R_m$ (c,d) and surface magnetic field intensity $B_P$ (e,f) of the BMO (a,c,e) and the core (b,d,f) to the exponent of the $Nu$-$Ra$ scaling in the mantle $\beta$, as a function of time $t$. Green dashed lines correspond to $E_\Phi=0$ (a,b) and $R_m=40$ (c,d). Dash-dotted lines indicate the time when a specific radius fraction of the core is convecting, as determined by the convective stability criterion. Pink dashed lines correspond to the value of the parameter used in the nominal model. \\
    We conduct the parameter sweep for $\beta \leq 0.6$, which covers uncertainties on the value of $\beta$ \citep{jaupart_2010}. The minimum value of the sweep, \textit{i.e.} $\beta=0.32$, corresponds to the value below which the thickness of the BMO, and the temperature at the CMB, would initially increase. Such a small $\beta$ would produce a thick upper boundary layer, leading to an insufficient release of energy to obtain a BMO decreasing in size, and reduce the CMB temperature.
    An increasing $\beta$ has an influence on mantle convection comparable to that of a decreasing reference viscosity of the mantle $\mu_{m_0}$. As a result, $E_\Phi$, $R_m$, and $B_P$ vary with $\beta$ as they would do with $\mu_{m_0}$ (see §4.2).}
    \label{fig:beta_mantle_sweep}
\end{figure}

\begin{figure}[!ht]
    \centering
    \includegraphics[width=\linewidth]{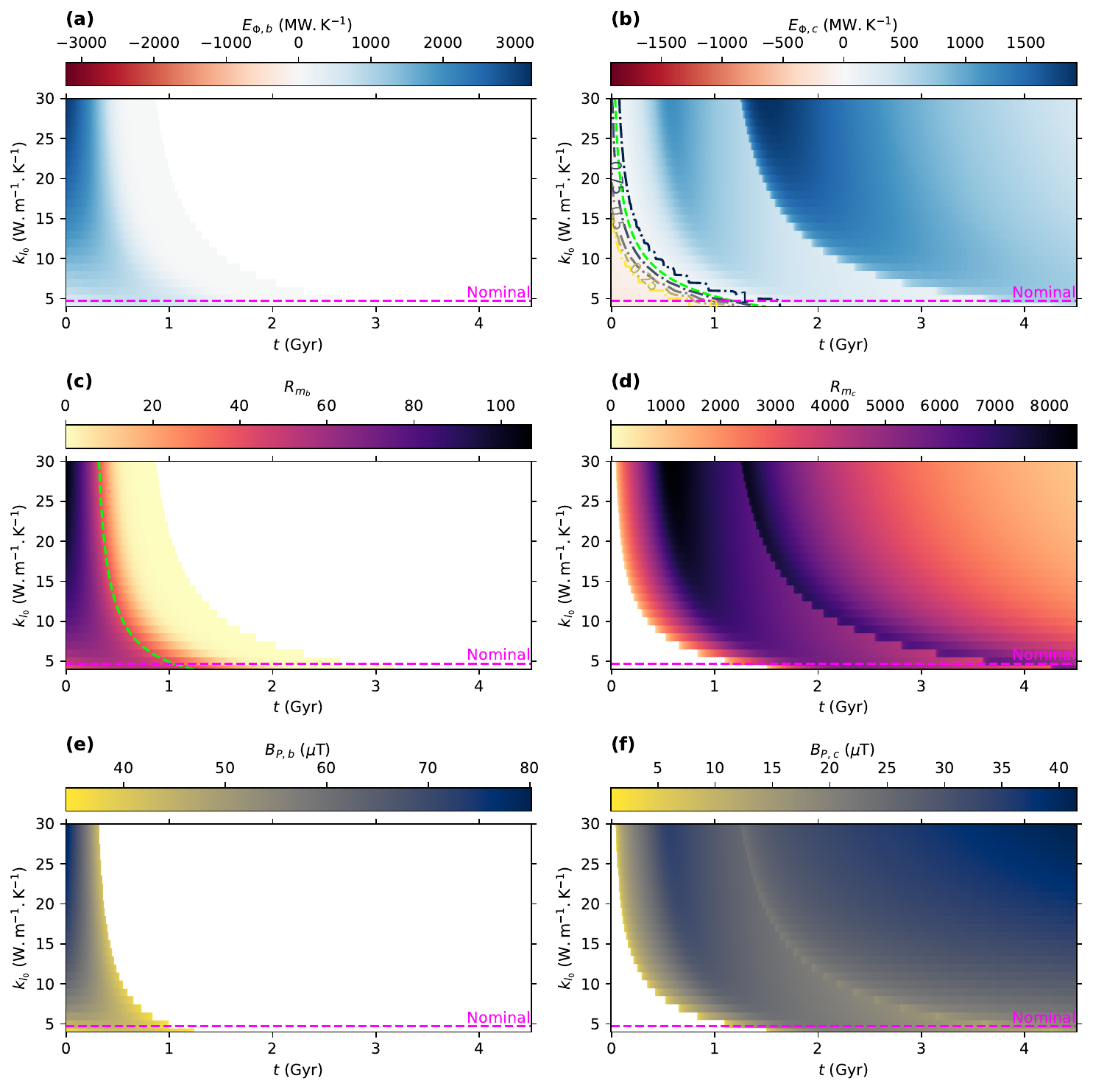}
    \caption{Sensitivity of the entropy of dissipation $E_\Phi$ (a,b), magnetic Reynolds number $R_m$ (c,d) and surface magnetic field intensity $B_P$ (e,f) of the BMO (a,c,e) and the core (b,d,f) to the phonon prefactor of the mantle thermal conductivity $k_{l_0}$ (Appendix C), as a function of time $t$. Green dashed lines correspond to $E_\Phi=0$ (a,b) and $R_m=40$ (c,d). Dash-dotted lines indicate the time when a specific radius fraction of the core is convecting, as determined by the convective stability criterion. Pink dashed lines correspond to the value of the parameter used in the nominal model. \\
    We conduct the parameter sweep for $k_{l_0} \leq 30~\mathrm{W.m^{-1}.K^{-1}}$, which corresponds to an average mantle thermal conductivity of $13.2~\mathrm{W.m^{-1}.K^{-1}}$. This allows to cover uncertainties on the value of the thermal conductivity \citep{hirose_2015,hofmeister_2015}.
    The minimum value of the sweep, \textit{i.e.} $k_{l_0}=4~\mathrm{W.m^{-1}.K^{-1}}$, which is equivalent to an average mantle thermal conductivity of $3.9~\mathrm{W.m^{-1}.K^{-1}}$, corresponds to the value below which the thickness of the BMO, and the temperature at the CMB, would initially increase. Such a small $k_{l_0}$ would produce a low heat flow at the surface, leading to an insufficient release of energy to obtain a BMO decreasing in size, and reduce the CMB temperature.
    An increasing $k_{l_0}$ has an influence on mantle convection comparable to that of a decreasing reference viscosity of the mantle $\mu_{m_0}$. As a result, $E_\Phi$, $R_m$, and $B_P$ vary with $\beta$ as they would do with $\mu_{m_0}$ (see §4.2).}
    \label{fig:k_0_sweep}
\end{figure}

\begin{figure}[!ht]
    \centering
    \includegraphics[width=\linewidth]{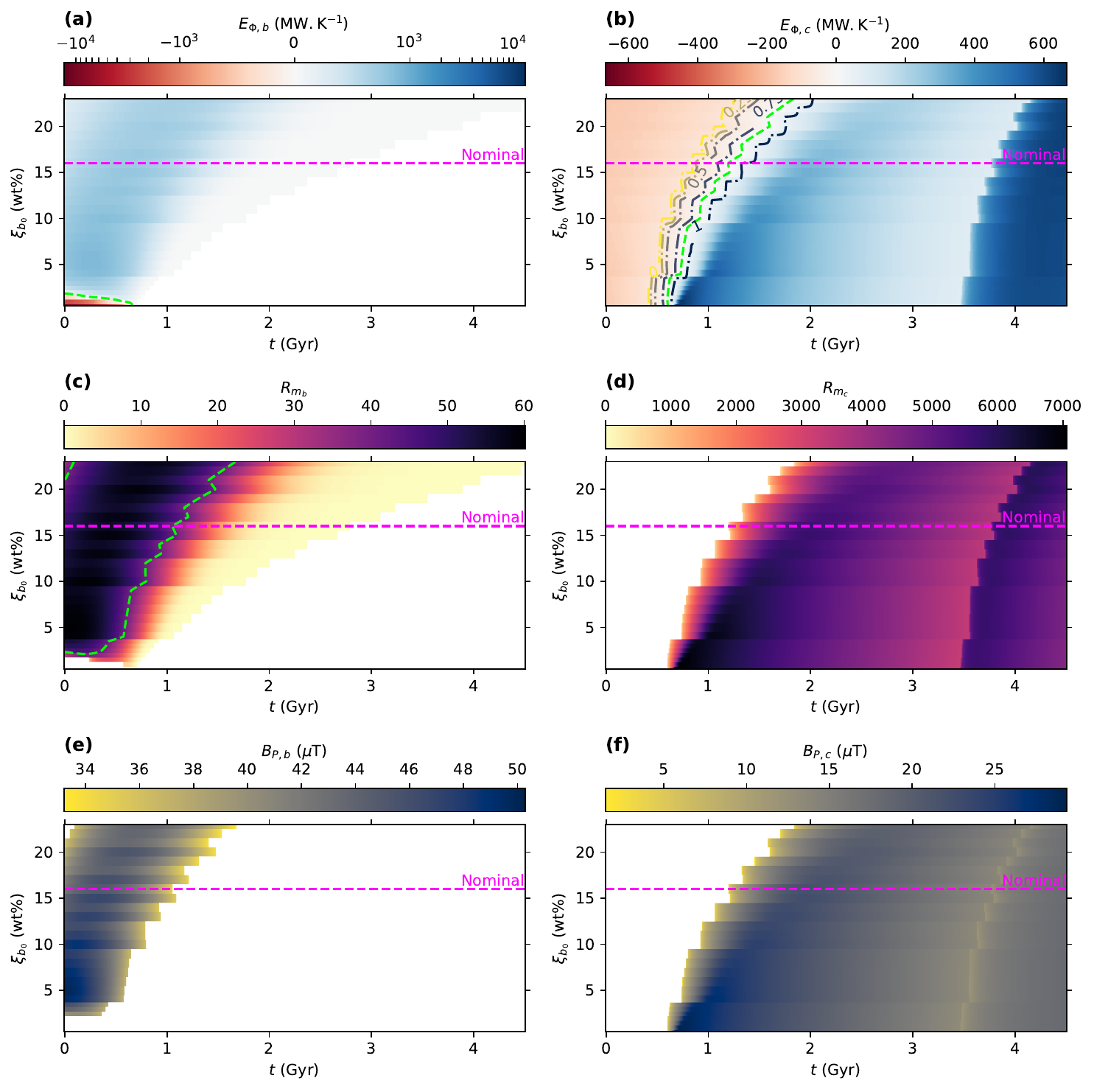}
    \caption{Sensitivity of the entropy of dissipation $E_\Phi$ (a,b), magnetic Reynolds number $R_m$ (c,d) and surface magnetic field intensity $B_P$ (e,f) of the BMO (a,c,e) and the core (b,d,f) to the initial mass fraction of $\mathrm{FeO}$ in the BMO $\xi_{b_0}$, as a function of time $t$. Green dashed lines correspond to $E_\Phi=0$ (a,b) and $R_m=40$ (c,d). Dash-dotted lines indicate the time when a specific radius fraction of the core is convecting, as determined by the convective stability criterion. Pink dashed lines correspond to the value of the parameter used in the nominal model. \\
    We conduct the parameter sweep for $\xi_{b_0} > 0$. The maximum value, \textit{i.e.} $\xi_{b_0}=23~\mathrm{wt\%}$, corresponds to the value above which the thickness of the BMO, and the temperature at the CMB, would initially increase, leading to a fully molten mantle instead of the intended BMO. Such a large $\xi_{b_0}$ would be responsible for a low mantle liquidus temperature, leading to a low temperature difference across the mantle, as well as a thick upper thermal boundary layer, leading to an insufficient release of energy to obtain a BMO decreasing in size, and reduce the CMB temperature.
    An increasing $\xi_{b_0}$ has an influence on composition comparable to that of a decreasing partition coefficient $D_{BMO}$. As a result, $E_\Phi$, $R_m$, and $B_P$ vary with $\xi_{b_0}$ as they would do with $D_{BMO}$ (see §4.3).}
    \label{fig:Xi_BMO_initial_sweep}
\end{figure}

\begin{figure}[!ht]
    \centering
    \includegraphics[width=\linewidth]{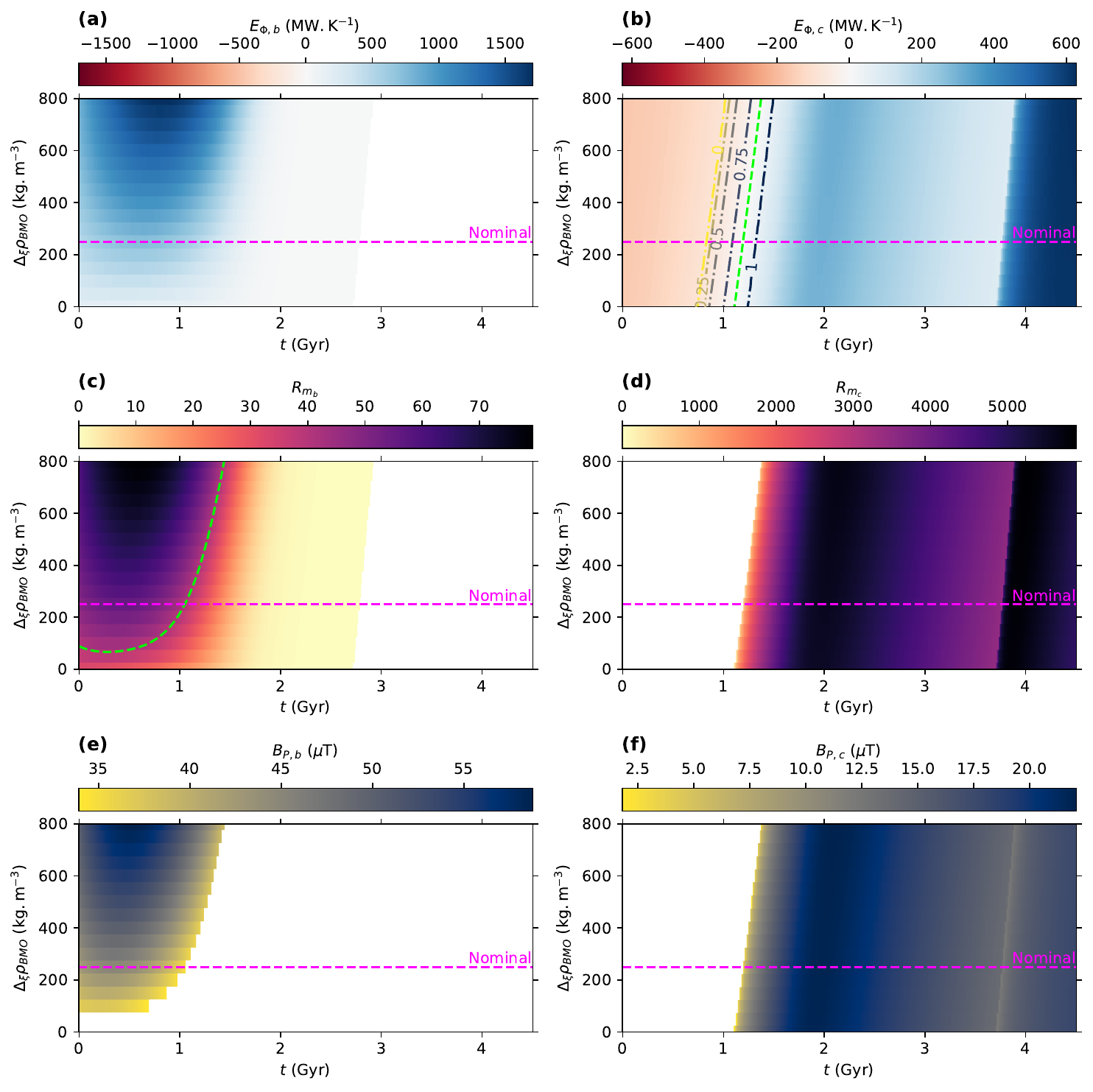}
    \caption{Sensitivity of the entropy of dissipation $E_\Phi$ (a,b), magnetic Reynolds number $R_m$ (c,d) and surface magnetic field intensity $B_P$ (e,f) of the BMO (a,c,e) and the core (b,d,f) to the compositional density jump of the BMO $\Delta_\xi\rho_{BMO}$, as a function of time $t$. Green dashed lines correspond to $E_\Phi=0$ (a,b) and $R_m=40$ (c,d). Dash-dotted lines indicate the time when a specific radius fraction of the core is convecting, as determined by the convective stability criterion. Pink dashed lines correspond to the value of the parameter used in the nominal model. \\
    We conduct the parameter sweep for $0 \leq \Delta_\xi\rho_{BMO} \leq 800~\mathrm{kg.m^{-3}}$, which covers uncertainties on the value of $\Delta_\xi\rho_{BMO}$ \citep{caracas_2019}. When $\Delta_\xi\rho_{BMO}$ increases, the BMO crystallizes later, dynamo operation is protracted, and $E_\Phi$, $R_m$, and $B_P$ increase. This may be explained by an increased gravitational energy in the BMO. In the core, the increase of $\Delta_\xi\rho_{BMO}$ promotes a lower heat flow at the CMB, explaining a delayed dynamo onset and inner core nucleation. When corrected for the time shift related to inner core nucleation, $E_\Phi$, $R_m$, and $B_P$ are relatively independent of $\Delta_\xi\rho_{BMO}$.
    }
    \label{fig:Drho_BMO_sweep}
\end{figure}

\begin{figure}[!ht]
    \centering
    \includegraphics[width=\linewidth]{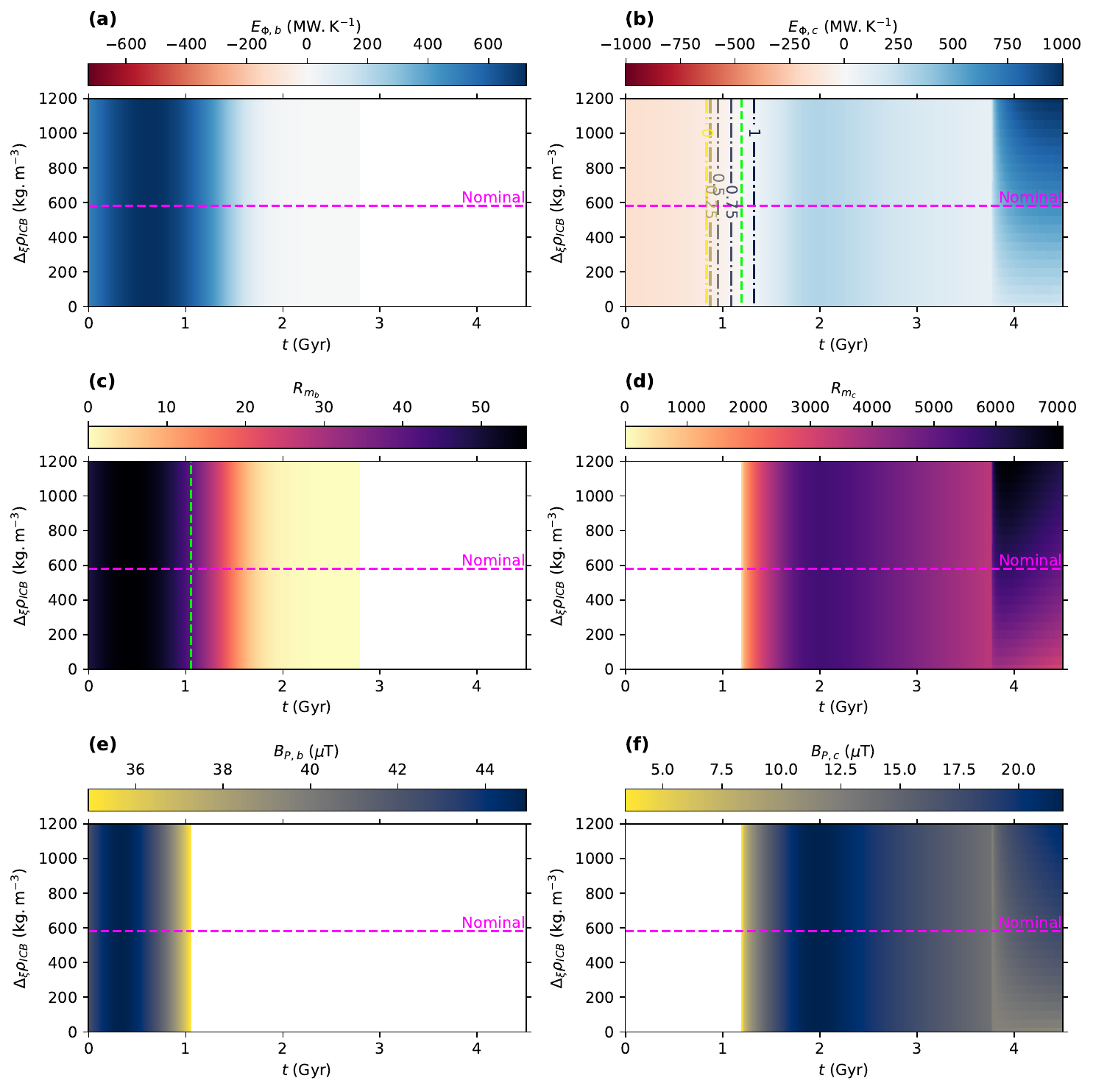}
    \caption{Sensitivity of the entropy of dissipation $E_\Phi$ (a,b), magnetic Reynolds number $R_m$ (c,d) and surface magnetic field intensity $B_P$ (e,f) of the BMO (a,c,e) and the core (b,d,f) to the compositional density jump of the core $\Delta_\xi\rho_{ICB}$, as a function of time $t$. Green dashed lines correspond to $E_\Phi=0$ (a,b) and $R_m=40$ (c,d). Dash-dotted lines indicate the time when a specific radius fraction of the core is convecting, as determined by the convective stability criterion. Pink dashed lines correspond to the value of the parameter used in the nominal model. \\
    We conduct the parameter sweep for $0 \leq \Delta_\xi\rho_{ICB} \leq 1200~\mathrm{kg.m^{-3}}$, which covers uncertainties on the value of $\Delta_\xi\rho_{ICB}$ \citep{alfe_1999,masters_2003}. The BMO crystallizes before inner core nucleation and is therefore independent of $\Delta_\xi\rho_{ICB}$. In the core, the increase of $\Delta_\xi\rho_{ICB}$ promotes a larger gravitational energy due to the release of light elements. This explains why $E_\Phi$, $R_m$, and $B_P$ increase with $\Delta_\xi\rho_{ICB}$ after inner core nucleation.}
    \label{fig:Drho_ICB_sweep}
\end{figure}

\begin{figure}[!ht]
    \centering
    \includegraphics[width=\linewidth]{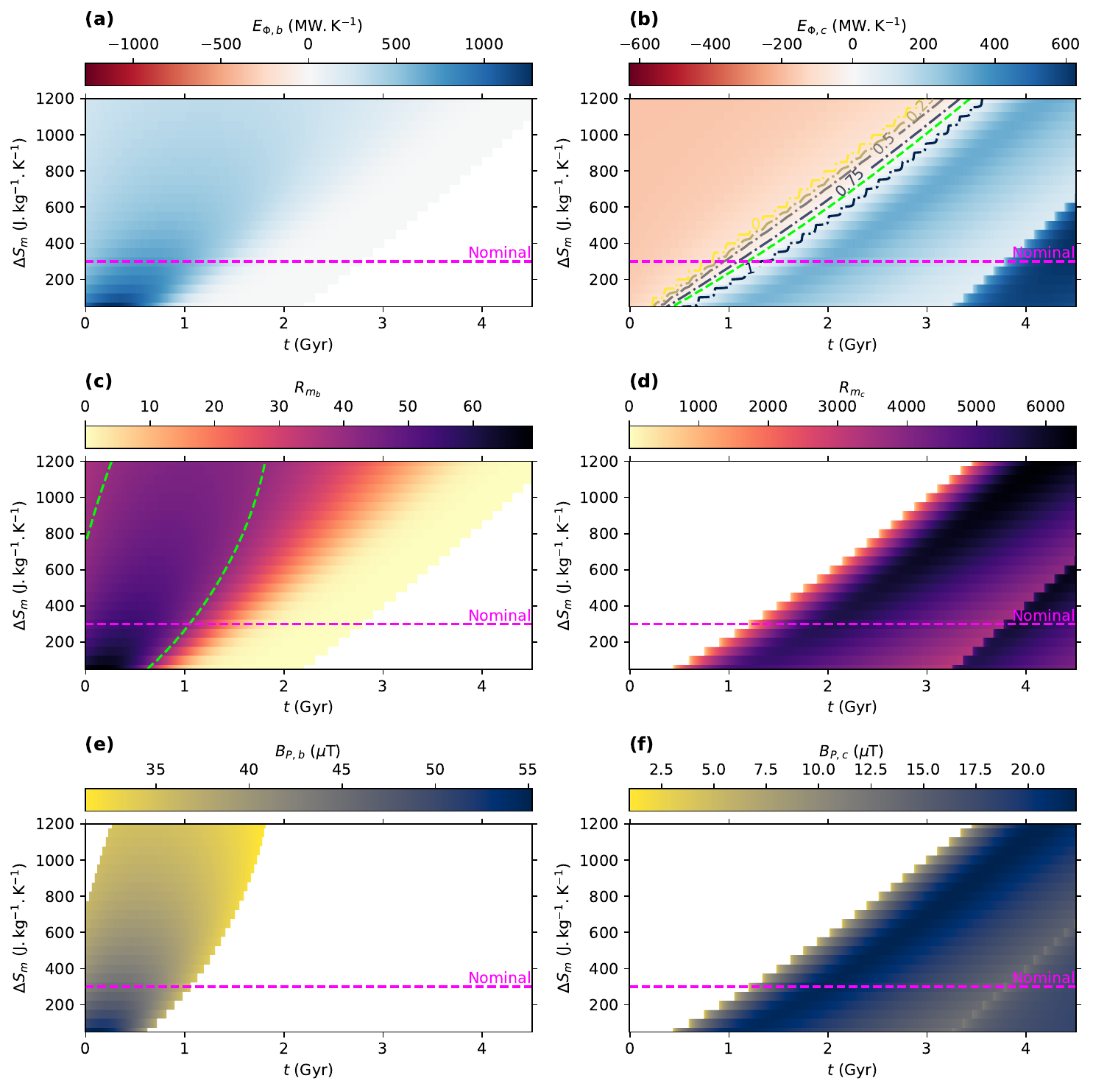}
    \caption{Sensitivity of the entropy of dissipation $E_\Phi$ (a,b), magnetic Reynolds number $R_m$ (c,d) and surface magnetic field intensity $B_P$ (e,f) of the BMO (a,c,e) and the core (b,d,f) to the entropy of fusion of the mantle $\Delta_{S_m}$, as a function of time $t$. Green dashed lines correspond to $E_\Phi=0$ (a,b) and $R_m=40$ (c,d). Dash-dotted lines indicate the time when a specific radius fraction of the core is convecting, as determined by the convective stability criterion. Pink dashed lines correspond to the value of the parameter used in the nominal model. \\
    We conduct the parameter sweep for $0 \leq \Delta_{S_m} \leq 1200~\mathrm{J.kg^{-1}.K^{-1}}$, which covers uncertainties on the value of $\Delta_{S_m}$ \citep[\textit{e.g.}][]{labrosse_2007,stixrude_2009,blanc_2020,orourke_2020}. When $\Delta_{S_m}$ increases, the BMO crystallizes later and dynamo operation is protracted, but $E_\Phi$, $R_m$, and $B_P$ decrease. This may be explained by an increased latent heat released at the top the BMO, leading to a reduction of the cooling and crystallization rates of the BMO. In the core, the increase of $\Delta_{S_m}$ promotes a lower heat flow at the CMB, explaining a delayed dynamo onset and inner core nucleation. When corrected for the time shift related to inner core nucleation, $E_\Phi$, $R_m$, and $B_P$ are relatively independent of $\Delta_{S_m}$.}
    \label{fig:DS_mantle_sweep}
\end{figure}

\begin{figure}[!ht]
    \centering
    \includegraphics[width=\linewidth]{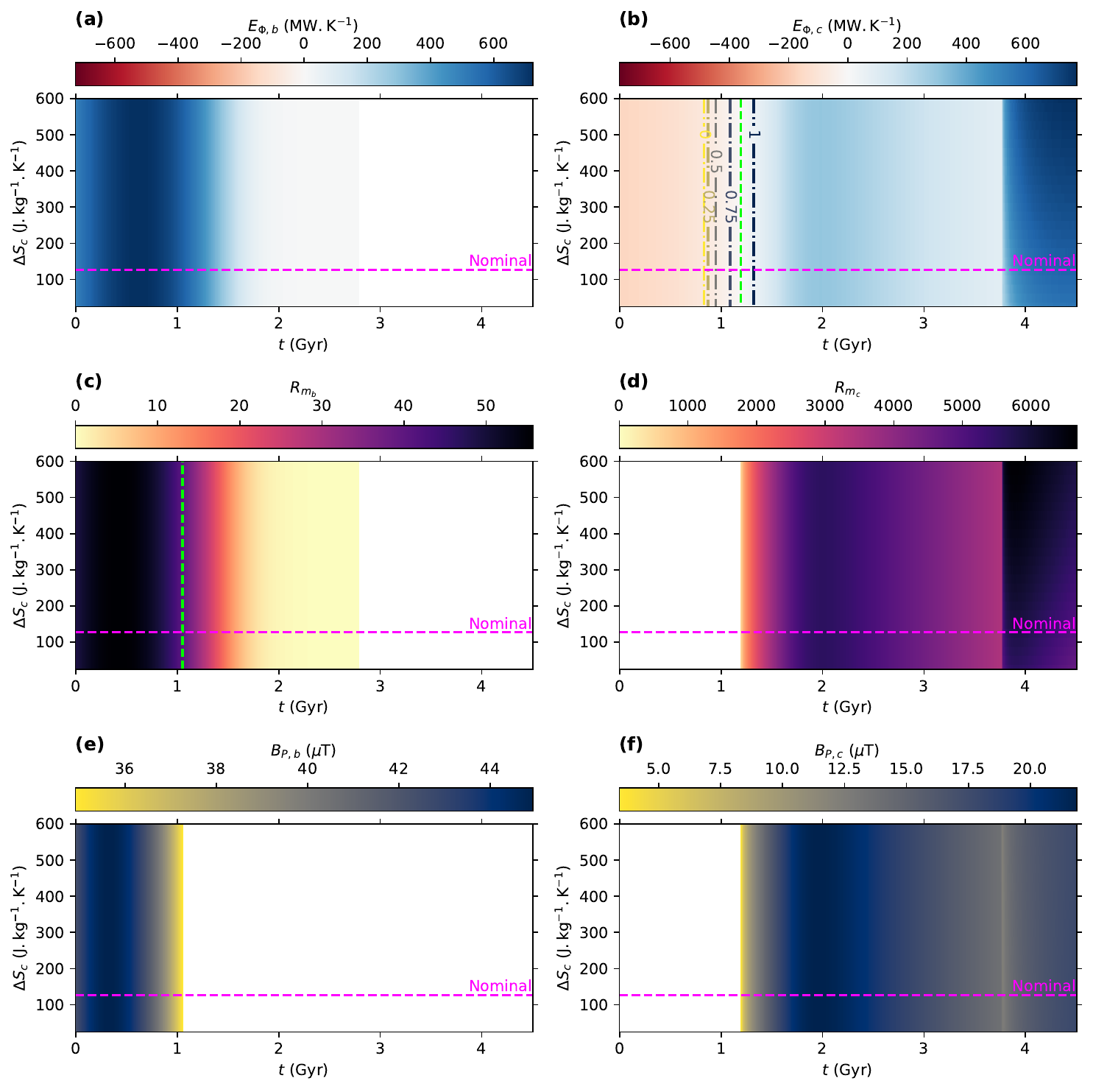}
    \caption{Sensitivity of the entropy of dissipation $E_\Phi$ (a,b), magnetic Reynolds number $R_m$ (c,d) and surface magnetic field intensity $B_P$ (e,f) of the BMO (a,c,e) and the core (b,d,f) to the entropy of fusion of the core $\Delta_{S_c}$, as a function of time $t$. Green dashed lines correspond to $E_\Phi=0$ (a,b) and $R_m=40$ (c,d). Dash-dotted lines indicate the time when a specific radius fraction of the core is convecting, as determined by the convective stability criterion. Pink dashed lines correspond to the value of the parameter used in the nominal model. \\
    We conduct the parameter sweep for $0 \leq \Delta_{S_c} \leq 600~\mathrm{J.kg^{-1}.K^{-1}}$, which covers uncertainties on the value of $\Delta_{S_c}$ \citep[\textit{e.g.}]{hirose_2013}. The BMO crystallizes before inner core nucleation and is therefore independent of $\Delta_{S_c}$. In the core, the increase of $\Delta_{S_c}$ promotes a higher release of latent heat. This explains why $E_\Phi$, $R_m$, and $B_P$ increase with $\Delta_{S_c}$ after inner core nucleation.}
    \label{fig:DS_core_sweep}
\end{figure}

\begin{figure}[!ht]
    \centering
    \includegraphics[width=\linewidth]{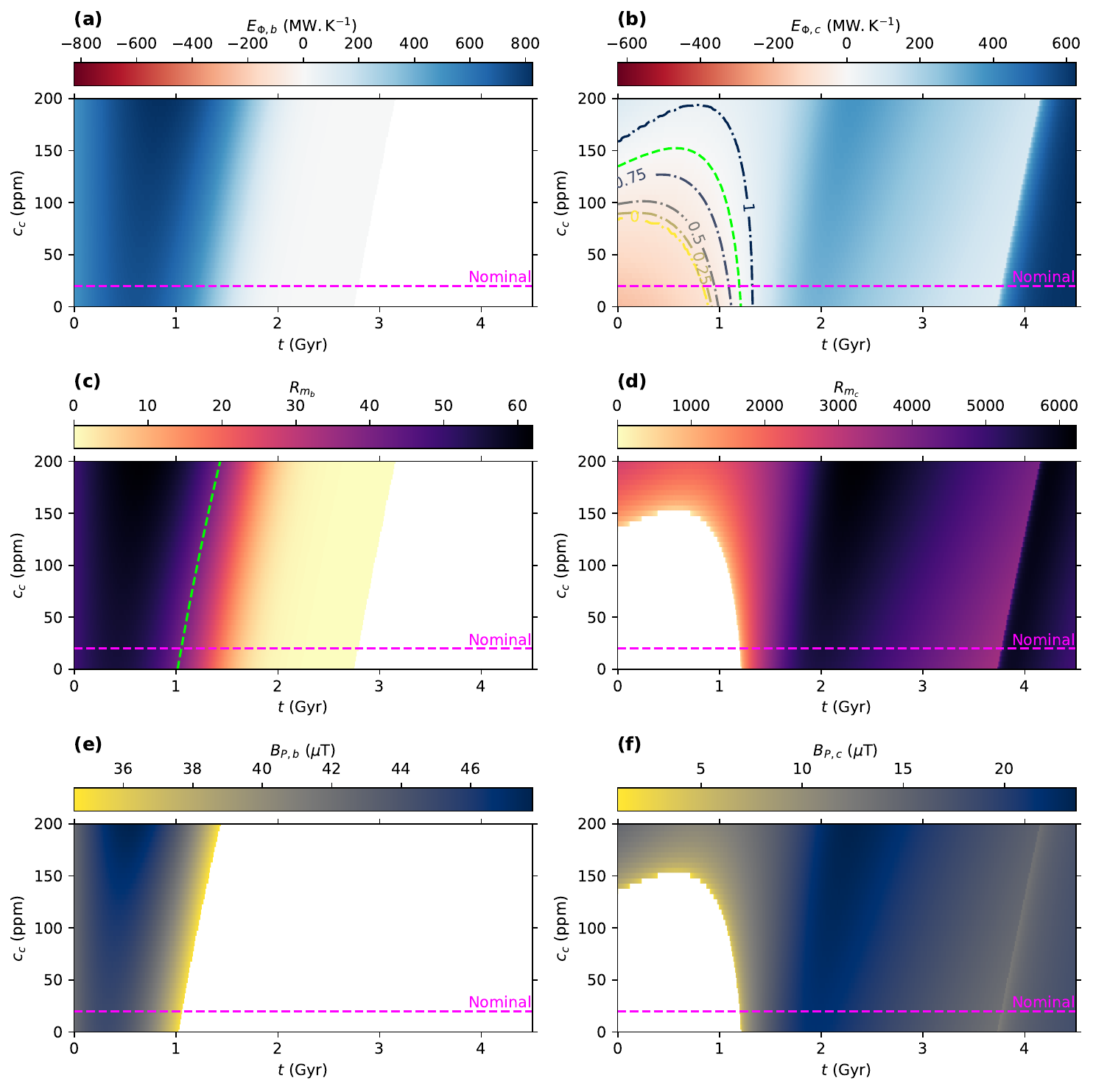}
    \caption{Sensitivity of the entropy of dissipation $E_\Phi$ (a,b), magnetic Reynolds number $R_m$ (c,d) and surface magnetic field intensity $B_P$ (e,f) of the BMO (a,c,e) and the core (b,d,f) to the $\mathrm{^{40}K}$ concentration of the core $c_c$, as a function of time $t$. Green dashed lines correspond to $E_\Phi=0$ (a,b) and $R_m=40$ (c,d). Dash-dotted lines indicate the time when a specific radius fraction of the core is convecting, as determined by the convective stability criterion. Pink dashed lines correspond to the value of the parameter used in the nominal model. \\
    We conduct the parameter sweep for $0 \leq c_c \leq 200~\mathrm{ppm}$, which covers uncertainties on the value of $c_c$ \citep{hirose_2013}. When $c_c$ increases, the BMO crystallizes later, dynamo operation is protracted, and $E_\Phi$, $R_m$, and $B_P$ increase. This may be explained by an increased heat flow at the CMB. In the core, the increase of $c_c$ promotes a larger radiogenic heat flow, explaining an earlier dynamo onset and a delayed inner core nucleation. When corrected for the time shift related to inner core nucleation, $E_\Phi$, $R_m$, and $B_P$ are relatively independent of $c_c$.}
    \label{fig:c_K40_core_sweep}
\end{figure}

\begin{figure}[!ht]
    \centering
    \includegraphics[width=\linewidth]{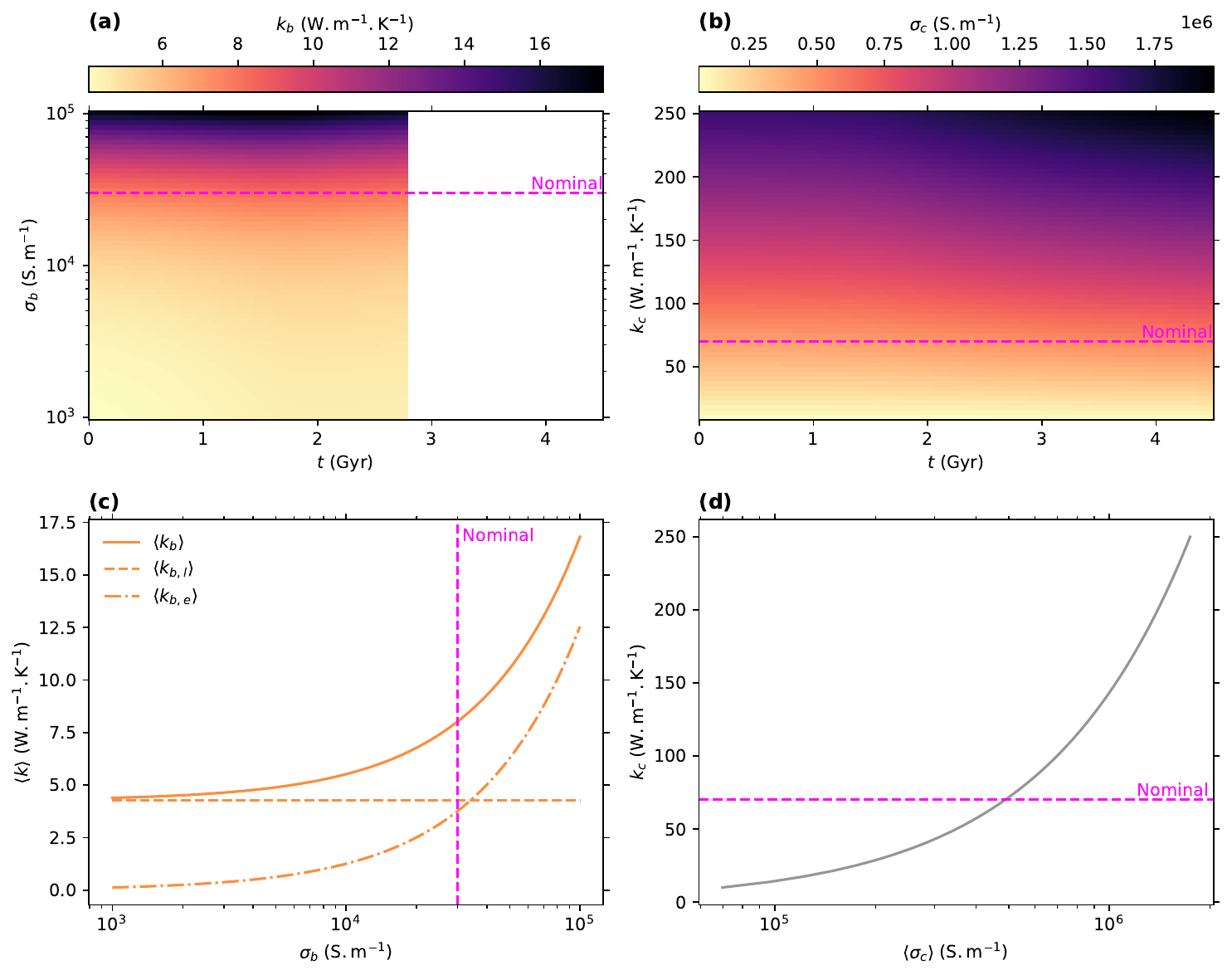}
    \caption{(a) Time evolution of the thermal conductivity $k_b$ of the BMO, as a function of the electrical conductivity $\sigma_b$. (b) Time evolution of the electrical conductivity $\sigma_c$ of the core, as a function of the thermal conductivity $k_c$. (c) Time-averaged total thermal conductivity $k_b$, ionic contribution $k_{b,l}$, and electronic contribution $k_{b,e}$ of the BMO as a function of the electrical conductivity $\sigma_b$. (d) Thermal conductivity $k_c$ as a function of the time-averaged electrical conductivity $\sigma_c$ of the core.}
    \label{fig:conductivity_sweep}
\end{figure}

\clearpage
\bibliographystyle{cas-model2-names}
\bibliography{biblio}